 %
\input harvmac.tex
%
%
 %
\catcode`@=11
\def\rlx{\relax\leavevmode}                   
 %
 %
 %
\font\eightrm=cmr8 \font\eighti=cmmi8 \font\eightsy=cmsy8
\font\eightbf=cmbx8 \font\eightit=cmti8 \font\eightsl=cmsl8
\skewchar\eighti='177 \skewchar\eightsy='60
\def\eightpoint{\def\rm{\fam0\eightrm}
\textfont0=\eightrm \scriptfont0=\sixrm \scriptscriptfont0=\fiverm
\textfont1=\eighti \scriptfont1=\sixi \scriptscriptfont1=\fivei
\textfont2=\eightsy \scriptfont2=\sixsy \scriptscriptfont2=\fivesy
\textfont\itfam=\eighti
\def\it{\fam\itfam\eightit}\def\sl{\fam\slfam\eightsl}%
\textfont\bffam=\eightbf \def\bf{\fam\bffam\eightbf}\rm}
 %
\font\tenmib=cmmib10
\font\sevenmib=cmmib10 at 7pt 
\font\fivemib=cmmib10 at 5pt  
\font\tenbsy=cmbsy10
\font\sevenbsy=cmbsy10 at 7pt 
\font\fivebsy=cmbsy10 at 5pt  
\def\BMfont{\textfont0\tenbf \scriptfont0\sevenbf
                              \scriptscriptfont0\fivebf
            \textfont1\tenmib \scriptfont1\sevenmib
                               \scriptscriptfont1\fivemib
            \textfont2\tenbsy \scriptfont2\sevenbsy
                               \scriptscriptfont2\fivebsy}
\def\BM#1{\rlx\ifmmode\mathchoice
                      {\hbox{$\BMfont#1$}}
                      {\hbox{$\BMfont#1$}}
                      {\hbox{$\scriptstyle\BMfont#1$}}
                      {\hbox{$\scriptscriptstyle\BMfont#1$}}
                 \else{$\BMfont#1$}\fi}
 %
 %
 %
 %
\def\inbar{\vrule height1.5ex width.4pt depth0pt}
\def\sinbar{\vrule height1ex width.35pt depth0pt}
\def\ssinbar{\vrule height.7ex width.3pt depth0pt}
\font\cmss=cmss10
\font\cmsss=cmss10 at 7pt
\def\ZZ{\rlx\leavevmode
             \ifmmode\mathchoice
                    {\hbox{\cmss Z\kern-.4em Z}}
                    {\hbox{\cmss Z\kern-.4em Z}}
                    {\lower.9pt\hbox{\cmsss Z\kern-.36em Z}}
                    {\lower1.2pt\hbox{\cmsss Z\kern-.36em Z}}
               \else{\cmss Z\kern-.4em Z}\fi}
\def\Ik{\rlx{\rm I\kern-.18em k}}  
\def\IC{\rlx\leavevmode
             \ifmmode\mathchoice
                    {\hbox{\kern.33em\inbar\kern-.3em{\rm C}}}
                    {\hbox{\kern.33em\inbar\kern-.3em{\rm C}}}
                    {\hbox{\kern.28em\sinbar\kern-.25em{\sevenrm C}}}
                    {\hbox{\kern.25em\ssinbar\kern-.22em{\fiverm C}}}
             \else{\hbox{\kern.3em\inbar\kern-.3em{\rm C}}}\fi}
\def\IP{\rlx{\rm I\kern-.18em P}}
\def\IR{\rlx{\rm I\kern-.18em R}}
\def\Ione{\rlx{\rm 1\kern-2.7pt l}}
 %
 %
\def\boxit#1#2{\hbox{\vrule\vbox{
  \hrule\vskip#1\hbox{\hskip#1\vbox{#2}\hskip#1}%
        \vskip#1\hrule}\vrule}}
\def\ytem{\par\hangindent\parindent\hangafter1\noindent\ignorespaces}
\def\intem#1{\par\leavevmode%
              \llap{\hbox to\parindent{\hss{#1}\hfill~}}\ignorespaces}
 %


 %
\newskip\humongous \humongous=0pt plus 1000pt minus 1000pt   
\def\caja{\mathsurround=0pt}
\newif\ifdtup
 %
\def\cmath#1{\,\vcenter{\openup2\jot \caja
     \ialign{\strut \hfil$\displaystyle{##}$\hfil\crcr#1\crcr}}\,}
 %
\def\eqalign#1{\,\vcenter{\openup2\jot \caja
     \ialign{\strut \hfil$\displaystyle{##}$&$
      \displaystyle{{}##}$\hfil\crcr#1\crcr}}\,}
 %
\def\twoeqsalign#1{\,\vcenter{\openup2\jot \caja
     \ialign{\strut \hfil$\displaystyle{##}$&$
      \displaystyle{{}##}$\hfil&\hfill$\displaystyle{##}$&$
       \displaystyle{{}##}$\hfil\crcr#1\crcr}}\,}
 %
\def\panorama{\global\dtuptrue \openup2\jot \caja
     \everycr{\noalign{\ifdtup \global\dtupfalse
      \vskip-\lineskiplimit \vskip\normallineskiplimit
      \else \penalty\interdisplaylinepenalty \fi}}}
 %
\def\cmathno#1{\panorama \tabskip=\humongous
     \halign to\displaywidth{\hfil$\displaystyle{##}$\hfil
       \tabskip=\humongous&\llap{$##$}\tabskip=0pt\crcr#1\crcr}}
 %
\def\eqalignno#1{\panorama \tabskip=\humongous
     \halign to\displaywidth{\hfil$\displaystyle{##}$
      \tabskip=0pt&$\displaystyle{{}##}$\hfil
       \tabskip=\humongous&\llap{$##$}\tabskip=0pt\crcr#1\crcr}}
 %
\def\eqalignnotwo#1{\panorama \tabskip=\humongous
     \halign to\displaywidth{\hfil$\displaystyle{##}$
      \tabskip=0pt&$\displaystyle{{}##}$
       \tabskip=0pt&$\displaystyle{{}##}$\hfil
        \tabskip=\humongous&\llap{$##$}\tabskip=0pt\crcr#1\crcr}}
 %
\def\twoeqsalignno#1{\panorama \tabskip=\humongous
     \halign to\displaywidth{\hfil$\displaystyle{##}$
      \tabskip=0pt&$\displaystyle{{}##}$\hfil
       \tabskip=0pt&\hfil$\displaystyle{##}$
        \tabskip=0pt&$\displaystyle{{}##}$\hfil
         \tabskip=\humongous&\llap{$##$}\tabskip=0pt\crcr#1\crcr}}
 %

 %
 %
 %
 %
   \let\SS=\S       
\let\ii=\i          
\def\,{\hskip1.5pt}           
 %
\let\a=\alpha
\let\b=\beta
\let\c=\chi
\let\d=\delta       \let\vd=\partial             
\let\e=\epsilon     
\let\f=\phi         \let\vf=\varphi              \let\F=\Phi
\let\g=\gamma                                    \let\G=\Gamma

\let\i=\iota
\let\j=\psi                                      

\let\l=\lambda                                   \let\L=\Lambda
\let\m=\mu
\let\n=\nu
\let\p=\pi          \let\vp=\varpi               \let\P=\Pi
\let\q=\theta       \let\vq=\vartheta            \let\Q=\Theta
\let\r=\rho         \let\vr=\varrho
\let\s=\sigma       \let\vs=\varsigma            \let\S=\Sigma
\let\t=\tau
                                    \let\W=\Omega
\let\x=\xi                                       \let\X=\Xi
                                  \let\Y=\Upsilon

 %
 %
\def\Box{{\sqcap\mkern-12mu\sqcup}}
\def\lapp{\lower.4ex\hbox{\rlap{$\sim$}} \raise.4ex\hbox{$<$}}
\def\gapp{\lower.4ex\hbox{\rlap{$\sim$}} \raise.4ex\hbox{$>$}}
\def\con{\ifmmode\raise.1ex\hbox{\bf*}
          \else\raise.1ex\hbox{\bf*}\fi}
\let\iff=\leftrightarrow

\let\from=\leftarrow
\let\To=\Rightarrow

\def\Imm{\mathop{\Im m}}
\def\Ree{\mathop{\Re e}}

\def\dual{\relax\leavevmode\lower.9ex\hbox{\titlerms*}}
\def\define{\buildrel\rm def\over =}
\let\id=\equiv
\let\8=\otimes
 %
 %
 %
 %
\let\ba=\overline
\let\2=\underline

\let\Tw=\widetilde
 %
\def\dt#1{{\buildrel{\smash{\lower1pt\hbox{.}}}\over{#1}}}
\def\pd#1#2{{\partial#1\over\partial#2}}

\def\6(#1){\relax\leavevmode\hbox{\eightrm(}#1\hbox{\eightrm)}}
\def\0#1{\relax\ifmmode\mathaccent"7017{#1}     
                \else\accent23#1\relax\fi}      
\def\7#1#2{{\mathop{\null#2}\limits^{#1}}}      
\def\5#1#2{{\mathop{\null#2}\limits_{#1}}}      
 %

 %

 %

 %

 %
\newbox\t@b@x
\def\rightarrowfill{$\m@th \mathord- \mkern-6mu
     \cleaders\hbox{$\mkern-2mu \mathord- \mkern-2mu$}\hfill
      \mkern-6mu \mathord\rightarrow$}
\def\tooo#1{\setbox\t@b@x=\hbox{$\scriptstyle#1$}%
             \mathrel{\mathop{\hbox to\wd\t@b@x{\rightarrowfill}}%
              \limits^{#1}}\,}
\def\leftarrowfill{$\m@th \mathord\leftarrow \mkern-6mu
     \cleaders\hbox{$\mkern-2mu \mathord- \mkern-2mu$}\hfill
      \mkern-6mu \mathord-$}
\def\froo#1{\setbox\t@b@x=\hbox{$\scriptstyle#1$}%
             \mathrel{\mathop{\hbox to\wd\t@b@x{\leftarrowfill}}%
              \limits^{#1}}\,}
 %
\def\frac#1#2{{#1\over#2}}
\def\frc#1#2{\relax\ifmmode{\textstyle{#1\over#2}} 
                    \else$#1\over#2$\fi}           
\def\inv#1{\frc{1}{#1}}                            
 %
\def\Claim#1#2#3{\bigskip\begingroup%
                  \xdef #1{\secsym\the\meqno}%
                   \writedef{#1\leftbracket#1}%
                    \global\advance\meqno by1\wrlabeL#1%
                     \noindent{\bf#2}\,#1{}\,:~\sl#3\vskip1mm\endgroup}

\def\QED{\rlx\hfill$\Box$\kern-7pt\raise3pt\hbox{$\surd$}\bigskip}
 %
 %
\def\1{\raise1pt\hbox{,}}     
\def\Tr{\mathop{\rm Tr}}
\def\:{\buildrel!\over=}
\def\ex#1{\hbox{$\>{\rm e}^{#1}\>$}}
\def\CP#1{\rlx\ifmmode\IP^{#1}\else\IP$^{#1}$\fi}
\def\cP#1{\rlx\ifmmode\IC{\rm P}^{#1}\else$\IC{\rm P}^{#1}$\fi}

\def\sll#1{\rlx\rlap{\,\raise1pt\hbox{/}}{#1}}
\def\Sll#1{\rlx\rlap{\,\kern.6pt\raise1pt\hbox{/}}{#1}\kern-.6pt}
\let\sss=\scriptscriptstyle
\let\SSS=\scriptstyle
\let\ttt=\textstyle

 %
\def\eg{\hbox{\it e.g.}}        
\def\ie{\hbox{\it i.e.}}        
\def\etc{\hbox{\it etc.}}       
\def\topic#1{\bigskip\noindent$\2{\hbox{#1}}$\nobreak\vglue0pt%
              \noindent\ignorespaces}

\def\CY{Calabi-\kern-.2em Yau}

\def\3{\ifmmode\ldots\else$\ldots$\fi}
\def\\{\hfill\break}
\def\Z{\hfil\break\rlx\hbox{}\quad}
\def\3{\ifmmode\ldots\else$\ldots$\fi}
\def\ping{\nobreak\par\centerline{---$\circ$---}\goodbreak\bigskip}
 %
 %

 %

 %

\def\NP#1{{\it Nucl.\,Phys.\,}{\bf#1\,}}
\def\PL#1{{\it Phys.\,Lett.\,}{\bf#1\,}}

\def\CQG#1{{\it Class.\,Quant.\,Grav.\,}{\bf#1\,}}

\baselineskip=13.0861pt plus2pt minus1pt
\parskip=\medskipamount
\let\ft=\foot
\noblackbox
 %
\def\Afour{\ifx\answ\bigans
            \hsize=16.5truecm\vsize=24.7truecm
             \else
              \hsize=24.7truecm\vsize=16.5truecm
               \fi}
 %
 %
\def\SaveTimber{\abovedisplayskip=1.5ex plus.3ex minus.5ex
                \belowdisplayskip=1.5ex plus.3ex minus.5ex
                \abovedisplayshortskip=.2ex plus.2ex minus.4ex
                \belowdisplayshortskip=1.5ex plus.2ex minus.4ex
                \baselineskip=12pt plus1pt minus.5pt
 \parskip=\smallskipamount
 \def\ft##1{\unskip\,\begingroup\footskip9pt plus1pt minus1pt\setbox%
             \strutbox=\hbox{\vrule height6pt depth4.5pt width0pt}%
              \global\advance\ftno by1
               \footnote{$^{\the\ftno)}$}{\ninepoint##1}%
                \endgroup}}
\catcode`@=12
%
%
%
\input epsf.tex
\SaveTimber  
 %
 %
\def\rd{{\rm d}}
\def\hc{\hbox{\it h.c.}}
\def\Zp{\big|}
\def\?{\,\hbox{--}}

\def\@{{\ttt\cdot}}
\def\T{{}^{\sss\rm T}}
\def\pp{{\mathchar'75\mkern-9mu|\mkern3mu}} 
\def\mm{{=}}                                
\def\Pp{{\id\mkern-9.5mu|\mkern3mu}}        
\def\Mm{{\id}}                              
\def\dvd{\!{\mathop{\smash\partial\vrule height1.3ex width0pt}%
          \limits^{\,_{\SSS\iff}}}\mkern-4mu}
\def\B#1{\mathaccent"7616{#1}}
\def\]#1{\mkern#10mu}
\def\[#1{\mkern-#10mu}
\def\N{\nabla}
\def\bN{\overline\nabla}
 %
\def\ab{{\mathaccent"7616\alpha}}
\def\ra{{\rm a}}

\def\ad{{\dot\a}}

\def\Ab{{\mathaccent"7616{A}}}
\def\bA{{\bf A}}
\def\bAb{\relax\leavevmode\hbox{{\bf A}\kern-.7em
               \vrule height1.9ex depth-1.8ex width5pt}\kern2pt}
\def\bb{{\mathaccent"7616\beta}}

\def\Bb{{\mathaccent"7616{B}}}
\def\bB{{\bf B}}
\def\bBb{\relax\leavevmode\hbox{{\bf B}\kern-.7em
               \vrule height1.9ex depth-1.8ex width5pt}\kern2pt}
\def\cb{{\mathaccent"7616\chi}}

\def\Cb{\relax\leavevmode\hbox{$C$\kern-.53em
               \vrule height1.9ex depth-1.8ex width5pt}\kern.5pt}
\def\bC{{\bf C}}
\def\bCb{\relax\leavevmode\hbox{{\bf C}\kern-.65em
               \vrule height1.9ex depth-1.8ex width5pt}\mkern2mu}

\def\bCC{\relax\leavevmode\hbox{\BM{\cal C}\kern-.5em
               \vrule height1.9ex depth-1.8ex width5pt}\kern.8pt}
\def\db{{\mathaccent"7616\partial}}

\def\Db{\relax\leavevmode\hbox{$D$\kern-.6em
               \vrule height1.9ex depth-1.8ex width5pt}\kern.5pt}
\def\bD{{\bf D}}

\def\cDb{\relax\leavevmode\hbox{$\cal D$\kern-.6em
               \vrule height1.9ex depth-1.8ex width5pt}\kern.5pt}
\def\rD{{\rm D}}
\def\rDb{\relax\leavevmode\hbox{D\kern-.7em
               \vrule height1.9ex depth-1.8ex width5pt}\kern1.5pt}
\def\CE{\BM{\cal E}}
\def\bCE{\relax\leavevmode\hbox{\BM{\cal E}\kern-.5em
               \vrule height1.9ex depth-1.8ex width4pt}\kern.8pt}
\def\tCE{\skew4\tilde{\BM{\cal E}}}
\def\Fb{\relax\leavevmode\hbox{$F$\kern-.55em
               \vrule height1.9ex depth-1.8ex width4.5pt}\kern1pt}
\def\FB{\relax\leavevmode\hbox{$\Phi$\kern-.6em
               \vrule height1.9ex depth-1.8ex width4.5pt}\kern1.5pt}
\def\bF{{\bf F}}
\def\rF{{\rm F}}
\def\Bf{\BM{f}}
\def\Bfb{\skew5\B{\BM{f}}}
\def\gb{{\mathaccent"7616\gamma}}
\def\bg{{\mkern4mu\mathchar"2D\mkern-10mu\gamma\mkern-1mu}}

\def\rg{{\rm g}}
\def\Gb{\relax\leavevmode\hbox{$\Gamma$\kern-.55em
               \vrule height1.9ex depth-1.8ex width4.5pt}\kern1pt}

\def\bG{{\mathchar"2D\mkern-7mu\Gamma\mkern-1mu}}
\def\GB{{\bf\Gamma}}
\def\bGB{\overline{\bf\Gamma}}
\def\iGb{\relax\leavevmode\hbox{$\mit\Gamma$\kern-.55em
               \vrule height1.9ex depth-1.8ex width4.5pt}\kern1pt}
\def\cG{{\cal G}}
\def\CG{\BM{\cal G}}
\def\tCG{\skew4\tilde{\BM{\cal G}}}
\def\CH{\BM{\cal H}}
\def\bCH{\relax\leavevmode\hbox{\BM{\cal H}\kern-.7em
               \vrule height1.9ex depth-1.8ex width6pt}\kern.8pt}
\def\fb{{\mathaccent"7616\phi}}
\def\bi{{\bar\imath}}

\def\bj{{\bar\jmath}}
\def\jb{{\B\j}}

\def\JB{\relax\leavevmode\hbox{$\Psi$\kern-.7em
               \vrule height1.9ex depth-1.8ex width6pt}\kern1.5pt}

\def\bCK{\relax\leavevmode\hbox{\BM{\cal K}\kern-.7em
               \vrule height1.9ex depth-1.8ex width6pt}\kern.8pt}

\def\bl{{\mathchar"16\mkern-10mu\lambda}} 

\def\LB{\relax\leavevmode\hbox{$\Lambda$\kern-.6em
               \vrule height1.9ex depth-1.8ex width5pt}\kern1.5pt}
\def\mb{{\mathaccent"7616\mu}}

\def\Mb{\relax\leavevmode\hbox{$M$\kern-.8em
               \vrule height1.9ex depth-1.8ex width7pt}\kern.8pt}

\def\nb{{\mathaccent"7616\nu}}

\def\cO{{\cal O}}

\def\pB{\relax\leavevmode\hbox{$\BMfont p$\kern-.45em
               \vrule height1.35ex depth-1.25ex width4pt}\kern1pt}
\def\Pb{\relax\leavevmode\hbox{$P$\kern-.55em
               \vrule height1.9ex depth-1.8ex width4.5pt}\kern.5pt}
\def\PB{\relax\leavevmode\hbox{$\P$\kern-.6em
               \vrule height1.9ex depth-1.8ex width4.5pt}\kern.5pt}

\def\qB{\relax\leavevmode\hbox{$\BMfont q$\kern-.45em
               \vrule height1.35ex depth-1.25ex width4pt}\kern1pt}

\def\Qb{\relax\leavevmode\hbox{$Q$\kern-.6em
               \vrule height1.9ex depth-1.8ex width5.5pt}\kern.5pt}
\def\QB{\relax\leavevmode\hbox{$\Q$\kern-.65em
               \vrule height1.9ex depth-1.8ex width5.0pt}\kern.5pt}

\def\rQb{\relax\leavevmode\hbox{Q\kern-.65em
               \vrule height1.9ex depth-1.8ex width5pt}\kern1.5pt}
\def\bQb{\relax\leavevmode\hbox{{\bf Q}\kern-.65em
               \vrule height1.9ex depth-1.8ex width5pt}\kern1.5pt}

\def\sb{{\mathaccent"7616\sigma}}
\def\vsb{\mathaccent"7616\varsigma\mkern2mu}

\def\Tb{\relax\leavevmode\hbox{$T$\kern-.55em
               \vrule height1.9ex depth-1.8ex width4.5pt}\kern.5pt}

\def\bV{{\bf V}}
\def\bVb{\relax\leavevmode\hbox{{\bf V}\kern-.65em
               \vrule height1.9ex depth-1.8ex width5pt}\kern1.5pt}
\def\CV{\BM{\cal V}}
\def\Wb{\relax\leavevmode\hbox{$W$\kern-.9em
               \vrule height1.9ex depth-1.8ex width7pt}\kern1.5pt}
\def\bW{{\bf W}}
\def\bWb{\relax\leavevmode\hbox{{\bf W}\kern-.95em
               \vrule height1.9ex depth-1.8ex width7pt}\kern1.5pt}

\def\xb{\mathaccent"7616\xi}
\def\bx{\skew3\B{x}}
\def\Xb{\relax\leavevmode\hbox{$X$\kern-.67em
               \vrule height1.9ex depth-1.8ex width6pt}\kern.5pt}
\def\bXB{\relax\leavevmode\hbox{\BM{X}\kern-.77em
               \vrule height1.9ex depth-1.8ex width6pt}\kern1.5pt}
\def\XB{\relax\leavevmode\hbox{$\X$\kern-.65em
               \vrule height1.95ex depth-1.85ex width6pt}\kern.5pt}
\def\YB{\relax\leavevmode\hbox{$\Upsilon$\kern-.6em
               \vrule height1.9ex depth-1.8ex width4.5pt}\kern1.5pt}

\def\ssl#1{\rlx\rlap{\,\raise1pt\hbox{$\backslash$}}{#1}}
\def\Ssl{\rlap{\kern1.2pt\raise1pt\hbox{\rm/}}{\hbox{$S$}}}
\def\sT{{/\mkern-9mu\vs}}
\def\sW{{\backslash\mkern-9mu\vs}}
\def\sL{{\vs_{{}_L}}}
\def\sR{{\vs_{{}_R}}}
\def\LL{{\!_R}}
\def\RR{{\!_L}}

\font\ff=cmff10 at 11pt
\def\FF#1{\relax\hbox{\ff#1}}
 %
\def\PixCap#1#2#3{\midinsert\vbox{\centerline{\epsfbox{#1}}%
                   \noindent\narrower{\bf Figure~#2}.~#3}\endinsert}
 %
 %
 %
\Title{\rightline{hep-th/9910007}}
      {\vbox{\centerline{Gauging Yang-Mills Symmetries}
              \vskip3mm
             \centerline{In 1+1-Dimensional Spacetime}}}
\centerline{\titlerms Raja Q.~Almukahhal
            and Tristan H\"ubsch\footnote{$^{\spadesuit}$}{On leave
            from the ``Rudjer Bo\v skovi\'c'' Institute, Zagreb, Croatia.}}
                                                             \vskip0mm
 \centerline{\it Department of Physics and Astronomy}        \vskip-.5mm
 \centerline{\it Howard University, Washington, DC~20059}    \vskip-.5mm
 \centerline{\tt thubsch\,@\,howard.edu}
\vfill

\centerline{ABSTRACT}\vskip2mm
\vbox{\narrower\narrower\baselineskip=12pt\noindent
We present a systematic and `from the ground up' analysis of the `minimal
coupling' type of gauging of Yang-Mills symmetries in (2,2)-supersymmetric
1+1-dimensional spacetime. Unlike in the familiar 3+1-dimensional $N=1$
supersymmetric case, we find several {\it distinct\/} types of minimal
coupling symmetry gauging, and so several distinct types of gauge
(super)fields, some of which entirely novel. Also, we find that certain
(quartoid) constrained superfields can couple to no gauge superfield at
all, others (haploid ones) can couple only very selectively, while still
others (non-minimal, \ie, linear ones) couple universally to all gauge
superfields. }

\Date{9/99. \hfill}  
\footline{\hss\tenrm--\,\folio\,--\hss}
 %
 %
 %
\lref\rBK{I.L.~Buchbinder and S.M.~Kuzenko: {\it Ideas and Methods of
        Supersymmetry and\Z Supergravity : Or a Walk Through Superspace},
        (IOP Publishing, Bristol, 1998).}

\lref\rTwJim{S.J.~Gates, Jr.: \PL{B352}(1995)43--49.}

\lref\rGGRS{S.J.~Gates, Jr., M.T.~Grisaru, M.~Ro\v cek and
       W.~Siegel: {\it Superspace}\Z (Benjamin/Cummings Pub.\ Co.,
       Reading, Massachusetts, 1983).}

\lref\rGGW{S.J.~Gates, Jr., M.T.~Grisaru and M.E.~Wehlau:
       \NP{B460}(1996)579--614.}

\lref\rGHR{S.J.~Gates, Jr., C.M.~Hull and M.~Ro\v cek: \NP{B248}(1984)157.}

\lref\rHPS{C.M.~Hull, G.~Papadopoulos and B.~Spencer:
       \NP{B363}(1991)593-621.}

\lref\rBeast{T.~H\"ubsch: {\it \CY\ Manifolds---A Bestiary for
      Physicists}\Z (World Scientific, Singapore, 1992).}

\lref\rHSS{T.~H\"ubsch: Haploid (2,2)-Superfields In 2-Dimensional
      Spacetime. hep-th/9901038, \NP{B}(in press).}

\lref\rChiLin{T.~H\"ubsch: Linear and Chiral Superfields are Usefully
      Inequivalent. hep-th/9903175, \CQG{}(in press).}

\lref\rSSYM{T.~H\"ubsch: Yang-Mills and Supersymmetry Covariance are
      Sometimes Mutually Exclusive. work in progress.}

\lref\rWB{J.~Wess and J.~Bagger: {\it Supersymmetry and Supergravity}\Z
      (Princeton University Press, Princeton NJ, 1983).}

\lref\rPW{P.~West: {\it Introduction to Supersymmetry and Supergravity}\Z
      (World Scientific, Singapore, 1990).}

\lref\rWAB{E.~Witten: 
      in {\it Essays on Mirror Manifolds}, p.120, Ed.~S.-T.~Yau
      (International Press, Hong Kong, 1992).}

\lref\rPhases{E.~Witten: \NP{B403}(1993)159--222.}

 %
 %
\newsec{Introduction}\noindent
\seclab\sIntro
Ref.~\refs{\rHSS} presented an intrinsically 1+1-dimensional analysis, `from
scratch', of the basic building blocks in (2,2)-supersymmetric theories:
the superconstrained superfields, most often used to represent `matter'.
This analysis re-established some earlier
results~\refs{\rPhases,\rHPS,\rTwJim,\rGGW}, generalized them in several
different aspects, and uncovered some hitherto unknown phenomena and
mechanisms leading to several open questions and new research topics.

 While comprehensive in its study of the variously constrained `matter'
superfields, Ref.~\refs{\rHSS} postponed the study of gauge (super)fields
and their couplings to `matter'. This important issue is analyzed herein,
again from an {\it intrinsically\/} 1+1-dimensional approach, `from
scratch', and loosely following the methodology of \SS\,4.2.b of
Ref.~\refs{\rGGRS} (p.\,169). We find {\it two new types of symmetry
gauging\/}, besides the one that descends from 3+1-dimensional, $N{=}1$
supersymmetric theories by dimensional reduction and its mirror-twisted
version~\refs{\rHPS,\rTwJim}.

This article is organized as follows: The remaining part of this
section presents the basics of 1+1-dimensional (2,2)-superspace and sets up
the notation; further details are found in Appendix~A.
 The gauge-covariant (super)derivatives are defined and the field strength
superfields calculated in Section~2, with the details of Jacobi identity
calculations deferred to the Appendix~B.
 Section~3 analyzes the {\it most general\/} minimal coupling of gauge
superfields with `matter'.
 Some special types of symmetry gauging are discussed in Section~4.
 Section~5 discusses the component field content within the gauge
superfields, with details deferred to appendix~C.
 The choice of the Lagrangian density for `matter' superfields with gauged
symmetries is discussed in Section~6, generalizing the already immense
palette presented in Ref.~\refs{\rHSS}.
 Finally, section~7 summarizes the presented results and discusses some
further topics.

We emphasize that our purpose here is to enlist all logical possibilities,
and explore their consistency and major features, but leave open the
details and issues of application.

\subsec{$(1,1|2,2)$ superspacetime}\noindent
\subseclab\ssSpT
(Super)field theory in 1+1-dimensional spacetime is crucially distinct from
that in higher dimensions because of its Lorentz group, $SO(1,1)$: being
Abelian, all of its  irreducible representations are 1-dimensional. For
example, the coordinate 2-vector $(\s^0,\s^1)$ decomposes into the
(light-cone) characteristic coordinates
$\s^\mm{\define}\inv2(\s^0-\s^1)$ and $\s^\pp{\define}\inv2(\s^0+\s^1)$. 
 These are eigenfunctions of the Lorentz (boost) generator\ft{The eigenvalue
of the Lorentz boost operator, extended to include total angular momentum
in the usual way, equals the spin projection, and will therefore be denoted
by $j_3$ although it {\it does not stem\/} from its 3+1-dimensional
namesake. Moreover, all representations of the 1+1-dimensional Lorentz
group being 1-dimensional, `spin' and `spin projection' are one and the
same, and we will call $j_3$ simply `spin'.} and the group elements of
$SO(1,1)$:
\eqn\eXXX{{\twoeqsalign{
 \BM{B}(\s^\mm,\s^\pp)&=(-\s^\mm,+\s^\pp)~, \quad&\quad
 \BM{B}&\define(\s^1\vd_0{+}\s^0\vd_1)~,\cr
 \BM{U}_\a(\s^\mm,\s^\pp)&=(\ex{-i\a}\s^\mm,\ex{+i\a}\s^\pp)~, \quad&\quad
 \BM{U}_\a&\define \ex{i\a\BM{B}}~.\cr}}}
All other tensors (spinors) decompose similarly into 1-component objects. 
  Upon the frequently practiced analytic continuation $\s^0\to i\s^0$,
$(\s^\pp,\s^\mm)\to(z,-\B{z})$: the light-cone structure becomes a
complex structure and $\BM{U}\!_\a$ becomes the winding number (holomorphic
homogeneity) operator.
 The sub- and superscripts ``$\pm$'' then simply denote the winding number
(spin in real time) in units of $\inv2$($\hbar$).
 Functions depending on only $\s^\pp$ or only $\s^\mm$ are called left- and
right-movers, respectively, and become holomorphic (complex-analytic) and
anti-holomorphic (complex-antianalytic) functions upon analytic
continuation to imaginary time.

\subsec{(2,2)-superbasics}\noindent
\subseclab\ssBasic
The (2,2)-supersymmetry algebra involves the supersymmetry charges
$Q_\pm$ and $\Qb_\pm$, which satisfy (adapting from
Refs.~\refs{\rWB,\rPhases}; comparison with Refs.~\refs{\rGGW,\rGGRS} is
provided in appendix~A):
\eqn\eSusyQ{ \big\{\, Q_- \,,\, \Qb_- \,\big\}~=~-2i\vd_\mm~,
      \qquad \big\{\, Q_+ \,,\, \Qb_+ \,\big\}~=~-2i\vd_\pp~. }
Here $\vd_\pp{\define}\pd{}{\s^\pp}{=}(\vd_0{+}\vd_1)$ and
$\vd_\mm{\define}\pd{}{\s^\mm}{=}(\vd_0{-}\vd_1)$.
$p{=}-i\vd_1{=}-\frc{i}2(\vd_\pp{-}\vd_\mm)$ is the linear momentum
operator and $H{=}-i\vd_0{=}-\frc{i}2(\vd_\pp{+}\vd_\mm)$ is the Hamiltonian
(energy) operator.

Equipping the world-sheet with anticommuting fermionic coordinates,
$\vs^\pm,\vsb^\pm$, the supercharges are realized as differential
operators on the (super) world-sheet:
\eqn\eQs{{\twoeqsalign{
 Q_-&\define \vd_-+i\vsb^-\vd_\mm
 \quad&\quad
 \Qb_-&\define -\db_--i\vs^-\vd_\mm~, \cr
 Q_+&\define \vd_++i\vsb^+\vd_\pp
 \quad&\quad
 \Qb_+&\define -\db_+-i\vs^+\vd_\pp~, \cr
 }}}
The spinorial derivatives
\eqn\eDs{{\twoeqsalign{
 D_-&\define \vd_--i\vsb^-\vd_\mm
 \quad&\quad
 \Db_-&\define -\db_-+i\vs^-\vd_\mm~, \cr
 D_+&\define \vd_+-i\vsb^+\vd_\pp
 \quad&\quad
 \Db_+&\define -\db_++i\vs^+\vd_\pp~, \cr
 }}}
are covariant with respect to supersymmetry transformations:
$\{Q_\pm,D_\pm\}=0=\{\Qb_\pm,D_\pm\}$, whereupon
$[\BM{U}\!_{\e,\B\e},D_\pm]=0=[\BM{U}\!_{\e,\B\e},\Db_\pm]$,
with $\BM{U}\!_{\e,\B\e} \define
 \exp\big\{i(\e^\pm Q_\pm+\B\e\,^\pm\Qb_\pm)\big\}$.

 Finally the $D,\Db$'s close virtually the same algebra~\eSusyQ,
as do the $Q,\Qb$'s:
\eqn\eSusyD{ \big\{\, D_- \,,\, \Db_- \,\big\}~=~2i\vd_\mm~,
      \qquad \big\{\, D_+ \,,\, \Db_+ \,\big\}~=~2i\vd_\pp~. }
All anticommutators among the $Q_\pm,\Qb_\pm,D_\pm,\Db_\pm$, other
than~\eSusyQ\ and~\eSusyD, vanish.

Berezin superintegrals are by definition equivalent to partial
superderivatives, and up to total (world-sheet) spacetime
derivatives\ft{Hereafter, `total derivative' will stand for `total
(world-sheet) spacetime derivative'.} (which we ignore, assuming
world-sheets without boundaries) equivalent to covariant
superderivatives~\refs{\rWB,\rGGRS,\rPW,\rBK}. Following Ref.~\rHSS, we
use (see appendix~A for further definitions and conventions):
\eqn\eDI{ \int\rd^4\vs~(\3) ~\define~
 \inv8\big\{[D_-,\Db_-],[D_+,\Db_+]\big\}(\3)\Zp ~\define~(D^4\3)\Zp~, }
where ``$|$'' means setting $\vs^\pm=0=\vsb^\pm$.
 Integration over a fermionic subspace is formally achieved by inserting a
fermionic Dirac delta-function (see Appendix~A), so that:
\eqna\eHFI
 $$ \eqalignno{
 \int\rd^2\vs~[\3]  &\id \inv2\big[D_-,D_+\big][\3]\Zp~,\quad
 \int\rd^2\vsb~[\3]  \id \inv2\big[\Db_+,\Db_-\big][\3]\Zp~,&\eHFI{a,b}\cr
 \int\rd^2\sT~[\3]  &\id \inv2\big[D_-,\Db_+\big][\3]\Zp~,\quad
 \int\rd^2\sW~[\3]   \id \inv2\big[D_+,\Db_-\big][\3]\Zp~,  &\eHFI{c,d}\cr
 \int\rd^2\sR\,[\3] &\id \inv2\big[D_-,\Db_-\big][\3]\Zp~,\quad
 \int\rd^2\sL\,[\3]  \id \inv2\big[D_+,\Db_+\big][\3]\Zp~,  &\eHFI{e,f}\cr}
 $$
are integrals over the various halves of the fermionic coordinates.

\newsec{Gauge-Covariant Superderivatives}\noindent
\seclab\sSSC
In the presence of a gauge symmetry, all derivatives, including those in
the Berezin integrals~\eDI\ and~\eHFI{}, need to be `covariantized'. This
modifies the supersymmetry algebra~\eSusyD, and this modification is used
here as the starting point for the analysis.
 Somewhat formally, the Berezin integrals are easily turned into
gauge-covariant ones, by replacing $D$'s$~\to\N$'s; see appendix~A for
details.

\subsec{Definitions}\noindent
\subseclab\ssDefs
Much of the subsequent analysis follows the procedures used in
Ref.~\refs{\rGGRS}, although the notation will be adjusted to conform with
Refs.~\refs{\rPhases,\rWB}.
 We start by defining the covariant (super)derivatives
\eqna\eCDs
 $$\twoeqsalignno{
 \N_\pm&=D_\pm-i\GB_\pm~, \quad&\quad
 \bN_\pm&=\Db_\pm-i\bGB_\pm~, &\eCDs{a}\cr
 \N_\mm&=\vd_\mm-i\GB_\mm~, \quad&\quad
 \N_\pp&=\vd_\pp-i\GB_\pp~. &\eCDs{b}\cr
}$$
 The $\GB$'s are Lie algebra valued gauge {\it super\/}fields and are, in
general, linear combinations of gauge superfields for the direct summands
in the possibly non-simple Lie algebra.

 For brevity and convenience, the gauge coupling constants were
absorbed in the definition of the gauge superfields $\GB$; these can be
reinserted later by replacing $\GB\to g\GB$.
 Also, recall that the superderivatives~\eDs\ already include a connection
`$\inv2$-form coefficient'. For example, $D_-$ and $\Db_-$ include
$\vsb^-\vd_\mm$ and $\vs^-\vd_\mm$ as the ({\it derivative-valued!\/})
`$\inv2$-form coefficients': spacetime derivatives here play the r\^ole of
the generators of the spacetime translation group, and the (fermionic)
`gauge superfields' here are simply the supercoordinates $\vsb^-$ and
$\vs^-$.

 The (super)derivatives~\eCDs{} are all covariant with respect to the
general gauge transformation:
\eqn\eCov{ \N' ~=~ \CG\,\N\,\CG^{-1}~,\qquad
           \BM{X}' ~=~ \CG\BM{X}~, }
where \BM{X} represents any (homogeneously transforming) `matter'
(super)field, and \CG\ is the operator implementing the gauge
transformation. Typically, we assume the gauge transformation operator, \CG,
to be unitary, so that the transformation of the `matter' superfield~\eCov\
implies
\eqn\eCoC{ \bXB' ~=~ \bXB\CG^{\dag} ~=~ \bXB\CG^{-1}~, }
whence preserving the norm (squared),
\eqn\eNrm{ \|\BM{X}\|^2~\define~\Tr\big(\bXB_i\BM{X}^i\big)
            ~=~\Tr\big(\BM{X}^i\bXB_i\big)~, }
where $i$ simply counts the superfields $\BM{X}^i$, {\it each\/} of which
is a collection of superfields forming a given representation of the gauge
group.
 Because of the transposition involved in~\eCoC, the action of
gauge-covariant derivatives on $\bXB$ is awkward: the derivative part of
$\N$ should act from the left as usual, but the gauge superfield part
should act from the right. We therefore calculate (implicitly) using a
double transposition:
\eqn\eTTr{ (\cO\bXB)~\define~(\cO\T\bXB\T)\T~=~(\B\cO\BM{X})^\dagger~, }
where $\cO$ denotes {\it any\/} gauge-covariant operator,
$\cO^*$ and $\cO\T$ its {\it complex\/} conjugate and transpose,
respectively; over-bar and dagger interchangeably denote {\it Hermitian\/}
conjugation: $\B\cO\id\cO^\dagger$. In practical calculation, and in cases
when above `matrix' notation would be ambiguous or confusing, we resort to
the explicit gauge group index notation. Assuming that the matter fields
form a representation of the gauge group the elements of which are indexed
by $\a,\b,\3$, Eqs.~\eCov,
\eCoC\ and~\eTTr\ become:
\eqn\eCovInd{ \N'_\a{}^\b ~=~ \CG_\a^\g\,\N_\g{}^\d\,\CG^{-1}{}_\d^\b~,
               \qquad\BM{X}'{}^\a ~=~ \CG_\b^\a\BM{X}^\b~, }
\eqn\eCoCInd{ \bXB_\a' ~=~ \bXB_\b\CG^{\dag}{}_\a^\b
                       ~=~ \bXB_\b\CG^{-1}{}_\a^\b~, }
and
\eqn\eTTrInd{ (\cO\bXB)_\a~\define~(\cO^\b_\a\bXB_\b)
              ~=~(\B\cO{}^\a_\b\BM{X}^\b)^\dagger~, }
respectively. This disentangles ordering issues and the `matrix' action of
the gauge fields on the matter fields: re-ordering now solely depends on
the spin/statistics of the involved superfields and operators. For most of
this article, however, we suppress explicit gauge group indices, hoping
that the Reader will always be able to discern the implied meaning of the
more compact implicit `matrix' notation.
 \ping

The gauge superfields, $\GB$, of course transform inhomogeneously:
\eqn\eInh{ \GB' = \CG\,\GB\,\CG^{-1} -i\CG^{-1}(D\CG)~. }
 Field strength superfields and torsion superfields, $\bF$ and
$\bf T$, are defined by (anti)commutation of the covariant
derivatives~\eCDs{}, according the general formula\ft{Here ``$[~{,}~\}$''
denotes the (anti)commutator, as appropriate for the (anti)commuted
quantities; see appendix~A for precise definition and use.}
\eqn\eFTs{ \big[\, \N \,,\, \N \,\big\}~=~{\bf T}{\cdot}\N -i\bF~, }
which determines $\bF,\bf T$ in terms of the gauge superfields $\GB$ upon
using Eqs.~\eCDs{}. Because of their definition~\eFTs, field strength
superfields and torsion superfields are also covariant: with respect to the
gauge transformation $\CG$, they transform just like the $\N$'s do~\eCov.

Next, we generalize the standard supersymmetry algebra by inserting the
so far unrestricted field strength superfields (choosing numerical
coefficients for later convenience):
\eqna\eTRs
 $$\twoeqsalignno{
 \{\N_-,\bN_+\}&\define\bAb~, \quad&\quad
 \{\bN_-,\N_+\}&\define\bA~,                                   &\eTRs{a}\cr
 \{\N_-,\N_+\}&\define\bBb~, \quad&\quad
 \{\bN_-,\bN_+\}&\define\bB~,                                  &\eTRs{b}\cr
 \{\bN_-,\bN_-\}&\define2\bC_\mm~, \quad&\quad
 \{\bN_+,\bN_+\}&\define2\bC_\pp~,                             &\eTRs{c}\cr
 \{\N_-,\bN_-\}&=2i\N_\mm~,\quad&\quad
 \{\N_+,\bN_+\}&=2i\N_\pp~,                                    &\eTRs{d}\cr
 [\N_-,\N_\mm]&\define-\bWb_\Mm~,\quad&\quad
 [\bN_-,\N_\mm]&\define-\bW_\Mm~,                              &\eTRs{e}\cr
 [\N_+,\N_\mm]&\define-\bWb_-~, \quad&\quad
 [\bN_+,\N_\mm]&\define-\bW_-~,                                &\eTRs{f}\cr
 [\N_-,\N_\pp]&\define+\bWb_+~, \quad&\quad
 [\bN_-,\N_\pp]&\define+\bW_+~,                                &\eTRs{g}\cr
 [\N_+,\N_\pp]&\define+\bWb_\Pp~, \quad&\quad
 [\bN_+,\N_\pp]&\define+\bW_\Pp~,                              &\eTRs{h}\cr
 [\N_\mm,\N_\pp]&\define-i\bF~, \quad&\quad
 \{\N_\mm,\N_\pp\}&\define+2\Box~,                             &\eTRs{i}\cr
 }$$
 where $\Box$ is the {\it gauge-covariant\/} d'Alembertian (wave operator).
Notice that the field strengths $\bA,\bB,\bC,\bW$ are all complex\ft{Being
first order bosonic derivatives, the $\N_\mm,\N_\pp$ are {\it
antihermitian\/}; the above definitions of the $\bWb$'s then ensure them to
be the hermitian conjugates of the $\bW$'s.}, the $\bCb$'s being defined by
the conjugates of Eqs.~\eTRs{c}, and no new field strength has been
introduced in Eqs.~\eTRs{d}, as it would merely redefine the gauge
superfields $\GB_\mm,\GB_\pp$. Also, all torsion vanishes, except for
$T_{--}^\mm=2i=T_{++}^\pp$ as it appears already in Eqs.~\eTRs{d}. This is
because the only derivative-valued connection 1-forms are within the
superderivatives, $D_\pm,\Db_\pm$, and so the torsion content is unchanged
by the covariantization~\eCDs{}.

The Reader familiar with the 3+1-dimensional supersymmetry algebra modified
by a gauge symmetry and its dimensional reduction to 1+1-dimensional
spacetime will realize that $\bA$ in~\eTRs{a} contains the two gauge
(super)field components that are transversal to the 1+1-dimensional
subspace. Naturally, the anticommutators in~\eTRs{a} yield no
torsion-and-spacetime derivative term, ${\bf T}{\cdot}\vd$, on the right
hand side, as these 3+1-dimensional spacetime derivatives are transversal
to the 1+1-dimensional spacetime of our interest and by the assumption of
dimensional reduction, everything is a constant in these transversal
directions.

Finally, note that the Berezin integration formulae~\eDI{} and~\eHFI{} also
need to be `covariantized'; see Appendix~A.

\subsec{Potential superfields}\noindent
\subseclab\ssPres
The covariant (super)derivatives~\eCDs{} must satisfy the graded Jacobi
identities. After straightforward algebra (see Appendix~B), we obtain the
following relationships:

Eight of the Jacobi identities involving the covariant
superderivatives~\eCDs{a} completely determine the fermionic field strength
superfields, $\bW$ and $\bWb$, in terms of the (gauge-covariant
superderivatives\ft{We follow the practice of writing $[\N,\BM{Y}\}$ for a
gauge-covariant (super)derivative a Lie algebra-valued superfield \BM{Y},
and $(\N\BM{X})$ for a gauge-covariant (super)derivative of a superfield
\BM{X} in any representation of the Lie group other than the adjoint.} of)
$\bA,\bAb,\bB,\bBb,\bC_\mm,\bC_\pp$:
\eqna\eWs
 $$\twoeqsalignno{
 \bW_{-}&=-\frc i2\big([\bN_-,\bAb]+[\N_-,\bB]\big)~,  \quad&\quad
 \bW_\Mm&=-\frc i2[\N_-,\bC_\mm]~,                        &\eWs{a}\cr
 \bWb_{-}&=-\frc i2\big([\N_-,\bA]+[\bN_-,\bBb]\big)~, \quad&\quad
 \bWb_\Mm&=-\frc i2[\bN_-,\bCb_\mm]~;                     &\eWs{b}\cr
 \bW_+&=\frc i2\big([\bN_+,\bA]+[\N_+,\bB]\big)~,      \quad&\quad
 \bW_\Pp&=\frc i2[\N_+,\bC_\pp]~,                         &\eWs{c}\cr
 \bWb_+&=\frc i2\big([\N_+,\bAb]+[\bN_+,\bBb]\big)~,   \quad&\quad
 \bWb_\Pp&=\frc i2[\bN_+,\bCb_\pp]~.                      &\eWs{d}\cr
}$$
We therefore call $\bA,\bAb,\bB,\bBb,\bC_\mm,\bC_\pp,\bCb_\mm,\bCb_\pp$ the
{\it gauge potential superfields\/}.

The four Jacobi identities involving three equal covariant
superderivatives impose the constraints
\eqna\eThC
 $$\twoeqsalignno{
 [\N_-,\bCb_\mm]&=0 \quad&\quad [\bN_-,\bC_\mm] &=0~, &\eThC{a,b}\cr
 [\N_+,\bCb_\pp]&=0 \quad&\quad [\bN_+,\bC_\pp] &=0~. &\eThC{c,d}\cr
}$$

Finally, the remaining eight Jacobi identities among the spinorial
superderivatives~\eCDs{a} imply the following curious relationships:
\eqna\eTwA
 $$ \twoeqsalignno{
 [\N_+,\bA]&=-[\bN_-,\bCb_\pp]~, &\qquad
 [\bN_-,\bA]&=-[\N_+,\bC_\mm]~,             &\eTwA{a}\cr
 [\N_-,\bAb]&=-[\bN_+,\bCb_\mm]~, &\qquad
 [\bN_+,\bAb]&=-[\N_-,\bC_\pp]~.            &\eTwA{b}\cr
 }$$
and:
\eqna\eChB
 $$\twoeqsalignno{
 [\bN_-,\bB]&=-[\bN_+,\bC_\mm]~, &\qquad
 [\bN_+,\bB]&=-[\bN_-,\bC_\pp]~, &\eChB{a}\cr
 [\N_-,\bBb]&=-[\N_+,\bCb_\mm]~, &\qquad
 [\N_+,\bBb]&=-[\N_-,\bCb_\pp]~, &\eChB{b}\cr
 }$$

Most of the Jacobi identities among two spinorial and one vectorial
covariant derivative are identically satisfied upon using the previous
identities\ft{They do provide (super)differential relations between field
strength superfields, such as
$[\N_\mm,\bA]=\{\bN_-,\bWb_-\}+\{\N_+,\bW_\Mm\}$,
$[\N_\mm,\bB]=\{\bN_-,\bW_-\}+\{\bN_+,\bW_\Mm\}$,
$[\N_\mm,\bCb_\mm]=\{\N_-,\bWb_\Mm\}$ and so on; for a complete listing,
see appendix~B.}. Two of them, however, express the bosonic field strength
superfield $\bF$ in terms of the spinorial field strengths, and in two {\it
different\/} ways:
\eqna\eFinW
 $$ \eqalignno{ 
 \bF &=+\inv2\big(\{\N_+,\bW_-\}+\{\bN_+,\bWb_-\}\big)~,    &\eFinW{a}\cr
 \bF &=+\inv2\big(\{\N_-,\bW_+\}+\{\bN_-,\bWb_+\}\big)~.    &\eFinW{b}\cr
 }$$
The half-sum and the difference of these produce the standard
results~\refs{\rGGRS}:
 $$ \eqalignno{ \bF 
 &=\inv4\big(\{\N_{(-},\bW_{+)}\}+\{\bN_{(-},\bWb_{+)}\}\big)~,
                                                   &\eFinW{c}\cr
 0&=\{\N_{[-},\bW_{+]}\}+\{\bN_{[-},\bWb_{+]}\}~,  &\eFinW{d}\cr
 }$$
where grouping of indices in parentheses indicates their symmetrization,
while bracketing indicates antisymmetrization. Eq.~\eFinW{d}, also known as
the `bisection formula' (Ref.~\refs{\rGGRS}, p.~158), is easy to rewrite in
the more familiar form, $\N^\a\bW_\a=\bN_\ad\bWb^\ad$ (Eq.~(6.12) in
Ref.~\rWB).

 As another consequence of the Jacobi identities, the spin $\pm\frc32$
field strengths $\bW_\Pp,\bW_\Mm$ are also related to their conjugates
through `bisection' formulae
 $${\eqalign{
 \{\N_-,\bW_\Mm\}+\{\bN_-,\bWb_\Mm\}&=0~, \cr
 \{\N_+,\bW_\Pp\}+\{\bN_+,\bWb_\Pp\}&=0~, \cr
 }} \eqno\eFinW{e}
 $$
Note that the summands in the bisection formula~\eFinW{d} have (total) spin
0, while those in~\eFinW{e} have (total) spin $\pm2$.

After using Eqs.~\eWs{}, \eFinW{c} becomes:
\eqna\eF
 $$\eqalignno{
 \bF         &=\inv2\big(\Bf+\Bfb\big)
              = \Ree(\Bf)~,                                    &\eF{a}\cr
 \Bf      &=\frc i4\Big(\big\{\N_-,[\bN_+,\bA]\big\}
                           -\big\{\bN_+,[\N_-,\bA]\big\}
                            +\big\{\N_{[-},[\N_{+]},\bB]\big\}\Big)~,\cr
             &~~~\id\frc i4\Big(\big([\N_-,\bN_+]\bA\big)
                            +\big([\N_-,\N_+]\bB\big)\Big)~,      &\eF{b}\cr
 \Bfb &=\frc i4\Big(\big([\bN_-,\N_+]\bAb\big)
                           +\big([\bN_-,\bN_+]\bBb\big)\Big)~.    &\eF{c}\cr
 }$$
Note that $\big([\N_-,\bN_+]\bA\big)$ is merely an abbreviation for the
antisymmetrized second gauge-covariant superderivative of the Lie
algebra-valued superfield $\bA$, written out in `long-hand' as
$\big\{\N_-,[\bN_+,\bA]\big\}-\big\{\bN_+,[\N_-,\bA]\big\}$.

We see that this field strength too is thus expressed entirely in terms of
the gauge potential superfields $\bA,\bAb,\bB,\bBb$. Notice, however, that
$\bF$ is not related to the $\bC,\bCb$'s, not even through the `curious
relations'~\eTwA{} and~\eChB{}! In fact, the `usual' bosonic field strengths
for the $\bC,\bCb$'s, the spin-$\pm2$ field strength superfields,
$\bF_{\mm\1\mm}$ and $\bF_{\pp\1\pp}$, vanish {\it identically\/}; see
appendix~B. From the definition of $\bF$ in Eq.~\eTRs{i}, it is clear that
its lowest component field is the standard Yang-Mills type field strength
(a.k.a.\ curvature). The independence of $\bF$ from the $\bC,\bCb$'s then
means that the type of extension of the supersymmetry algebra~\eTRs{c} does
not modify the standard Yang-Mills field strength.

The Jacobi identities among one spinorial and two vectorial covariant
derivatives are all identically satisfied upon using the earlier
identities. Finally, the Jacobi identities among three vectorial covariant
derivatives are identically satisfied as there are only two such
derivatives.

\subsec{Gauge potential superfields}\noindent
\subseclab\ssPP
Having started by introducing fermionic and bosonic gauge
superfields~\eCDs{}, we now determine the relationship between these and
the gauge potentials $\bA,\bB,\bC$ and their conjugates by expanding the
covariant derivatives in~\eTRs{}. We find:
\eqna\ePPs
 $$\eqalignno{
 \bA &= -i\big[\{\Db_-,\GB_+\}+\{D_+,\bGB_-\}
          -i\{\bGB_-,\GB_+\}\big]~, &\ePPs{a}\cr
 \bAb &= -i\big[\{D_-,\bGB_+\}+\{\Db_+,\GB_-\}
          -i\{\GB_-,\bGB_+\}\big]~, &\ePPs{b}\cr
 \bB &= -i\big[\{\Db_-,\bGB_+\}+\{\Db_+,\bGB_-\}
          -i\{\bGB_-,\bGB_+\}\big]~, &\ePPs{c}\cr
 \bBb &= -i\big[\{D_-,\GB_+\}+\{D_+,\GB_-\}
          -i\{\GB_-,\GB_+\}\big]~, &\ePPs{d}\cr
 \bC_\mm &= -i\big[\{\Db_-,\bGB_-\}
          -\frc{i}2\{\bGB_-,\bGB_-\}\big]~, &\ePPs{e}\cr
 \bC_\pp &= -i\big[\{\Db_+,\bGB_+\}
          -\frc{i}2\{\bGB_+,\bGB_+\}\big]~, &\ePPs{f}\cr
 \bCb_\mm &= -i\big[\{D_-,\GB_-\}
          -\frc{i}2\{\GB_-,\GB_-\}\big]~, &\ePPs{g}\cr
 \bCb_\pp &= -i\big[\{D_+,\GB_+\}
          -\frc{i}2\{\GB_+,\GB_+\}\big]~. &\ePPs{h}\cr
}$$
Finally, Eqs.~\eTRs{d} imply that
 $$\eqalignno{
 \GB_\mm &= -\frc{i}2\big[\{D_-,\bGB_-\}+\{\Db_-,\GB_-\}
          -i\{\GB_-,\bGB_-\}\big]~,  &\ePPs{i}\cr
 \GB_\pp &= -\frc{i}2\big[\{D_+,\bGB_+\}+\{\Db_+,\GB_+\}
          -i\{\GB_+,\bGB_+\}\big]~.  &\ePPs{j}\cr
}$$
Thus, both the vectorial gauge superfields $\GB_\mm,\GB_\pp$ from
Eqs.~\eCDs{b}, and all the gauge potential superfields $\bA,\bB,\bC$ and
their conjugates from Eqs.~\eTRs{a\?c} are completely defined in terms of
the fermionic gauge superfields $\GB_\pm,\bGB_\pm$ from
Eqs.~\eCDs{a}. Note that the lowest components of $\GB_\pm,\bGB_\pm$ appear
in the gauge potential and vectorial gauge superfields only through
the non-linear terms produced if the gauge Lie group is nonabelian. Also,
the lowest components of $\bA,\bB,\bC$, their conjugates and of
$\GB_\mm,\GB_\pp$ are all related only through these non-linear terms.

Since all gauge superfields are in the adjoint representation of the gauge
group, we can expand them over the generators, $T_i$, which satisfy
\eqn\eLie{ [T_j,T_k]~=~if_{jk}{}^lT_l~, }
where $f_{jk}{}^l$ are the structure constants. That is, $\GB=\GB^iT_i$,
where the $\GB^i$ are now simply (anti)commutative gauge fields, as
appropriate in~\eCDs{}. The expressions~\ePPs{} now become:
 $$\eqalignno{
 \bA &= -i\big[\{\Db_-,\GB_+^l\}+\{D_+,\bGB_-^l\}
          +\bGB_-^j\GB_+^kf_{jk}{}^l\big]T_l~, &\ePPs{a}'\cr
 \bAb &= -i\big[\{D_-,\bGB_+^l\}+\{\Db_+,\GB_-^l\}
          +\GB_-^i\bGB_+^kf_{jk}{}^l\big]T_k~, &\ePPs{b}'\cr
}$$
and so on.

Furthermore, it should be clear that any field strength superfield added to
the r.h.s\ of Eqs.~\eTRs{d} would again appear on the r.h.s\ of
Eqs.~\ePPs{i,j}, so that these additional field strength superfields {\it
and\/} the bosonic gauge superfields become two independent degrees of
freedom|for each gauge transformation. Standard wisdom (see Ref.~\rGGRS,
p.\,170) has it that this redundant duplication of gauge fields must be
avoided|as we did do by introducing no field strength superfield on the
r.h.s\ of Eqs.~\eTRs{d}.

In retrospect, the modifications of the standard supersymmetry algebra shown
in Eqs.~\eTRs{} and parametrized by the superfields
$\bA,\bB,\bC_\mm,\bC_\pp$, their conjugates and $\GB_\mm,\GB_\pp$ represent
{\it the most general\/} `minimal coupling' covariantization of
(2,2)-supersymmetric theories in 1+1-dimensional spacetime.

\newsec{Coupling to Constrained Matter}\noindent
\seclab\sCoupling
The above definitions provide gauge superfields, each of which contains
component fields which are gauge fields for some symmetry.
 One of the primary interests in gauge (super)fields is of course their
coupling to `matter', typically represented by some sort of constrained
(super)fields. A systematic list of such `matter' superfields was given in
Ref.~\refs{\rHSS}, defined to satisfy a (system of) superdifferential
equation(s).

Their coupling to the gauge superfields is induced by the `minimal
coupling' modification~\eCDs{}. Lagrangian densities (for a comprehensive
listing, see Ref.~\refs{\rHSS}) are defined as Berezin integrals, which in
turn are equivalent to (multiple) superderivatives. The
`covariantization'~\eCDs{} then indices a `covariantization' of the
Lagrangian densities. Upon expansion into component fields, these will then
exhibit explicit coupling terms between `matter' and gauge component fields.

However, in the presence of gauged symmetries, the defining
(super)constraints themselves must be modified, and we first explore the
consequences of this modification.

\subsec{Minimal gauge-covariantly haploid superfields}\noindent
\subseclab\ssMHSF
Adapting from Ref.~\refs{\rHSS}, we recall the definition of the minimal
(first-order constrained) {\it gauge-covariantly haploid\/} superfields:
\eqna\eHSF
 $$ \twoeqsalignno{
 \[61. & \hbox{\bf Chiral}:
 &\qquad  & (\bN_+\F)~ = ~0~ = ~(\bN_-\F)~,    &\eHSF{a} \cr
 \[62. & \hbox{\bf Antichiral}:
 &\qquad  & (\N_+\FB)~ = ~0~ = ~(\N_-\FB)~,    &\eHSF{b} \cr
 \[63. & \hbox{\bf Twisted-chiral}:
 &\qquad  & (\N_+\X)~ = ~0~ = ~(\bN_-\X)~,     &\eHSF{c} \cr
 \[64. & \hbox{\bf Twisted-antichiral}:
 &\qquad  & (\bN_+\XB)~= ~0~ =~(\N_-\XB)~,     &\eHSF{d} \cr
 \[65. & \hbox{\bf Lefton}:
 &\qquad  & (\N_-\L)~ = ~0~ = ~(\bN_-\L)~,     &\eHSF{e} \cr
 \[66. & \hbox{\bf Righton}:
 &\qquad  & (\N_+\Y)~ = ~0~ = ~(\bN_+\Y)~.     &\eHSF{f} \cr}
 $$
It is easy to show that the equations~\eHSF{} are gauge-covariant:
\eqn\eXXX{ (\bN_\pm\F)\to(\bN'_\pm\F')
 =(\cG\bN_\pm\cG^{-1}\cG\F)=\cG(\bN_\pm\F)~, }
so that
\eqn\eXXX{ (\bN_\pm\F)=0~\to~\cG\,(\bN_\pm\F)=0~. }
Note that the above constraints {\it imply\/} that gauge-covariantly haploid
superfields commute with the covariant derivatives which annihilate them;
for example:
\eqn\eXXX{ (\bN_+\F)~ = ~0~\qquad\To\qquad [\bN_+,\F]~=~0~. }
This `commutator form' of the constraints will be necessary when
reconsidering the above superfields as multiplicative {\it operators\/} in
quantum theory, in which case also the second of Eqs.~\eCov\ needs to read
$\BM{X}'=\cG\BM{X}\cG^{-1}$, where \BM{X} stands for any multiplicative
superfield {\it operator\/}. Of course, when \BM{X} is in the adjoint
representation, it makes perfect sense to regard it as an operator even as
a classical superfield, and expand it over the gauge group generators,
$T_i$.

A simple argument~\refs{\rTwJim} shows that a gauge-covariantly chiral
superfield, $\F$, cannot be charged with respect to that part of
the gauged Lie algebra in which the $\bB$'s take value:
\eqna\eCNB
 $$\twoeqsalignno{ 
 \matrix {(\bN_-\F)=0\cr\noalign{\vglue2mm}(\bN_+\F)=0\cr}\bigg\}&~\To~
 &\big(\{\bN_-,\bN_+\}\,\F\big)&=~0~,              &\eCNB{a}\cr
 \hbox{thus, using Eq.}&\eTRs{b}:& (\bB\,\F)&=~0~. &\eCNB{b}\cr
}$$
That is, the Lie group generators in which the $\bB$'s take values
annihilate gauge-covariantly (anti)chiral superfields. But, as the
generators of a Lie group are Hermitian, the $\bBb$'s are valued in the
same generators as are the $\bB$'s, and we obtain:
 $$
 (\bB\,\F)=0=(\bBb\,\F)~,\quad\hbox{and}\quad(\bBb\,\FB)=0=(\bB\,\FB)~.
                                                             \eqno\eCNB{c}
 $$

Similarly, gauge-covariantly twisted-(anti)chiral superfields are
annihilated by those Lie group generators in which the $\bA,\bAb$'s take
values:
\eqna\eTNA
 $$\twoeqsalignno{
 \matrix {(\bN_-\X)=0\cr \noalign{\vglue2mm} (\N_+\X)=0\cr}\bigg\}&~\To
 &\big(\{\bN_-,\N_+\}\,\X\big)&=~0~,               &\eTNA{a}\cr
 \hbox{thus, using Eq.}&\eTRs{a}:& (\bA\,\X)&=~0~. &\eTNA{b}\cr
}$$
Again, as for Eqs.~\eCNB{c}, it follows that
 $$
 (\bA\,\X)=0=(\bAb\,\X)~, \quad\hbox{and}\quad (\bAb\,\XB)=0=(\bA\,\XB)~.
\eqno\eTNA{c}
 $$

By a similar calculation, covariant leftons are annihilated by the
$\bC_\mm,\bCb_\mm$'s:
\eqna\eLNC
 $$\twoeqsalignno{
 (\N_-\,\L)&=0~, \quad&\To\quad
  (\N_-^{~2}\,\L)=(\bCb_\mm\,\L)&=~0~, &\eLNC{a}\cr
 (\bN_-\,\L)&=0~, \quad&\To\quad
  (\bN_-^{~2}\,\L)=(\bC_\mm\,\L)&=~0~, &\eLNC{b}\cr
}$$
and covariant rightons by the $\bC_\pp,\bCb_\pp$'s:
\eqna\eRNC
 $$\twoeqsalignno{
 (\N_+\,\Y)&=0~, \quad&\To\quad
  (\N_+^{~2}\,\Y)=(\bCb_\pp\,\Y)&=~0~, &\eRNC{a}\cr
 (\bN_+\,\Y)&=0~, \quad&\To\quad
  (\bN_+^{~2}\,\Y)=(\bC_\pp\,\Y)&=~0~. &\eRNC{b}\cr
}$$
In fact, the $\bC_\mm,\bCb_\mm$'s annihilate all but the covariant rightons:
\eqna\eANC
 $$\twoeqsalignno{
 (\bN_-\,\F)&=0 \quad\To\quad (\bC_=\,\F)\,&=~0&=(\bCb_\mm\,\F)~,   
&\eANC{a}\cr
 (\N_-\,\FB)&=0 \quad\To\quad (\bCb_=\,\FB)\,&=~0&=(\bC_\mm\,\FB)~, 
&\eANC{b}\cr
 (\bN_-\,\X)&=0 \quad\To\quad (\bC_=\,\X)\,&=~0 &=(\bCb_\mm\,\X)~,  
&\eANC{c}\cr
 (\N_-\,\XB)&=0 \quad\To\quad (\bCb_=\,\XB)\,&=~0 &=(\bC_\mm\,\XB)~.
&\eANC{d}\cr }$$
In all of these, the middle (second) equality is obtained by re-applying the
superderivative from the left (first) equality, and the right (third)
equality follows from the middle (second) one since the $\bC$'s and the
$\bCb$'s take values in the same, Hermitian, Lie group generators.
Similarly, the $\bC_\pp,\bCb_\pp$'s annihilate all but the covariant
leftons.

The last (third, left) equality in Eqs.~\eANC{} would seem to imply
additional constraints on the `matter' superfields, such as $\N_-^2\F=0$.
This, however, does not in the least restrict the gauge-covariantly haploid
superfields. In the vanishing gauge coupling constant limit, $\N_-^2\to
D_-^2\id0$, and the (would-be) additional constraints are vacuous. For any
non-zero gauge coupling constant, the superfields $\F$ are simply
$\bC_\mm$-chargeless. In particular, the first order superderivatives
$\N_\pm\F,\N_-\X,\bN_+\X$ and their conjugates remain unrestricted.

We have therefore proven that the minimal coupling type of interaction
between the gauge superfields~\eCDs{} and gauge-covariantly haploid `matter'
superfields~\eHSF{} is highly selective, as summarized in Table~1.

\midinsert
\vbox{\vglue2mm\noindent\hfill
 \vbox{\offinterlineskip
  \halign{
   &\vrule width0pt#&\strut~\hfil#~&\vrule width1pt#
                     &\strut~\hfil#\hfil~&\vrule#
                      &\strut~\hfil#\hfil~&\vrule#
                       &\strut~\hfil#\hfil~&\vrule#
                        &\strut~\hfil#\hfil~&\vrule width0pt#\cr
height2pt&\omit&&\omit&\omit&\omit&\omit&\omit&\omit&\omit&\cr
& &&$\bA,\bAb$&&$\bB,\bBb$&&$\bC_\pp,\bCb_\pp$&&$\bC_\mm,\bCb_\mm$&\cr
height2pt&\omit&&\omit&&\omit&&\omit&&\omit&\cr
\noalign{\hrule height1pt}
height3pt&\omit&&\omit&&\omit&&\omit&&\omit&\cr
&$\F,\FB$&
& $\SSS\surd$ && -- && -- && -- &\cr
height3pt&\omit&&\omit&&\omit&&\omit&&\omit&\cr
\noalign{\hrule}
height3pt&\omit&&\omit&&\omit&&\omit&&\omit&\cr
&$\X,\XB$&
& -- && $\SSS\surd$ && -- && -- &\cr
height3pt&\omit&&\omit&&\omit&&\omit&&\omit&\cr
\noalign{\hrule}
height3pt&\omit&&\omit&&\omit&&\omit&&\omit&\cr
&$\L,\LB$&
& $\SSS\surd$ && $\SSS\surd$ && $\SSS\surd$ && -- &\cr
height3pt&\omit&&\omit&&\omit&&\omit&&\omit&\cr
\noalign{\hrule}
height3pt&\omit&&\omit&&\omit&&\omit&&\omit&\cr
&$\Y,\YB$&
& $\SSS\surd$ && $\SSS\surd$ && -- && $\SSS\surd$ &\cr
height3pt&\omit&&\omit&&\omit&&\omit&&\omit&\cr}
}\hfill\nobreak\vglue1mm
\vbox{\narrower\noindent
{\bf Table 1}: The minimal coupling of gauge (super)fields~\eTRs{} to
               gauge-covariantly haploid `matter' superfields~\eHSF{} is
               highly selective: the entry `$\SSS\surd\,$' indicates that
               the minimal coupling type interaction is possible, and `--'
               that it is impossible.}}
\endinsert

The results summarized in Table~1 may be `translated' into the gauge group
index notation of Eqs.~\eCovInd--\eTTrInd\ as follows. The
gauge-covariantly chiral superfields, $\F^\m$ with $\m=1,\3,N_c$ must form
a representation\ft{We assume that the collections of the various
gauge-covariantly haploid superfields all form {\it irreducible\/}
representations. Reducible representations are then obtained simply by
including formal sums of the irreducible ones.} of $\cG_\bA$, the (factor
of the) gauge group gauged by the $\bA$'s. The gauge-covariantly twisted
chiral superfields, $\X^\a$ with $\a=1,\3,N_t$ must form a representation
of $\cG_\bB$, the (factor of the) gauge group gauged by the $\bB$'s. The
gauge-covariant leftons, $\L^a$ with $a=1,\3,N_L$ must form a
representation of $\cG_\bA$, of $\cG_\bB$ and of $\cG_\pp$, while the
gauge-covariant rightons, $\Y^i$ with $i=1,\3,N_R$ must form a
representation of $\cG_\bA$, of $\cG_\bB$ and of $\cG_\mm$. In turn, this
implies that the $\bA$'s must have three matrix representations:
$\bA_\n^\m$, $\bA_b^a$ and $\bA_j^i$, as must the $\bB$'s: $\bB_\b^\a$,
$\bB_b^a$ and $\bB_j^i$. The $\bC$'s on the other hand must only have one
each: $\bC_\pp{}_b^a$ and $\bC_\mm{}_j^i$. 

Thus, for example, $(\bA\F)^\m=\bA^\m_\n\F^\n$, but
$(\bA\L)^a=\bA^a_b\L^b$, and $(\bA\Y)^i=\bA^i_j\L^j$. Also,
$(\bA\FB)^\nb\define(\d^{\m\nb}\bA_\m^\r\d_{\r\sb})\F^{\sb}
 =\d^{\m\nb}\big(\bA_\m^\r(\d_{\r\sb}\F^{\sb})\big)$. 
Note that this is in perfect agreement with (and in fact clarifies) the
`matrix' notation~\eTTr. Raising and lowering indices with $\d^{\m\nb}$ and
$\d_{\r\sb}$, respectively, precisely corresponds to transposition:
\eqn\eXXX{ \FB\T~\iff~\FB_\r\define\d_{\r\sb}\F^\sb~,\quad\hbox{and}\quad
           (\3)\T~\iff~\d^{\m\nb}(\3)_\m~. }
As the matrix multiplication in $\bA_\m^\r\FB_\r$ is the transpose of that
in $\bA^\m_\r\F^\r$,
\eqn\eTTrIdx{
(\bA\FB)^\nb\define\d^{\m\nb}\big(\bA_\m^\r(\d_{\r\sb}\F^{\sb})\big)
 ~\iff~ (\bA\FB)\define(\bA\T\FB\T)\T~, }
re-derives Eq.~\eTTrInd.

\subsec{Covariantly quartoid superfields}\noindent
\subseclab\ssQSF
Besides the minimal (first-order constrained) gauge-covariantly haploid
superfields defined in Eqs.~\eHSF{}, Ref.~\refs{\rHSS} also lists
superfields which are constrained by three rather than
two first-order superderivatives. Adapting for gauge covariance, we then
define {\it gauge-covariantly quartoid\/} superfields:
\eqna\eQSF
$$ \twoeqsalignno{
 1.&\hbox{\bf Chiral Lefton}:
   &\qquad&(\bN_+\,\F_\LL)=(\bN_-\,\F_\LL)=(\N_-\,\F_\LL)=~0~,   
&\eQSF{a}\cr
 2.&\hbox{\bf Antichiral Lefton}:
   &\qquad&(\N_+\,\FB_\LL)=(\bN_-\,\FB_\LL)=(\N_-\,\FB_\LL)=~0~, 
&\eQSF{b}
\cr
 3.&\hbox{\bf Chiral Righton}:
   &\qquad&(\bN_-\,\F_\RR)=(\bN_+\,\F_\RR)=(\N_+\,\F_\RR)=~0~,   
&\eQSF{c}\cr
 4.&\hbox{\bf Antichiral Righton}:
   &\qquad&(\N_-\,\FB_\RR)=(\bN_+\,\FB_\RR)=(\N_+\,\FB_\RR)=~0~. 
&\eQSF{d} \cr}
$$
Notice that all of these may be regarded as gauge-covariantly chiral
superfields (or their hermitian conjugates) which so happen to satisfy an
additional gauge-covariant superconstraint.

All of these being (anti)chiral, they must have no charges with respect to
the gauge groups generated by the $\bB,\bC$'s and their conjugates.
However, each gauge-covariantly quartoid superfield obeys an additional
constraint, which then implies that the $\bA$-charges must be zero too. For
example, a chiral lefton satisfies both the constraints of a chiral
superfield, and those of a twisted-antichiral superfield. The former
preclude coupling to all but the type-A gauge superfields, while the latter
precludes even that.

\subsec{Non-minimal gauge-covariantly haploid superfields}\noindent
\subseclab\ssNHSF
Unlike the highly selective coupling of the minimal gauge-covariantly
haploid superfields to the various gauge (super)fields as displayed in
Table~1, non-minimal gauge-covariantly haploid superfields couple
universally~\rChiLin~!

Adapting from Ref.~\refs{\rHSS}, we recall the definitions of the
non-minimal (second-order constrained) {\it gauge-covariantly\/} haploid
superfields:
\eqna\eNMF
 $$ \twoeqsalignno{
 \[61. & \hbox{\bf NM-Chiral}:
 &\qquad  & \big([\bN_+,\bN_-]\,\Q\big)~ = ~0~, &\eNMF{a} \cr
 \[62. & \hbox{\bf NM-Antichiral}:
 &\qquad  & \big([\N_-,\N_+]\,\QB\big)~ = ~0~~, &\eNMF{b} \cr
 \[63. & \hbox{\bf NM-Twisted-chiral}:
 &\qquad  & \big([\N_+,\bN_-]\,\P\big)~ = ~0~,   &\eNMF{c} \cr
 \[64. & \hbox{\bf NM-Twisted-antichiral}:
 &\qquad  & \big([\N_-,\bN_+]\,\PB\big)~ = ~0~, &\eNMF{d} \cr
 \[65. & \hbox{\bf NM-(Almost)-Lefton}:
 &\qquad  & \big([\N_-,\bN_-]\,\BM{A}\big)~ =~0~,     &\eNMF{e} \cr
 \[66. & \hbox{\bf NM-(Almost)-Righton}:
 &\qquad  & \big([\N_+,\bN_+]\,\BM{U}\big)~ =~0~.     &\eNMF{f} \cr}
 $$
It is absolutely essential to realize that the quadratic superderivatives
in these defining constraints {\it must\/} be commutators, as this|and not
the anticommutator|has the correct vanishing coupling limit: when
$\N,\bN\to D,\Db$, the symmetric products of the superderivatives vanish
identically, except for Eqs.~\eSusyD.

It is indeed a trivial observation|antisymmetric and symmetric products
being linearly independent|that Eqs.~\eNMF{} imply {\it no exclusion\/} on
the minimal coupling type interaction between any of the $\bA,\bB,\bC$ and
their conjugates to any of the $\Q,\P,\BM{A},\BM{U}$ and their conjugates.
So, although in the {\it massless free-field limit\/} the physical component
fields of non-minimal superfields occur in a 1--1 correspondence with those
of the minimal ones (see p.200 of Ref.~\refs{\rGGRS}), their differing
couplings to the minimal coupling gauge (super)fields as defined in
Eqs.~\eCDs{} prove the physical inequivalence of the `minimal' haploid
superfields~\eHSF{} from their non-minimal brethren~\eNMF{}~\refs{\rChiLin}.

Similarly, the peculiar inhomogeneous (gauge)
transformation which enables the minimal/non-minimal 1--1 identification
even without a choice of the Lagrangian density~\refs{\rHSS} is manifestly
broken\ft{The definitions of the non-minimal {\it gauge-covariantly\/}
haploid superfields~\eNMF{} are invariant under this peculiar inhomogeneous
(gauge) transformation only in the limit of no gauge coupling.} by the
couplings to the gauge (super)fields defined in Eqs.~\eCDs{}.
 After all, what really matters in {\it nontrivial\/} physics models are
the types of {\it interactions\/} the various involved (super)fields can
have.

\ping
Unlike the $N{=}1$ supersymmetric 3+1-dimensional spacetime,
$(2,2)$-supersymmetric 1+1-dimensional spacetime admits different types of
symmetry gauging, and gauge-covariantly constrained superfields which
represent:
\item{1.} `matter' with no minimal-coupling gauge interaction:
gauge-covariantly quartoid superfields~\eQSF{};
\item{2.} `matter' with highly selective minimal-coupling gauge
interactions: minimal gauge-covariantly haploid superfields~\eHSF{} | see
Table~1;
\item{3.} `matter' with universal minimal-coupling gauge interactions:
non-minimal gauge-covariantly quartoid superfields~\eNMF{}.

Note that the above discussion focuses on the so-called `minimal coupling',
wherein the gauge (super)fields enter as connection superfields, through the
modification of the (super)derivative, \ie, (super)momentum
operator~\eCDs{}. Additional types of symmetry gauging are possible through
introduction of `Pauli-terms' or other higher-derivative terms; their
study is left to the interested Reader.

\newsec{Special Symmetry Gauging Types}\noindent
\seclab\sCases
With the consistency of the modifications~\eTRs{} established, and the
resulting constraints~\eTwA{}, \eChB{}, and relations~\eWs{}, \eFinW{}
and~\eF{} in hand, we now seek interesting special cases.

The selectivity of couplings between gauge-covariantly haploid superfields
and various gauge potentials shown in Table~1 suggests the following
simplification. Assume that the gauge group is of the form
$\cG=\cG_A{\times}\cG_B{\times}\cG_\mm{\times}\cG_\pp$, and
let the gauge potentials $\bA,\bB,\bC_\mm,\bC_\pp$ each take values in the
generators of the accordingly named factor. It is then possible to
project on any one direct product factor, effectively setting the generators
(charges) of all other factors to zero. This annihilates every (super)field
valued only in the generators of the complementary factor of the gauge
group. Therefore, projecting on a factor is equivalent to setting all
(super)fields valued in the complementary factor to zero, and this is how
we proceed.

\subsec{Type-A gauging}\noindent
\subseclab\ssA
Let us first project on the $\cG_A$ factor in the gauge group, \ie, set
$\bB=0=\bC$. The peculiar relations~\eChB{} now imply that
\eqn\eATW{ [\bN_-,\bA]=0=[\N_+,\bA]~,\quad\hbox{and}\quad
           [\N_-,\bAb]=0=[\bN_+,\bAb]~. }
That is, $\bA$ is gauge-covariantly twisted-chiral and $\bAb$
gauge-covariantly twisted-antichiral. Reassuringly, the conclusion of
Eqs.~\eTNA{} does not apply here. That is, Eqs.~\eTNA{} notwithstanding,
Eqs.~\eATW\ {\it do not imply\/} that $\bA$ must itself be chargeless with
respect to gauge group it generates which would preclude type-A gauging of
nonabelian groups. Repeating the calculation of Eqs.~\eTNA{}, however with
$\X\to\bA$, we now obtain
\eqn\eXXX{ 0\:[\bA,\bA] = i\bA^j\bA^kf_{jk}{}^lT_l~, }
which vanishes on account of the antisymmetry $f_{jk}{}^l=-f_{kj}{}^l$ and
the commutivity of the $\bA^j$'s. This allows $\bA$ to have nonzero
$\cG_A$-charges, as is indeed necessary when the gauge group $\cG_A$
is nonabelian.

\topic{Superfield strengths}
The formulae~\eWs{} and~\eF{} simplify:
\eqna\eWFA
 $$\cmathno{
 \bW_- = -\frc i2[\bN_-,\bAb]~,\qquad
 \bW_+ = \frc i2[\bN_+,\bA]~,  &\eWFA{a}\cr
 \bWb_- = -\frc i2[\N_-,\bA]~,\qquad
 \bWb_+ = \frc i2[\N_+,\bAb]~,  &\eWFA{b}\cr
}$$
and $\bF=\Ree(\Bf)$, with
 $$
 \Bf=\frc i4\big([\N_-,\bN_+]\bA\big)~,
 \eqno\eWFA{c}$$
and the $\bW_\Mm,\bW_\Pp$ and their conjugates vanish. The Jacobi
identities involving two spinorial and one vectorial $\N$ then produce the
familiar relations:
\eqna\eChW
 $$\twoeqsalignno{
 \{\bN_-,\bW_\pm\}&=0~, \quad&\quad \{\bN_+,\bW_\pm\}&=0~, &\eChW{a}\cr
 \{\N_-,\bWb_\pm\}&=0~, \quad&\quad \{\N_+,\bWb_\pm\}&=0~, &\eChW{b}\cr
}$$
proving that $\bW_\pm$ are gauge-covariantly chiral and $\bWb_\pm$ are
gauge-covariantly antichiral, related by the bisection
formula~\eFinW{d}|just as in
$N{=}1$ supersymmetric 3+1-dimensional spacetime theories.

We will refer to $\{\bA,\bAb;\bW_\pm,\bWb_\pm;\bF\}$, subject to the
superdifferential relations~\eWFA{}, as the type-A, or {\it twisted\/} gauge
multiplet, since the gauge potential superfields $\bA$ are gauge-covariantly
{\it twisted-chiral\/}. Notice however that the spinorial gauge field
strength superfields $\bW_\pm$ are gauge-covariantly {\it chiral\/}. Recall
also that these gauge fields {\it do not couple\/} to gauge-covariantly
twisted-chiral `matter' superfields and their conjugates~\eHSF{c,d}.

\topic{Dimensional reduction}
The type-A gauge supermultiplet, $\{\bA,\bAb;\bW_\pm,\bWb_\pm;\bF\}$, was
called VM-I in Ref.~\refs{\rTwJim} and was identified there with the gauge
supermultiplet obtained by dimensional reduction from $N{=}1$
supersymmetric 3+1-dimensional spacetime. Recall that all field strength
superfields with two fermionic indices are conventionally set to zero to
prevent the duplication of gauge field degrees of freedom per gauge symmetry
(see p.171--172 of Ref.~\refs{\rGGRS}). Appearances to the contrary, this
is not violated by the inclusion of $\bA,\bAb$ on the r.h.s.\ of
Eqs.~\eTRs{a}. 

To see this, consider the 1+1-dimensional spacetime (world-sheet) embedded
in a 3+1-dimensional one. Let $\N_\mm,\N_\pp$ be locally tangent to the
world-sheet, and $\N_j=\vd_j-i\GB_j$, with $j=2,3$, be transversal to it.
Dimensional reduction then implies that all (super)fields are set to be
annihilated by $\vd_j$, whereupon $\N_j\to-i\GB_j$. Finally, since
\eqn\eXXX{ \{D_\mp,\Db_\pm\}=2i(\N_2\mp i\N_3)~, }
a comparison with Eqs.~\eTRs{a} sets $\bA=2(\GB_2{+}i\GB_3)$. So, the
superfields $\bA$, which are scalars in the 1+1-dimensional sense, are in
fact a linear combination of the `transversal' components of the gauge
3+1-vector potential superfields. Also, although $\bA,\bAb$ are
gauge-covariant (being defined as the anticommutator of gauge-covariant
superderivatives), they are identified as gauge potential superfields,
and not as (super)field strengths.

 Furthermore, although $\bA,\bAb,\bW_\pm,\bWb_\pm,\bF$ are separate
(2,2)-superfields related through the superdifferential equations~\eWFA{},
it is convenient to regard them as jointly forming a gauge superfield
multiplet. Finally, the $\bA,\bAb$ are thus seen to be rightly regarded
as gauge potential superfields, which determine the field strength
superfields through Eqs.~\eWFA{}.

\topic{Abelian case}
A further simplification is easy to demonstrate when $\cG_A$ is abelian,
so that $\bA$ are $\cG_A$-chargeless, being Lie-algebra valued, \ie, in the
adjoint representation of $\cG_A$. Thus|{\it while acting on\/}
$\bA,\bAb$|the $\cG_A$-covariant (super)derivatives in Eqs.~\eATW\ act as
ordinary (super)derivatives~\eDs{}, $\bA$ and $\bAb$ become `plain'
twisted-chiral and twisted-antichiral superfields. These are then easily
expressed in terms of an unconstrained, {\it prepotential\/} superfield:
\eqn\eAinV{ \bA = \big([D_+,\Db_-]\,\bV^{(A)}\big)~,\quad\hbox{and}\quad
           \bAb = \big([D_-,\Db_+]\,\bVb^{(A)}\big)~. }

This produces
\eqn\eFAV{{\eqalign{\bF
 &=\frc i8\Big([D_-,\Db_+][D_+,\Db_-]\bV^{(A)}
              +[\Db_-,D_+][D_-,\Db_+]\bVb^{(A)}\Big)~,\cr
 &=\frc i2\big(D_-\Db_-\Db_+D_+\bV^{(A)}
              -\Db_-D_-D_+\Db_+\bVb^{(A)}\big)~, }}}
and
\eqna\eWAV
 $$\cmathno{
 \bW_- = -i\big(\Db_+\Db_-D_-\,\bVb^{(A)}\big)~,\qquad
 \bW_+ = -i\big(\Db_+\Db_-D_+\,\bV^{(A)}\big)~,  &\eWAV{a}\cr
 \bWb_- = -i\big(D_-D_+\Db_-\,\bV^{(A)}\big)~,\qquad
 \bWb_+ = -i\big(D_-D_+\Db_+\,\bVb^{(A)}\big)~,  &\eWAV{b}\cr
}$$
which are consistent\ft{Unlike \bF\ and the \bW's, the $\GB$'s transform
{\it inhomogeneously\/} under gauge transformations, and so remain
defined only up to additive terms stemming from this.} with
\eqn\eGAV{{\twoeqsalign{
 \GB_-&=-i(D_-\bVb^{(A)})~, \quad&\quad
 \bGB_-&=i(\Db_-\bV^{(A)})~, \cr
 \GB_+&=-i(D_+\bV^{(A)})~, \quad&\quad
 \bGB_+&=i(\Db_+\bVb^{(A)})~. \cr
 }}}
Finally, we must ensure that $\bB,\bBb$ do vanish as assumed. To this end,
substituting Eqs.~\eGAV\ into~\ePPs{a,b}, we see that $\bVb^{(A)}=\bV^{(A)}$
must be ensured, and no further restrictions transpire from the remaining
equations~\ePPs{}. This reproduces the well-known reality of the gauge
vector prepotential superfield, $\bV$. After a little $D$-algebra, we also
obtain
 $$
 \bF=2(\vd_\pp\bV^{(A)}_\mm-\vd_\mm\bV^{(A)}_\pp)~, \eqno\eFAV'
 $$
where
\eqn\eXXX{ \bV^{(A)}_\mm\define\inv4\big([D_-,\Db_-]\bV^{(A)}\big)~,\qquad
           \bV^{(A)}_\pp\define\inv4\big([D_+,\Db_+]\bV^{(A)}\big)~, }
so that the lowest component of $\bF$ in~\eFAV$'$ is the usual (abelian)
Yang-Mills field strength, in terms of the 2-vector potential obtained as
the lowest components of $\bV^{(A)}_\mm,\bV^{(A)}_\pp$.

\subsec{Type-B gauging}\noindent
\subseclab\ssB
Let us next project on the $\cG_B$ factor in the gauge group, \ie, set
$\bA=0=\bC$. The peculiar relations~\eTwA{} now imply that
\eqn\eBCH{ [\bN_\pm,\bB]=0~,\quad\hbox{and}\quad [\N_\pm,\bBb]=0~. }
That is, $\bB$ is {\it gauge-covariantly\/} chiral and $\bBb$
gauge-covariantly antichiral. Again, the conclusion of Eqs.~\eCNB{} does
not apply here. Repeating the calculation of Eqs.~\eCNB{}, with $\F\to\bB$,
we obtain
\eqn\eXXX{ 0\:[\bB,\bB] = i\bB^j\bB^kf_{jk}{}^lT_l~, }
which again vanishes on account of the antisymmetry of the $f_{jk}{}^l$'s
and the commutivity of the $\bB^j$'s, allowing $\bB$ to have nonzero
$\cG_B$-charges.

\topic{Superfield strengths}
The formulae~\eWs{} and~\eF{} again simplify:
\eqna\eWFB
 $$\cmathno{
 \bW_\mp =\mp\frc i2[\N_\mp,\bB]~,\qquad
 \bWb_\mp =\mp\frc i2[\bN_\mp,\bBb]~,  &\eWFB{a}\cr
 \bF=\Ree(\Bf)~,\qquad
 \Bf=\frc i4\big([\N_-,\N_+]\bB\big)~, &\eWFB{b}\cr
}$$
and $\bW_\Mm,\bW_\Pp$ and their conjugates vanish. The Jacobi
identities involving two spinorial and one vectorial $\N$ then produce the
`mirror' of the familiar relations~\eChW{}:
\eqna\eTwW
 $$\twoeqsalignno{
 \{\bN_-,\bW_+\}&=0~, \quad&\quad \{\N_+,\bW_+\}&=0~,   &\eTwW{a}\cr
 \{\bN_-,\bWb_-\}&=0~, \quad&\quad \{\N_+,\bWb_-\}&=0~, &\eTwW{b}\cr
 \noalign{\noindent and}
 \{\N_-,\bW_-\}&=0~, \quad&\quad \{\bN_+,\bW_-\}&=0~,   &\eTwW{c}\cr
 \{\N_-,\bWb_+\}&=0~, \quad&\quad \{\bN_+,\bWb_+\}&=0~, &\eTwW{d}\cr
}$$
proving that now $\bW_+,\bWb_-$ are gauge-covariantly twisted-chiral and
$\bW_-,\bWb_+$ are gauge-covariantly twisted-antichiral, again related by
the bisection formula~\eFinW{d}. The abelian version of this type-B gauge
multiplet was labeled VM-II in Ref.~\refs{\rTwJim} and was identified as
the `mirror' of the usual vector multiplet. Notice that now the spinorial
field strength superfields $\bW_+,\bWb_-$ are gauge-covariantly
{\it twisted-chiral\/}, while the gauge potentials $\bB$ are
gauge-covariantly {\it chiral\/}.

We will refer to $\{\bB,\bBb;\bW_\pm,\bWb_\pm;\bF\}$, subject to the
superdifferential relations~\eWFB{}, as the type-B or chiral gauge
multiplet. Recall that these gauge fields {\it do not couple\/} to
gauge-covariantly chiral superfields and their conjugates~\eHSF{a,b}.

\topic{Dimensional reduction}
The type-B gauge supermultiplet, $\{\bB,\bBb;\bW_\pm,\bWb_\pm;\bF\}$,
was called VM-II and identified with the `mirror'-twisted
cousin of the type-A gauge superfield multiplet~\refs{\rTwJim}. This gauge
multiplet has no counterpart in the rather more familiar 3+1-dimensional
models, where the anticommutator of any two (un)conjugate gauge-covariant
superderivatives vanishes.

However, one may regard the superfields $\bB$ as the gauge potential
superfield introduced to covariantize the (spin-0) central charges|were
such to have been introduced in further generalizing the gauge-covariant
supersymmetry algebra~\eTRs{}. That the present case~\eTRs{} contains no
such central charges may be understood merely as the statement that all
involved superfields are chargeless with respect to these central charges.

\topic{Abelian case}
Again, a further simplification is easy to demonstrate when $\cG_B$ is
abelian, so that $\bB$ are $\cG_B$-chargeless. Again|{\it while acting
on the\/} $\bB,\bBb$|the $\cG_B$-covariant (super)derivatives in Eqs.~\eBCH\
act as ordinary (super)derivatives~\eDs{}, $\bB$ and $\bBb$ become `plain'
chiral and antichiral superfields, and so can be expressed in terms of an
unconstrained prepotential superfield:
\eqn\eBinV{ \bB = \big([\Db_+,\Db_-]\,\bV^{(B)}\big)~,\quad\hbox{and}\quad
           \bBb = \big([D_-,D_+]\,\bVb^{(B)}\big)~. }

Easily then,
\eqn\eFBV{{\eqalign{\bF
 &=\frc i8\Big([D_-,D_+][\Db_+,\Db_-]\bV^{(B)}\}
           + [\Db_-\Db_+][D_-,D_+]\bVb^{(B)}\big\}\Big)~,\cr
 &=\frc i2\big(D_-\Db_-D_+\Db_+\bV^{(B)}
              -\Db_-D_-\Db_+D_+\bVb^{(B)}\big)~, }}}
and
\eqn\eWBV{
 \bW_\mp = \mp i\big(D_\mp\Db_+\Db_-\,\bV^{(B)}\big)~,\qquad
 \bWb_\mp = \mp i\big(\Db_\mp D_-D_+\,\bVb^{(B)}\big)~,  }
which are consistent with
\eqn\eGaB{ \GB_\pm = \pm i(D_\pm\bVb^{(B)})~,\quad\hbox{and}\quad
           \bGB_\pm = \mp i(\Db_\pm\bV^{(B)})~. }
Again, to ensure that $\bA,\bAb$ do vanish as assumed, substitute
Eqs.~\eGaB\ into~\ePPs{c,d}, to show that $\bVb^{(B)}=\bV^{(B)}$; no
further restrictions transpire from the remaining equations~\ePPs{}. So,
the mirror of the gauge vector prepotential superfield, $\bV$, must also be
real. After a little $D$-algebra, this produces
 $$
 \bF=2(\vd_\pp\bV^{(B)}_\mm-\vd_\mm\bV^{(B)}_\pp)~, \eqno\eFBV'
 $$
where
\eqn\eXXX{ \bV^{(B)}_\mm\define\inv4\big([D_-,\Db_-]\bV^{(B)}\big)~,\qquad
           \bV^{(B)}_\pp\define\inv4\big([D_+,\Db_+]\bV^{(B)}\big)~, }
so that the lowest component of $\bF$ in~\eFBV$'$ is the usual (abelian)
Yang-Mills field strength, in terms of the 2-vector potential obtained as
the lowest components of $\bV^{(B)}_\mm,\bV^{(B)}_\pp$.

\subsec{Type-C$_\mm$ gauging}\noindent
\subseclab\ssCm
Let us project on the $\cG_\mm$ factor in the gauge group, \ie, set
$\bB=\bA=0=\bC_\pp$. The peculiar relations~\eTwA{} and~\eChB{} now imply
that
\eqn\eCmL{{\twoeqsalign{
 [\bN_+,\bC_\mm]&=0~, \quad&\quad [\N_+,\bCb_\mm]&=0~, \cr
 [\N_+,\bC_\mm]&=0~, \quad&\quad [\bN_+,\bCb_\mm]&=0~, \cr }}}
while Eqs.~\eThC{} further constrain
\eqn\eCmC{ [\bN_-,\bC_\mm]=0~, \quad\hbox{and}\quad [\N_-,\bCb_\mm]=0~. }
That is, $\bC_\mm$'s are gauge-covariantly chiral rightons~\eQSF{c}, while
$\bCb_\mm$'s are gauge-covariantly antichiral rightons~\eQSF{d}; in fact,
they are gauge-covariantly {\it quartoid\/} superfields. That is, they
(effectively) depend on only one quarter of the four supercoordinates,
$\vs^\pm,\vsb^\pm$~\refs{\rHSS}.

As with the $\bB$'s and the $\bA$'s, the conclusion of \SS\,\ssMHSF\
on the high selectivity of gauge couplings does not apply to the $\bC_\mm$'s
themselves. In fact, we find that the $\bC_\mm$ do couple to the
$\cG_\mm$-gauge superfields, since 
\eqn\eXXX{ 0\:[\bC_\mm,\bC_\mm] = i\bC_\mm^i\bC_\mm^jf_{ij}{}^kT_k~, }
vanishes on account of the antisymmetry of the $f_{ij}{}^k$'s and the
commutivity of the $\bC$'s. This allows $\bC_\mm$ and its conjugate to have
nonzero $\cG_\mm$-charges, as is necessary for nonabelian $\cG_\mm$.

\topic{Superfield strengths}
Again, the formulae~\eWs{} and~\eF{} simplify:
\eqn\eWFCm{
 \bW_\Mm = -\frc i2[\N_-,\bC_\mm]~,\qquad
 \bWb_\Mm = -\frc i2[\bN_-,\bCb_\mm]~, }
and $\bW_\Pp,\bW_\pm$, their conjugates and $\bF$ all vanish. Recall that
the would-be `usual' bosonic field strength, $\bF_{\mm\1\mm}$, vanishes
identically; see appendix~B.
 The Jacobi identities involving two spinorial and one vectorial
$\N$ then produce the analogue of the familiar relations~\eChW{}:
\eqna\eRW
 $$\twoeqsalignno{
 \{\N_+,\bW_\Mm\}&=0~, \quad&\quad \{\bN_+,\bW_\Mm\}&=0~,   &\eRW{b}\cr
 \{\N_+,\bWb_\Mm\}&=0~, \quad&\quad \{\bN_+,\bWb_\Mm\}&=0~, &\eRW{a}\cr
}$$
proving that $\bW_\Mm,\bWb_\Mm$ are covariant rightons, related by the
`bisection formula'~\eFinW{e}. In fact, since
\eqn\eXXX{ -2i\{\N_-,\bW_\Mm\}=\big\{\N_-,[\N_-,\bC_\mm]\big\}
 =\inv2\big[\bC_\mm,\{\N_-,\N_-\}\big]=[\bC_\mm,\bCb_\mm]~, }
the $\bW_\Mm$ are gauge-covariantly {\it antichiral\/} rightons if
$\cG_\mm$ is abelian. Of course, the $\bWb_\Mm$ then are
gauge-covariantly {\it chiral\/} rightons.

We will refer to $\{\bC_\mm,\bCb_\mm;\bW_\Mm,\bWb_\Mm\}$, subject to
the superdifferential relations~\eWFCm, as the type-C$_\mm$, or chiral
righton gauge multiplet. These gauge fields {\it only couple\/}
to covariant rightons~\eHSF{f}, and to all non-minimal haploid
superfields~\eNMF{}.

\topic{Dimensional reduction}
The chiral righton gauge multiplet
$\{\bC_\mm,\bCb_\mm;\bW_\Mm,\bWb_\Mm\}$ also has no counterpart in the
rather more familiar 3+1-dimensional models. Somewhat in analogy with the
$\bB,\bBb$ potential superfields, the spin-1 gauge potential superfields
$\bC_\mm,\bCb_\mm$ also may be regarded as the gauge potentials introduced
to covariantize the (now spin-1) central charges|were such to have been
introduced on the r.h.s.\ of first of Eqs.~\eTRs{c}. Again, their absence
from the present case~\eTRs{} may be understood merely as the statement
that all involved superfield are chargeless with respect to these central
charges.

\topic{Abelian case}
Again, $\bC_\mm,\bCb_\mm$ are $\cG_\mm$-chargeless, and the $\cG_\mm$-gauge
covariant (super)derivatives act on the $\bC_\mm,\bCb_\mm$ as ordinary
(super)derivatives do. Now, as noted in Ref.~\refs{\rHSS}, it is not
possible to express chiral rightons, and so also the $\bC_\mm,\bCb_\mm$'s,
in terms of a superderivative of an ambidextrous superfield. It is possible,
however, to write $\bC_\mm\define(\Db_-{\cal C}_-)$, thereby specifying the
$\cG_\mm$-gauge potential in terms of a spin-$\inv2$ righton prepotential
superfield, ${\cal C}_-$.

Since now Eqs.~\ePPs{e} imply that $\bC_\mm=-i(\Db_-\GB_-)$, it
follows that the half of $\GB_-$ which is not annihilated by $\Db_-$ must be
equal to the analogous half of $i{\cal C}_-$. In the same fashion, the half
of $\bGB_-$ which is not annihilated by $\Db_-$ must be equal to the
analogous half of $i\ba{\cal C}_-$. The complementary halves of these
respective superfields remain unrelated. Of course, now $\GB_+=0=\bGB_+$.
Thus, unlike in the type-A and type-B gauging, the type-C$_\mm$ gauge
superfields $\GB_-,\bGB_-$ are {\it not\/} completely determined by the
prepotential superfields ${\cal C}_-,\ba{\cal C}_-$, and their introduction
is less useful.

\subsec{Type-C$_\pp$ gauging}\noindent
\subseclab\ssCp
Finally, let us project on the $\cG_\pp$ factor in the gauge group, \ie,
set $\bB=\bA=0=\bC_\mm$. The peculiar relations~\eTwA{} and~\eChB{} now
imply that
\eqn\eCpR{{\twoeqsalign{
 [\bN_-,\bC_\pp]&=0~, \quad&\quad [\N_-,\bCb_\pp]&=0~, \cr
 [\N_-,\bC_\pp]&=0~, \quad&\quad [\bN_-,\bCb_\pp]&=0~, \cr }}}
while Eqs.~\eThC{} further constrain
\eqn\eCpC{ [\bN_+,\bC_\pp]=0~, \quad\hbox{and}\quad [\N_+,\bCb_\pp]=0~. }
That is, $\bC_\pp$'s are gauge-covariantly chiral leftons~\eQSF{a}, while
$\bCb_\mm$'s are gauge-covariantly antichiral leftons~\eQSF{b}.

As with the $\bC_\mm$'s, the $\bC_\pp$'s do couple to the
$\cG_\pp$-gauge superfields, since 
\eqn\eXXX{ 0\:[\bC_\pp,\bC_\pp] = i\bC_\pp^i\bC_\pp^jf_{ij}{}^kT_k~, }
vanishes on account of the antisymmetry of the $f_{ij}{}^k$'s and the
commutivity of the $\bC$'s. This allows $\bC_\pp$ and its conjugate to have
nonzero $\cG_\pp$-charges.

\topic{Superfield strengths}
Again, the formulae~\eWs{} and~\eF{} simplify:
\eqn\eWFCp{
 \bW_\Pp = \frc i2[\N_+,\bC_\pp]~,\qquad
 \bWb_\Pp = \frc i2[\bN_+,\bCb_\pp]~, }
and $\bW_\Mm,\bW_\pm$, their conjugates and $\bF$ all vanish. Again,
recall that the would-be `usual' bosonic field strength, $\bF_{\pp\1\pp}$,
vanishes identically; see appendix~B.
The Jacobi identities involving two spinorial and one vectorial $\N$ then
produce the analogue of the familiar relations~\eChW{}:
\eqna\eLW
 $$\twoeqsalignno{
 \{\N_-,\bW_\Pp\}&=0~, \quad&\quad \{\bN_-,\bW_\Pp\}&=0~,   &\eLW{b}\cr
 \{\N_-,\bWb_\Pp\}&=0~, \quad&\quad \{\bN_-,\bWb_\Pp\}&=0~, &\eLW{a}\cr
}$$
proving that $\bW_\Pp,\bWb_\Pp$ are covariant leftons, related by the
`bisection formula'~\eFinW{e}. Again, since
\eqn\eXXX{ 2i\{\N_+,\bW_\Pp\}=\big\{\N_+,[\N_+,\bC_\pp]\big\}
 =\inv2\big[\bC_\pp,\{\N_+,\N_+\}\big]=[\bC_\pp,\bCb_\pp]~, }
the $\bW_\Pp$ are gauge-covariantly {\it antichiral\/} leftons if
$\cG_\pp$ is abelian. Of course, the $\bWb_\Pp$ then are
gauge-covariantly {\it chiral\/} leftons.

We will refer to $\{\bC_\pp,\bCb_\pp;\bW_\Pp,\bWb_\Pp\}$, subject to
the superdifferential relations~\eWFCp, as the type-C$_\pp$, or chiral
lefton gauge multiplet. These gauge fields {\it only couple\/}
to covariant leftons~\eHSF{e}, and to all non-minimal haploid
superfields~\eNMF{}.

\topic{Dimensional reduction}
The chiral lefton gauge multiplet
$\{\bC_\pp,\bCb_\pp;\bW_\Pp,\bWb_\Pp\}$ also has no counterpart in the
rather more familiar 3+1-dimensional models. Somewhat in analogy with the
$\bB,\bBb$ and $\bC_\mm,\bCb_\mm$ potential superfields, the spin-$(-1)$
gauge potential superfields $\bC_\pp,\bCb_\pp$ also may be regarded as the
gauge potentials introduced to covariantize the spin-$(-1)$ central
charges|were such to have been introduced on the r.h.s.\ of second of
Eqs.~\eTRs{c}. Again, their absence from the present case~\eTRs{} may be
understood merely as the statement that all involved superfield are
chargeless with respect to these central charges.

\topic{Abelian case}
Again, $\bC_\pp,\bCb_\pp$ are $\cG_\pp$-chargeless, and the $\cG_\pp$-gauge
covariant (super)derivatives act on the $\bC_\pp,\bCb_\pp$ as ordinary
(super)derivatives do. Now, as noted in Ref.~\refs{\rHSS}, it is not
possible to express chiral leftons, and so also the $\bC_\pp,\bCb_\pp$'s,
in terms of a superderivative of an ambidextrous superfield. It is possible,
however, to write $\bC_\pp\define(\Db_+{\cal C}_+)$, thereby specifying the
$\cG_\pp$-gauge potential in terms of a spin-$(-\inv2)$ lefton prepotential
superfield, ${\cal C}_+$.

Since now Eqs.~\ePPs{f} imply that $\bC_\pp=-i(\Db_+\GB_+)$, it
follows that the half of $\GB_+$ which is not annihilated by $\Db_+$ must be
equal to the analogous half of $i{\cal C}_+$. In the same fashion, the half
of $\bGB_+$ which is not annihilated by $\Db_-$ must be equal to the
analogous half of $i\ba{\cal C}_+$. The complementary halves of these
respective superfields remain unrelated. Of course, this time
$\GB_-=0=\bGB_-$. Again, unlike in the type-A and type-B gauging and just
like in the type-C$_\mm$ gauging, the type-C$_\pp$ gauge superfields
$\GB_+,\bGB_+$ are {\it not\/} completely determined by the prepotential
superfields ${\cal C}_+,\ba{\cal C}_+$.

\ping
Thus, for type-A and type-B gauging, both the spinorial field strength
superfields, $\bW,\bWb$, and the potential superfields, $\bA,\bAb,\bB,\bBb$,
are gauge-covariantly {\it haploid\/} superfields, and of relatively twisted
chirality: if the gauge potentials are (anti)chiral, the spinorial field
strengths are twisted (anti)chiral, and {\it vice versa\/}. For both type-C
gaugings, however, the spinorial field strengths, $\bW,\bWb$, are
gauge-covariantly unidexterous {\it haploid\/} superfields, whereas the
gauge potentials, $\bC,\bCb$, are gauge-covariantly {\it quartoid\/}
superfields.

Similarly, there are marked differences in the abelian case. While the
type-A and type-B gauging of an {\it abelian\/} symmetry both allow a
straightforward reduction of all gauge superfields to a real gauge
prepotential superfield, this is not the case in the two type-C gauging.
Here, only a pair of spinorial gauge prepotential superfields,
${\cal C}_-,\ba{\cal C}_-$ and ${\cal C}_+,\ba{\cal C}_+$, can be
introduced as easily, but they {\it do not\/} determine completely the gauge
superfields $\GB,\bGB$, and so are of limited use. Also, while the
spinorial field strengths for type-A and type-B gauging remain
gauge-covariantly {\it haploid\/} superfields, those for the two type-C
gaugings become gauge-covariantly {\it quartoid\/} superfields.

Finally, unlike the type-A and type-B gauging, neither of the type-C
gaugings contributes to the usual Yang-Mills field strength superfield,
$\bF$.

\subsec{Mixed type gauging}\noindent
\subseclab\ssMixG
The previous simple cases, when all but one (and its conjugate) of the
$\bA,\bB,\bC_\mm,\bC_\pp$ gauge superfields is zero, represent the most
restricted type of symmetry gauging.

It is also possible to have only some {\it two\/}, or only some {\it
three\/}, or indeed {\it all four\/} of the $\bA,\bB,\bC_\mm,\bC_\pp$'s
valued in the generators of the {\it same\/} irreducible factor of the
total gauge group. Thus, the most general type of gauge symmetry is, in
principle, of the form of a direct product of 15 factors:
\eqn\eTOT{ \cG_{Gen} ~=~ \Big(\bigotimes_{\cal I}\cG_{\cal I}\Big)~, }
where $\cal I$ is a multi-index, taking values in 1,- 2,- 3- and 4-element
subsets of the label-set $\{A,B,\mm,\pp\}$. $T_j$, the generators of
$\cG_{Gen}$ then have a block-diagonal matrix representation, such that the
matrix-generators of the $n^{th}$ factor in~\eTOT\ have only the
$n^{th}$ diagonal block non-zero. Normalizing these generators so that
\eqn\eXXX{ \Tr\big\{\,T_j\,,\,T_k\,\big\}~=~\d_{jk}~, }
we easily define projectors on any one of the 15 factors in~\eTOT:
\eqn\eXXX{ {\cal P}\!\!_{\cal{}_I}(\BM{X})~\define~
 \sum_{j\in\cal I} T_j\,\Tr\big\{\,T_j\,,\,\BM{X}\,\big\}~, }
where \BM{X} is an arbitrary superfield, assumed only to be expandable over
the generators of $\cG_{Gen}$: $\BM{X}=T_k\BM{X}^k$. Clearly then,
\eqn\eXXX{ {\cal P}\!\!_{\cal{}_I}(\BM{X})~=~
 \sum_{j\in\cal I}T_j\Tr\big\{T_j,T_k\big\}\BM{X}^k~=~
 \cases{\BM{X} & if $k\in\cal I$,\cr \noalign{\vglue2mm}
        0      & if $k\ne\in\cal I$.\cr} }

The $\bA,\bAb$ gauge potential superfields are expanded over the generators
of
\eqn\eXXX{{\eqalign{\cG_{Gen}\big|_{\ni A}
 &=\bigotimes_{{\cal I}\ni A}{\cal P}\!\!_{\cal{}_I}\big(\cG_{Gen}\big)~,\cr
 &= \cG_A\8\cG_{AB}\8\cG_{A\mm}\8\cG_{A\pp}\8
            \cG_{AB\mm}\8\cG_{AB\pp}\8\cG_{A\mm\pp}\8\cG_{AB\mm\pp}~,\cr
 }}}
and so on. The properties of the `mixed' gauging types can be deduced from
the above analysis of the `pure' gauging types, and shortly we turn to a few
sample cases.

Before that, however, a general remark is in order: all the `mixed' types of
gauging are beset with  a {\it common\/} property. As now more than one of
the gauge superfields $\bA,\bB,\bC_\mm,\bC_\pp$ is non-zero, there will
occur a {\it duplication\/} of degrees of freedom per gauge transformation.
This is against standard wisdom (see Ref.~\rGGRS, p.\,170), but need not be
deleterious in 1+1-dimensional spacetime.
 We will comment on this below, but defer a detailed study for a later
time.

\topic{Type-AB gauging}
An example of this `mixed' kind is provided by the gauge group which is
denoted by $\cG_{AB}$ in~\eTOT. Projecting the (total) gauge group to this
factor, we obtain $\bC=0$, but $\bA,\bB\neq0$. In this case, the
spin-$\pm\inv2$ superfield strengths are as given in Eqs.~\eWs{} and they
obey the `bisection formula'~\eFinW{d}, but the spin-$\pm\frc32$ superfield
strengths given in Eqs.~\eWs{} vanish. The bosonic superfield strength,
$\bF$, is as given in Eq.~\eFinW{c}. Since the $\bC$'s vanish, so do the
right hand sides of the `peculiar relations'~\eTwA{} and~\eChB{}, so that
the gauge potential superfields $\bA,\bB$ are twisted-chiral and chiral,
respectively, just as in the simple cases of \hbox{\SS\,\ssA}\ and~\ssB.

Reviewing the selectivity of coupling to matter (Table~1), it should be
clear that $\cG_{AB}$ couples neither to chiral nor to twisted-chiral
superfields (or their conjugates), but it can couple to leftons~\eHSF{e}
and rightons~\eHSF{f}, and of course to all non-minimal haploid
superfields~\eNMF{}.

The two non-zero gauge superfields, $\bA,\bB$, now provide {\it two\/}
independent complex gauge scalars, $a,b$, and {\it four\/} independent
gauginos, $\a_-,\ra_+,\b_\mp$; see below. Doubling the physical degrees of
freedom assigned to each gauge transformation, this will definitely lead to
incorrect (redundant) interaction with matter. The similar duplication of
the independent contributions into the bosonic field strength, $\bF$, is
less alarming, as it is an auxiliary field in 1+1-dimensions. 

The gauge symmetry in this context may be identified as the `diagonal'
subgroup $\cG_{AB}\subset(\cG_A\times\cG_B)$, where of course
$\cG_A\approx\cG_B$. A properly non-redundant gauging would then imply an
identification of the type-A gauge fields with the type-B ones, and in a
supersymmetric manner. To this end, one has to impose additional
(super)constraints. This approach however often has unexpected and
undesired consequences~\rHSS: 
$[\bN_+,\bA]{=}[\N_+,\bB]$, for example, induces both $\bA$ and $\bB$ to
become gauge-covariantly right-moving, \ie, $\N_\pp\bA=0=\N_\pp\bB$! In the
nonabelian case this would also produce non-linear relationships between
the component fields of $\bA,\bB$.
 Whether or not this redundancy of physically relevant gauge degrees of
freedom can be eliminated in a manifestly supersymmetric fashion then
remains an open question for now.

\topic{Type-AC$_\mm$ gauging}
Another example of this `mixed' kind is provided by the gauge group
$\cG_{A\mm}$: here $\bB,\bC_\pp=0$, but $\bA,\bC_\mm\neq0$. This time, the
spin-$\pm\inv2$ superfield strengths are as given in Eqs.~\eWFA{a,b} and
they obey the `bisection formula'~\eFinW{d}; the nonzero spin-$\pm\frc32$
superfield strengths are as given in Eqs.~\eWFCm. The bosonic superfield
strength, $\bF$, is as given in Eq.~\eWFA{c}. The gauge potential
superfields $\bA$ are no longer twisted-chiral, but instead
satisfy only the single constraint, $[\N_+,\bA]{=}0$, and are related to
$\bC_\mm$ through the `curious relation'~\eTwA{a}:
$[\bN_-,\bA]=-[\N_+,\bC_\mm]$. A combination of the remaining `curious
relations'~\eTwA{}, \eChB{} and the constraints~\eThC{} imply that
$\bC_\mm$ now is chiral.

 Reviewing again the selectivity of coupling to matter (Table~1), it should
be clear that $\cG_{A\mm}$ again couples neither to chiral nor to
twisted-chiral superfields (or their conjugates), but it can couple to
leftons~\eHSF{e}, rightons~\eHSF{f}, and of course to all non-minimal
haploid superfields~\eNMF{}.

 As compared to the type-A gauging, the gauge superfields $\bA,\bC_\mm$
now contribute additional spin-$(1,\frc32)$ degrees of freedom, none of
which are physical. They do however complicate the auxiliary field structure
of any model where this type of gauging is employed. In the quantum theory,
these additional, higher-spin component fields will also induce the
appearance of additional ghost degrees of freedom as compared to the
type-A gauging. It is possible that this leads to essentially and usefully
different dynamics, much as non-minimal haploid and minimal haploid
superfields are physically inequivalent~\rChiLin. A definite answer to this
question will however require a study of its own.

\topic{Universal, \ie, type-ABC gauging}
Our final example of this `mixed' kind is provided by the`universal' gauge
group $\cG_{AB\mm\pp}$: here all of $\bA,\bB,\bC$'s are nonzero, and valued
in the same irreducible Lie algebra. This is in fact the general case
considered in \SS\,\sSSC, and no modification is needed.

 The selectivity of coupling to matter (Table~1) now implies that
$\cG_{AB\mm\pp}$ couples to {\it none\/} of the gauge-covariantly haploid
superfields, but of course can couple to {\it all\/} of the non-minimal
haploid superfields~\eNMF{}. This yet again reinforces the distinction
between `minimal' gauge-covariantly haploid superfields~\eHSF{}, and their
`non-minimal' brethren~\eNMF{}.

Clearly, the issue of duplication of physical degrees of freedom becomes
most complicated here, and it combines the kind found in the type-AB
and the type-AC$\mm$ gaugings. Whether or not a suitable non-redundant
gauging of this type is possible remains an open question for now.

\subsec{Gauge transformation}\noindent
\subseclab\ssGTr
The gauge transformation operator, \CG, which appears in Eqs.~\eCov, has
so far remained unspecified. In general, of course, $\CG=\exp(i\CE)$, where
\CE\ is a \eTOT-Lie algebra valued superfield, \ie, a superfield expanded
over the generators of $\cG_{Gen}$ in~\eTOT. Notice that the definitions of
the constrained superfields~\eHSF{},
\eQSF{} and~\eNMF{} are all covariant with respect to the gauge
transformation~\eCov|regardless of the superfield type of \CE\,! For
example, \CE\ may be {\it chosen\/} to be an unconstrained superfield, yet
the transformed gauge-covariantly chiral superfield~\eHSF{a},
$\F'\define\CG\F$ satisfies the {\it transformed\/} covariant chirality
condition,
\eqn\eCovCons{ (\bN_\mp\F)=0 \qquad\To\qquad
 (\bN_\mp'\F')\define(\CG\bN_\mp\CG^{-1}\CG\F)=\CG(\bN_\mp\F)=0~. }
The same applies to all other superconstraints~\eHSF{}, \eQSF{} and~\eNMF{}.
Note that as the gauge transformation operator, \CG, is required to be
unitary, the gauge transformation generator, \CE, must be Hermitian. Other
that that, however, \CE\ remains an unconstrained general superfield.

Depending on the {\it choice\/} of superfield type for \CE, there
exist several distinct `representations'~\refs{\rGGRS}, and we now discuss
some of them in turn.

\topic{Chiral representation}
The infinitesimal version of $\F'\define\CG\F$ reads
\eqn\eSmF{ \d\F ~=~ i\CE\F~, }
and the infinitesimal version of~\eInh\ becomes
\eqn\eSmG{{\eqalign{
 \d\GB_\mp &= (D_\mp\CE)-i[\GB_\mp,\CE] ~=~\N_\mp\CE~,\cr
 \d\bGB_\mp &= (\Db_\mp\CE)-i[\bGB_\mp,\CE] ~=~\bN_\mp\CE~.\cr }}}
The infinitesimal transformation of, say, the chirality condition~\eHSF{a}
becomes
\eqn\eXXX{ \d(\bN_\pm\,\F) ~=~
 \big((\d\bN_\mp)\F\big)~+~\big(\bN_\mp\big(\d\F)\big)~. }
Now, since $\d\bN_\mp=-i\d\bGB_\mp$ and $\d\F=i\CE\F$, we have that for a
gauge-covariantly chiral superfield to remain so, it must be that
\eqn\eVAR{ -i\d\bGB_\mp\,\F~+~\bN_\mp\,i\CE\F~=~0~, }
whereupon
\eqn\eSmGC{ \d\bGB ~=~\bN_\mp\CE~, \quad\hbox{and}\quad
            \d\GB ~=~\N_\mp\CE~, }
in agreement with~\eSmG.

Frequently, however, one encounters a {\it weaker\/} notion of preserving
the type of constrained superfields. In comparison to the preceding
argument, let us regard the variation of the conjugate gauge superfields,
$\bGB$, as being of higher order. That is, the gauge parameter \CE\ is now
chosen to be a `slowly varying' superfield, so that $\bN\CE\ll\CE$, and
$\d\bGB\approx0$. In this limit, the condition~\eVAR\ for a
gauge-covariantly chiral superfield to remain so may be stated as
\eqn\eWkG{ \bN_\mp\tCE~=~0~, }
\ie, that $\tCE$ is a gauge-covariantly chiral superfield itself, and
with respect to the {\it untransformed\/} covariant derivative. This
necessarily contradicts the originally required hermiticity of \CE, and so
the unitarity of \CG. For, if $\tCE$ was also Hermitian, then the
Hermitian conjugate of~\eWkG\ would imply
\eqn\eCWkG{ \N_\mp\tCE^\dagger~=~\N_\mp\tCE~=~0~, }
whereupon $\N_\mm\tCE=0=\N_\pp\tCE$, and $\tCE$ would have to be
gauge-covariantly constant. Note that this {\it does not\/} necessarily
imply absolute constancy; nevertheless, this is often considered as too
restrictive a choice. One hence leaves $\tCE$ to be complex and chiral,
and so the gauge transformation operator
$\tCG\define\exp\{i\tCE\}$ is no longer unitary. 

This complicates things, since now $\tCG^{-1}\neq\tCG^\dagger$,
and expressions such as $\Tr(\F\FB)$, $\Tr(\N^4\F\FB)|$, \etc\ are no
longer gauge-invariant. The remedy (see, \eg, Ref.~\rBK, \SS\,3.6.4)
involves the introduction of two distinct covariant derivatives,
$\N^{(\pm)}$, such that $(\N^{(+)}\F)$ and $(\N^{(-)}\FB)$ transform
covariantly:
\eqn\eXXX{ (\N^{(+)}\F)'=\tCG(\N^{(+)}\F)~, \quad\hbox{and}\quad
           (\N^{(-)}\FB)'=(\N^{(-)}\FB)(\tCG^{-1})^\dagger~. }
We will return to this issue below.

This is the `simplest' choice, in that Eq.~\eSmF\ maintains chirality in
this weaker, but more familiar sense: gauge-covariantly chiral superfield
form a ring, and are in particular closed under multiplication|a product of
two gauge-covariantly chiral superfields is again gauge-covariantly chiral.
Note that this implies that, {\it in this (chiral) representation\/},
 $$ \d\GB_\mp ~=~ (\N_\mp\tCE)~, \quad\hbox{and}\quad
           \d\bGB_\mp ~=~ (\bN_\mp\tCE)=0~,
 \eqno\eSmGC' $$
which is rather asymmetric~\refs{\rGGRS}. In the ``antichiral''
representation, $\tCE$ of course obeys~\eHSF{b}, and the asymmetry
of~$\eSmGC'$ is reversed.

\topic{Other representations}
It is of course equally easy to choose a `twisted-chiral representation',
in which $\tCE$ obeys~\eHSF{c}, and is suitable for transforming
twisted-chiral superfields. Similarly, in the `lefton representation',
$\tCE$ is chosen to obey~\eHSF{e} and is suitable for transforming
leftons, and in the `righton representation', $\tCE$ is required to
obey~\eHSF{f} and is suitable for transforming rightons. In each of these
case, one uses the closure under multiplication of the `minimal' haploid
superfields~\eHSF{}:
\nobreak\vglue\medskipamount
\centerline{\vbox{\hsize=.75\hsize\boxit{4pt}{\vbox{\noindent
For any two superfields which both satisfy any one of the pairs of simple,
first order superdifferential constraints~\eHSF{}, so does {\it any analytic
function\/} thereof, and in particular, so does their product.}}}}\noindent
By contrast, `non-minimal' haploid superfields~\eNMF{} are not closed under
multiplication:
\nobreak\vglue\medskipamount
\centerline{\vbox{\hsize=.75\hsize\boxit{4pt}{\vbox{\noindent
For any two superfields which both satisfy any one of the simple, second
order superdifferential constraints~\eNMF{}, only their {\it linear
combinations\/} satisfy the same superconstraint.}}}}\noindent
However, since all `minimal' gauge-covariantly haploid superfields~\eHSF{}
also satisfy the `non-minimal' gauge-covariantly haploid
superconstraints~\eNMF{}, there {\it does\/} exist a choice of $\tCE$
which preserves the type of the `non-minimal' gauge-covariantly haploid
superfields~\eNMF{}. If the gauge parameter superfield $\tCE$ is chosen
to be a `minimal' gauge-covariantly haploid superfield, its product with the
corresponding `non-minimal' gauge-covariantly haploid remains a
`non-minimal' gauge-covariantly haploid.

For example, if $\tCE$ obeys~\eHSF{a} while $\Q$ obeys~\eNMF{a}, then
$\ex{i\tCE}\Q$ also obeys~\eNMF{a}, and so does $\d\Q=i\tCE\Q$.
Thus, any one of the `minimal' haploid representations of the gauge
transformation may also be used for the corresponding `non-minimal' haploid
superfields, and their superconstraint will remain preserved also in the
weaker sense~\eWkG.

\subsec{Non-unitary (de)covariantization and gauge prepotentials}\noindent
\subseclab\ssDeCov
On physical grounds, the covariantization $D\to\N$, as done in
Eqs.~\eCDs{}, relates the (super)derivatives with their gauge-covariant
counterparts in a continuous fashion. Indeed, the (suppressed) coupling
constant may be continuously turned off to recover $\N\to D$. This operation
resembles the gauge transformation process, and it is indeed
possible~\refs{\rGGRS,\rBK} to find operators, \CH, such that
\eqn\eDeCov{ \N_\mp = \bCH^{-1}D_\mp\bCH~, \quad\hbox{and}\quad
           \bN_\mp = \N_\mp^{~\dagger} = \CH \Db_\mp\CH^{-1}~. }
It should be clear that \CH\ must not be unitary: if it were, there would
always have to exist a gauge in which $\N\to D$, and where all the gauge
fields $\GB$ would vanish|contradicting the fact that (at least some) gauge
fields {\it are\/} physically relevant. As it is, \CH\ may be written as
the exponential with a complex exponent, the imaginary part of which can
always be annihilated by a suitable (unitary) gauge transformation,
$\CG=\ex{i\CE}$, where $\CE^\dagger=\CE$. In this gauge then, \CH\ is
Hermitian, being an exponential with a Hermitian exponent:
$\CH=\ex{-\CV}$, where $\CV^\dagger=\CV$ so also
$\CH^\dagger{\id}\bCH=\CH$. 

Note that the {\it non-unitary\/} transformation~\eDeCov\ induces a
corresponding transformation on superfields. For example, the
gauge-covariantly chiral superfields satisfy Eq.~\eHSF{a}, which now becomes
\eqn\eXXX{ 0=\bN_\mp\F=\CH \Db_\mp\CH^{-1}\F~. }
This prompts the definitions
\eqn\eNCCh{ \F=\CH\F^{(0)}~, \quad\hbox{and so}\quad \FB=\FB^{(0)}\bCH~, }
where now $\F^{(0)}$ ($\FB^{(0)}$) is
{\it simply\/}\ft{`Non-gauge-covariantly~\3'
being such a clumsy mouthful, we write `simply~\3' instead.} (anti)chiral:
\eqn\eXXX{ \Db_\mp\F^{(0)}=0~, \quad\hbox{and}\quad D_\mp\FB^{(0)}=0~. }

Using the {\it non-unitary\/} transformation~\eDeCov\ and~\eNCCh, it is
then always possible to rewrite a model involving gauge-covariantly
(anti)chiral superfields, $\F,\FB$, coupled to gauge (super)fields in terms
of the simply (anti)chiral superfields, $\F^{(0)},\FB^{(0)}$, coupled to
the same gauge (super)fields. For example,
\eqn\eXXX{ \Tr\big(\F\FB\big)
 =\Tr\big(\CH\F^{(0)}\FB^{(0)}\bCH\big)
 =\Tr\big(\FB^{(0)}\bCH\CH\F^{(0)}\big)~. }
Furthermore, upon the gauge transformation described above in which
$\CH=\ex{-\CV}$ is Hermitian, we have
\eqn\eStdNrm{ \Tr\big(\F\FB\big)
 =\Tr\big(\FB^{(0)}\ex{-2\CV}\F^{(0)}\big)~, }
which recovers (upon 4-fermionic integration) the standard expressions for
the gauge-covariant kinetic term in the Lagrangian for (anti)chiral
superfields~\refs{\rGGRS,\rWB,\rPW,\rBK}. Moreover, this shows the origin
of the gauge (pre)potential superfield, $\CV$, as the generator of the
coset $\cG^c/\cG$, where $\cG^c$ denotes the complexification of the gauge
group $\cG$. That is, having relaxed unitarity, \CH\ is an element of
$\cG^c$; but, when taken modulo gauge transformations (which live in $\cG$),
\CH\ belongs to the coset $\cG^c/\cG$. Indeed, this \CV\ becomes the
gauge prepotential ${\bf V}^{(A)}$, as found in Eqs.~\eAinV\ for the
abelian case of type-A gauging, and ${\bf V}^{(B)}$, as found in
Eqs.~\eBinV\ for the abelian case of type-B gauging.
 See \SS\,3.6 of Ref.~\rBK\ for the derivation of the analogous statement
from the standard (but opposite to our) approach, where one employs the
simple (super)derivatives and simply constrained superfields but inserts
explicitly gauge-covariantizing factors such as $\ex{-2\CV}$.

In the present approach, gauge-covariance is made manifest through the use
of gauge-covariant derivatives~\eCDs{} and gauge-covariantly constrained
superfields~\eHSF{}, \eQSF{} and~\eNMF{}. With the present level of
generality, this seems preferable to determining the precise form of all the
gauge-covariantizing insertions separately. Instead, the analogues of
$\ex{-2\CV}$ for other simply haploid superfields may be obtained by
applying to the above results: (1)~either of the (mirror map) discrete
transformations, $\BM{C}_\pm$, of Ref.~\rHSS\ for twisted-(anti)chiral
superfields, and (2)~either of $\BM{q},\qB$ of Ref.~\rHSS\ to map
(anti)chiral superfields into leftons (rightons). We leave this to the
interested Reader.

We do note, however, that the appearance of the prepotential superfield,
\CV, in expressions like~\eStdNrm\ proves its existence and uncovers its
geometric origins, for all types of symmetry gauging and all (compact) 
symmetry groups---not just the abelian cases shown explicitly in
\SS\,\sCases. A more complete discussion of such decovariantization(s) is
presented elsewhere~\rSSYM.

\newsec{Component field content}\noindent
\seclab\sCmps
The definition of component fields by means of projecting on the various
terms in the Taylor expansion~\refs{\rGGRS,\rBK,\rHSS} is easy to adapt for
the presence of gauge symmetries. One simply uses covariant
superderivatives~\eCDs{} in place of ordinary ones~\eDs{}, and
{\it commutators\/} for second order superderivatives. As usual, `$|$' will
denote setting $\vs^\mp=0=\vsb^\mp$.

\subsec{Matter fields and general convention}\noindent
\subseclab\sBasic
To begin, consider the `minimal' gauge-covariantly haploid
superfields~\eHSF{}.
\eqna\eCps
 $$
   \f\define\F|~,\qquad
   \j_\pm\define \inv{\sqrt{2}}\N_\pm\F|~,\qquad
   F\define \inv4[\N_-,\N_+]\F|~; \eqno\eCps{a}
 $$
 $$
   \fb\define\FB|~,\qquad
   \B\j_\pm\define \inv{\sqrt{2}}\bN_\pm\FB|~,\qquad
   \Fb\define \inv4[\bN_+,\bN_-]\FB|~; \eqno\eCps{b}
 $$
are the component fields of the gauge-covariantly chiral and the
gauge-covariantly antichiral spin-0 superfield~\eHSF{a,b}. We write
$\F=(\f;\j_\mp;F)$.

Next,
 $$
   x\define\X|~,\quad
   \x_-\define \inv{\sqrt{2}}\N_-\X|~,\quad
   \c_+\define \inv{\sqrt{2}}\bN_+\X|~,\quad
   X\define \inv4[\N_-,\bN_+]\X|~; \eqno\eCps{c}
 $$
 $$
  \bx\define\XB|~,\quad
   \xb_-\define \inv{\sqrt{2}}\bN_-\XB|~,\quad
   \cb_+\define \inv{\sqrt{2}}\N_+\XB|~,\quad
   \Xb\define \inv4[\N_+,\bN_-]\XB|~; \eqno\eCps{d}
 $$
are the component fields of the gauge-covariantly twisted-chiral and the
gauge-covariantly twisted-antichiral spin-0 superfield~\eHSF{c,d}. We write
$\X=(x;\x_-,\c_+;X)$.

Finally, we define
 $$
   \ell\define\L|~,\quad
   \l_+\define \inv{\sqrt{2}}\N_+\L|~,\quad
   \bl_+\define \inv{\sqrt{2}}\bN_+\L|~,\quad
   L_\pp\define \inv4\big[\N_+,\bN_+\big]\L|~; \eqno\eCps{e}
 $$
and
 $$
   r\define\Y|~,\quad
   \r_-\define \inv{\sqrt{2}}\N_-\Y|~,\quad
   \vr_+\define \inv{\sqrt{2}}\bN_-\Y|~,\quad
   R_\mm\define \inv4\big[\N_-,\bN_-\big]\Y|~; \eqno\eCps{f}
 $$
are the component fields in the gauge-covariantly lefton and the
gauge-covariantly righton superfields. We write
$\L=(\ell;\l_+,\bl_+;L_\pp)$ and
$\Y=(r;\r_-,\vr_-;R_\mm)$.

Since gauge-covariantly quartoid superfields cannot couple to any of the
minimally coupled gauge (super)fields, the definitions of their component
fields remain as given in Ref.~\rHSS.

The component field content of the `non-minimal' haploid superfields is
just as straightforward to obtain, only this time there are more component
fields than in Eqs.~\eCps{}. For example, 
\eqna\eNMCp
 $${\cmath{
 t\define\Q|~,\quad \q_\pm\define\inv{\sqrt2}\N_\pm\Q|~,\quad
 \vq_\pm\define\inv{\sqrt2}\bN_\pm\Q|~,\cr
 T\define\inv4[\N_-,\N_+]\Q|~,\quad
 T_\mp\define\inv4[\N_-,\bN_+]\Q|~,\quad
 T_\pm\define\inv4[\N_+,\bN_-]\Q|~,\cr
 T_\mm\define\inv4[\N_-,\bN_-]\Q|~,\quad
 T_\pp\define\inv4[\N_+,\bN_+]\Q|~,\cr
 \t_-\define\inv{4\sqrt2}[\N_-,\bN_-]\N_+\Q|~,\quad
 \t_+\define\inv{4\sqrt2}[\N_+,\bN_+]\N_-\Q|~.\cr
 }}\eqno\eNMCp{a}$$
Note: the $\pm$ and $\mp$ subscripts on $T_\pm,T_\mp$ indicate
{\it single\/} components, whereas the $\pm$ on the fermions
$\q_\pm,\vq_\pm,\t_\pm$ indicates a {\it choice\/} of spin $+\inv2$ and
$-\inv2$. We also write
$\Q=(t;\q_\mp,\vq_\mp;T_\mm,T,T_\pm,T_\mp,T_\pp;\t_\mp)$.
 Similarly,
 $$\cmath{
 p\define\P|~,\quad \p_\pm\define\inv{\sqrt2}\N_\pm\P|~,\quad
 \vp_\pm\define\inv{\sqrt2}\bN_\pm\P|~,\cr
 P\define\inv4[\N_-,\N_+]\P|~,\quad
 P_\mp\define\inv4[\N_-,\bN_+]\P|~,\quad
 \FF{P}\define\inv4[\bN_+,\bN_-]\P|~,\cr
 P_\mm\define\inv4[\N_-,\bN_-]\P|~,\quad
 P_\pp\define\inv4[\N_+,\bN_+]\P|~,\cr
 \Tw\vf_-\define\inv{4\sqrt2}[\N_-,\bN_-]\bN_+\P|~,\quad
 \vf_+\define\inv{4\sqrt2}[\N_+,\bN_+]\N_-\P|~.\cr
 }\eqno\eNMCp{b}$$
The component fields of $\QB$ and $\PB$ are then obtained by hermitian
conjugation. We write
$\P=(p;\p_\mp,\vp_\mp;P_\mm,P,P_\mp,\FF{P},P_\pp;\Tw\vf_-,\vf_+)$.

The components of the non-minimal gauge-covariantly (almost) lefton~\eNMF{e}
and (almost) righton~\eNMF{f} can be defined as easily. However, to avoid
confusion with the gauge potential superfield $\bA$ and since we will not
need these (almost) unidexterous superfields, we leave the definition of
the components of the non-minimal gauge-covariantly (almost) lefton~\eNMF{e}
and righton~\eNMF{f} to the diligent Reader.

\subsec{Component diagrammatics}\noindent
The Reader may find it convenient to use the diagram in Fig.~1 to find
the component field content of any superfield. All components are obtained
by acting on the superfield with one of the covariant operators from the
diagram, and then setting $\vs^\pm{=}0{=}\vsb^\pm$.
\PixCap{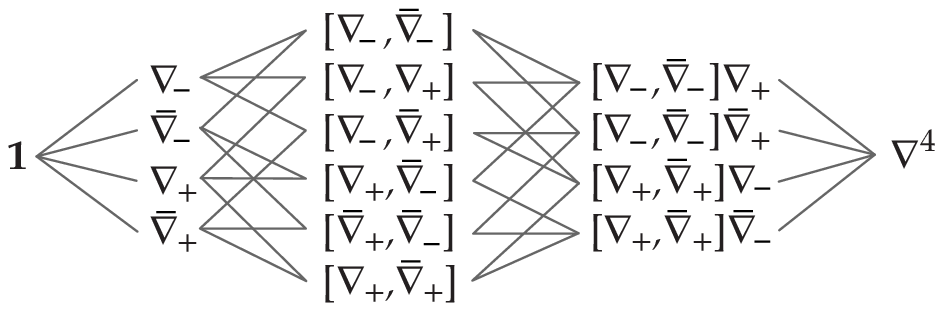}{1}{The sequence of multiple covariant superderivatives
used in defining component fields. Component fields of conjugate
superfields are found using the hermitian conjugates of the operators shown
here. The operator $\N^4$ is obtained by replacing all $D$'s in Eq.~\eDI\
with $\N$'s.}
Finally, throughout this article, we set the numerical coefficients as
in~\eCps{a}: $\inv{\sqrt2}$ for every superderivative, and $\inv2$ for
every commutator. The examples in~\eCps{} and~\eNMCp{} should suffice in
clarifying this.

Applying additional (super)derivatives merely produces total derivatives
of the already defined component fields. In this sense, the  component
fields, onto which the operators in the diagram in Fig.~1 project (upon
setting $\vs^\pm{=}0{=}\vsb^\pm)$, are considered {\it independent\/}.

\subsec{Gauge fields}\noindent
\subseclab\sGFields
As with any of the `matter' superfields in the preceding sections, the
component fields of the gauge superfields are also defined by applying the
operators from the sequence in Fig.~1, and setting $\vs=0=\vsb$'s. Bending
somewhat the typographical conventions of Ref.~\refs{\rHSS}, we list the
component fields of the gauge superfields $\GB_\mp$ as follows
\eqna\eGCp
 $$ \eqalignno{
 \GB_-
 &=\left(\quad\matrix{g_-\cr}\quad
         \matrix{\G_\mm\cr \bG_\mm\cr \G_\mp \cr\bG_\mp \cr}\quad
         \matrix{\g^\pp_-\cr \g_-\cr \g^\pm_-\cr
                 \bg^\mp_-\cr\bg_-\cr \bg^\mm_-\cr}\quad
         \matrix{G_\mm\cr \FF{G}_\mm\cr G_\mp\cr \FF{G}_\mp\cr}\quad
         \matrix{\rg_-\cr}\quad
         \right)~, &\eGCp{a}\cr
 \GB_+
 &=\left(\quad\matrix{g_+\cr}\quad
         \matrix{\G_\pm\cr \bG_\pm\cr \G_\pp \cr\bG_\pp \cr}\quad
         \matrix{\g^\pp_+\cr \g_+\cr \g^\pm_+\cr
                 \bg^\mp_+\cr\bg_+\cr \bg^\mm_+\cr}\quad
         \matrix{G_\pm\cr \FF{G}_\pm\cr G_\pp\cr \FF{G}_\pp\cr}\quad
         \matrix{\rg_+\cr}\quad
         \right)~, &\eGCp{b}\cr
 }$$
and similarly for their conjugates. Many of these component fields either do
not even show up in the Lagrangian densities of interest, or they show up
as auxiliary fields, \ie, such that their equations of motion are
algebraic, which allows for immediate elimination. Until however a
Lagrangian density is chosen for such an elimination to take place, one must
deal with all the component fields~\eGCp{}.

 It is fairly standard in
theories in 1+1-dimensional spacetime, to assign the coupling constant, $g$,
the canonical dimensions of $\vd_\mm,\vd_\pp$, \ie, of a mass parameter.
The lowest components of the superfields $\GB_\mm,\GB_\pp$, which through
using Eqs.~\ePPs{i,j} become
\eqn\eXXX{{\eqalign{
 \GB_\mm| &=-\frc{i}{\sqrt2}\big[\Gb^l_\mm+\bG^l_\mm
       +\inv{\sqrt2}g^j_-\B{g}^k_-\f_{jk}{}^l\big]T_l~,\cr
 \GB_\pp| &=-\frc{i}{\sqrt2}\big[\Gb^l_\pp+\bG^l_\pp
       +\inv{\sqrt2}g^j_+\B{g}^k_+\f_{jk}{}^l\big]T_l~,\cr
 }}}
then acquire the proper canonical dimension, 0, only {\it upon the
rescaling\/} $\GB\to g\GB$. This makes it possible for the canonically
rescaled superfields to appear in the `standard, flat' kinetic term
 $$
   -\inv4\int\rd^2\s~\Tr\big\|\,(\vd_{[\mm}\GB_{\pp]})|\,\big\|^2+\ldots
 $$
{\it Prior\/} to the $\GB\to g\GB$ rescaling, such action terms ought to be
premultiplied by $g^{-2}$.
 Through the $\GB\to g\GB$ rescaling, $g_\pm$, the lowest components of
$\GB_\pm,\bGB_\pm$, acquire the canonical dimension $-\inv2$; the
(potentially) physical fields there being the gauge bosons $\G,\bG$ and the
gauginos $\g,\bg$; the $G,\FF{G}$'s and the $\rg_\pm$ stand for bosonic
and fermionic `auxiliary' components.

Essentially the same fate befalls the gauge potential superfields
$\bA,\bB,\bC$'s and their conjugates~\ePPs{a\?h}. Their lowest
components|the gauge bosons|are linear combinations of the $\G,\bG$'s, and
in the nonabelian case, of $g{-}g$ anticommutators. Their next-to-lowest
components|the gauginos|are linear combinations of the $\g,\bg$'s and
derivatives of $g$'s, and in the nonabelian case, of $g{-}\G$ commutators,
and so on. 

\subsec{Field strengths}\noindent
Since the gauge superfields $\GB$ transform inhomogeneously under the
gauge transformations~\eInh|in the infinitesimal form~\eSmGC, so do their
components. This means that the gauge transformation parameter \CE\ may be
{\it chosen\/} so as to gauge away some of the degrees of component fields
in $\GB$. The Wess-Zumino gauge~\refs{\rGGRS,\rWB,\rPW,\rBK} is the best
known such choice.

On the other hand, the field strength superfields defined in Eqs.~\eTRs{}
all transform homogeneously, so that there exists no choice of the gauge
parameter superfield, \CE, that would `gauge away' a component field of
$\bA,\bB,\bC$. This then guarantees that all the component fields of these
superfields are unambiguously defined and independent of any gauge choice.

Furthermore, $\bF$ is defined as a linear combination of gauge covariant
superderivatives of the $\bW$'s, which in turn are all defined as gauge
covariant superderivatives of the $\bA,\bB,\bC$'s. This guarantees that the
lowest component field of $\bF$ is a linear combination of the
next-to-lowest component fields of the $\bW$'s. And in turn, the lowest
component fields of the $\bW$'s are linear combinations of the
next-to-lowest component fields of the $\bA,\bB,\bC$'s.

For this reason, all physically relevant (and unambiguously defined)
component fields in the gauge multiplets must be expressible in terms of the
component fields of the $\bA,\bB,\bC$ and their conjugates. Consequently,
we leave to the interested Reader the tracing of the physical degrees of
freedom in gauge multiplets all the way to the component fields of the
$\GB$'s and specification of the Wess-Zumino-like gauge(s) in which the
remaining component fields are annihilated. Herein, we proceed instead to
use the component fields of the $\bA,\bB,\bC$'s.

\topic{Type-A gauge fields}
In the simple type-A, -B and -C gauging, these latter superfields satisfy
important superconstraints~\eATW, \eBCH, \eCmL\ and~\eCmC, \eCpR\
and~\eCpC. Thus, the independent component fields of the type-A gauge
multiplet can be specified equivalently either as the component fields of
the gauge-covariantly twisted-chiral superfield $\bA$, or as the lowest
component fields of the type-A gauge multiplet $\{\bA;\bW_\pm;\bF\}$ given
in \SS\,\ssA\ft{Recall: $([\N_-,\bN_+]\bA)$ merely abbreviates
$\big\{\N_-,[\bN_+,\bA]\big\}-\big\{\bN_+,[\N_-,\bA]\big\}$; see
appendix~A.}:
\eqna\eCpA
 $$\twoeqsalignno{
 a&\define\bA|~, \quad&\quad \B{a}&\define\bAb|~,  &\eCpA{a}\cr
 \a_-&\define\inv{\sqrt2}[\N_-,\bA]|=\sqrt2i\bWb_-|~,\quad&\quad
 \ab_-&\define-\inv{\sqrt2}[\bN_-,\bAb]|=-\sqrt2i\bW_-|~, &\eCpA{b}\cr
 \ra_+&\define\inv{\sqrt2}[\bN_+,\bA]|=-\sqrt2i\bW_+|~,\quad&\quad
 \B\ra_+&\define-\inv{\sqrt2}[\N_+,\bAb]|=\sqrt2i\bWb_+|~, &\eCpA{c}\cr
 A&\define\inv4[\N_-,\bN_+]\bA|=-i\Bf|~,\quad&\quad
 \Ab&\define\inv4[\N_+,\bN_-]\bA|=i\Bfb|~. &\eCpA{d}\cr
}$$
Note the relative signs in the definition of the conjugate fermionic
components; they stem from the antisymmetry of the commutator. These are
the spin-(0;$\pm\inv2$;0) component fields of the scalar gauge-covariantly
twisted-chiral superfield $\bA=(a;\a_-,\ra_+;A)$ and its Hermitian
conjugate, $\bAb=(\B{a};\ab_-,\B\ra_+;\Ab)$. The bosonic field strength,
$\bF|$ then equals (the negative of) the imaginary part of
$A$:
\eqn\eXXX{ \rF_\bA\define\bF|=\Ree({\Bf})|=-\Imm(A)~. }
The real part of $A$ is easily shown to be the lowest component of the
superfield
\eqn\eThDA{{\cmath{
 \bD_\bA ~\define~\inv{2i}\{\N_{[-},\bW_{+]}\}_\bA=\Imm(\Bf)~,\cr
            \rD_\bA\define\bD_\bA|=\Imm(\Bf)|=\Ree(A)~.}}}
where the subscript indicates projection on type-A gauge fields.
Clearly both $\bF$ and $\bD_\bA$ are real. The lowest component of
$\bD_\bA$ is indeed, up to some conventional overall factor, the familiar
`auxiliary D field', descending by dimensional reduction from $N{=}1$
supersymmetric Yang-Mills theory in $3{+}1$-dimensions. This vindicates our
use of the {\it complex\/} superfield \Bf\ as compared to using only its
imaginary part,
$\bF$.

In the present incarnation, the lowest components of $\bF$ and $\bD_\bA$
appear as the `auxiliary' components of the gauge-covariantly twisted-chiral
superfield $\bA$. In particular, note that the canonical dimension of $\bF$
and $\bD_\bA$ is one more than that of $\bA$. Therefore, if a term in the
Lagrangian density is {\it chosen\/} so as to provide a kinetic term for
$a$|something like $h\,|\vd_0 a|^2+\3$ with $h$ an arbitrary analytic
function of the available fields, then the same term will also feature
$\bF$ and $\bD_\bA$ with no derivatives on them.

\topic{Type-B gauge fields}
The independent component fields of the type-B gauge multiplet can
similarly be specified equivalently either as the component fields of the
gauge-covariantly chiral superfield $\bB$, or as the lowest component fields of
the multiplet $\{\bB;\bW_\pm;\bF\}$ given in \SS\,\ssB:
\eqna\eCpB
 $$\twoeqsalignno{
 b&\define\bB|~, \quad&\quad \B{b}&\define\bBb|~,  &\eCpB{a}\cr
 \b_\mp&\define\inv{\sqrt2}[\N_\mp,\bB]|=\pm\sqrt2i\bW_\mp|~,\quad&\quad
 \bb_\mp&\define-\inv{\sqrt2}[\bN_\mp,\bBb]|=\mp\sqrt2i\bWb_\mp|~,
                                                     &\eCpB{b}\cr
 B&\define\inv4[\N_-,\N_+]\bB|=-i\Bf|~,\quad&\quad
 \Bb&\define\inv4[\bN_+,\bN_-]\bBb|=i\Bfb|~.  &\eCpB{c}\cr
}$$
These are the spin-(0;$\pm\inv2$;0) component fields of the scalar
gauge-covariantly chiral superfield $\bB=(b;\b_\mp;B)$ and its Hermitian
conjugate, $\bBb=(\B{b};\bb_\mp;\Bb)$. The bosonic field strength, $\bF|$
now equals (the negative of) the imaginary part of $B$:
\eqn\eXXX{ \rF_\bB\define\bF|=\Ree(\Bf)|=-\Imm(B)~. }
The real part of $B$ is easily shown now to be the lowest component of the
superfield
\eqn\eThDB{{\cmath{
 \bD_\bB~\define~\inv{2i}\big[\{\N_+,\bW_-\}-\{\bN_-,\bWb_+\}\big]_\bB
 =\Imm(\Bf)~,\cr
 \rD_\bB\define\bD_\bB|=\Imm(\Bf)=\Ree(B)~.\cr }}}
where the subscript indicates projection on type-B gauge fields. Again,
both $\bF$ and $\bD_\bA$ are real.

\topic{Type-C$_\mm$ gauge fields}
The type-C$_\mm$ gauge multiplet is equivalently described either as the
component fields of the gauge-covariantly chiral lefton $\bC_\mm$, or as the
lowest component fields of the gauge multiplet $\{\bC_\mm;\bW_\Mm\}$ of
\SS\,\ssCm:
\eqna\eCpCm
 $$\cmathno{
 C_\mm\define\bC_\mm|~, \qquad \Cb_\mm\define\bCb_\mm|~,       &\eCpCm{a}\cr
 \g^\pp_-\define\inv{\sqrt2}[\N_-,\bC_\mm]|=\sqrt2i\bW_\Mm|~,&\eCpCm{b}\cr
 \gb^\pp_-\define-\inv{\sqrt2}[\bN_-,\bCb_\mm]|=-\sqrt2i\bWb_\Mm|~,
                                                               &\eCpCm{c}\cr
}$$
These are the spin-$(1;\frc32)$ component fields of the spin-($+1)$
gauge-covariantly chiral lefton superfield $\bC_\mm=(C_\mm;\g^\pp_-)$ and
its Hermitian conjugate, $\bCb_\mm=(\Cb_\mm;\gb^\pp_-)$. There are no $\bF$
and no $\bD$ field strength superfields in type-C$_\mm$ gauging.

\topic{Type-C$_\pp$ gauge fields}
The component fields of the type-C$_\pp$ gauge multiplet either as the
component fields of the gauge-covariantly chiral righton $\bC_\pp$, or as the
lowest component fields of the gauge multiplet $\{\bC_\pp;\bW_\Pp\}$ of
\SS\,\ssCp:
\eqna\eCpCp
 $$\cmathno{
 C_\pp\define\bC_\pp|~, \qquad \Cb_\pp\define\bCb_\pp|~,       &\eCpCp{a}\cr
 \g^\mm_+\define\inv{\sqrt2}[\N_+,\bC_\pp]|=\sqrt2i\bW_\Pp|~,&\eCpCm{b}\cr
 \gb^\mm_+\define-\inv{\sqrt2}[\bN_+,\bCb_\pp]|=-\sqrt2i\bWb_\Pp|~,
                                                               &\eCpCm{c}\cr
}$$
These are the spin-$(-1;-\frc32)$ component fields of the spin-($-1$)
gauge-covariantly chiral lefton superfield $\bC_\pp=(C_\pp;\g^\mm_+)$ and
its Hermitian conjugate, $\bCb_\pp=(\Cb_\pp;\gb^\mm_+)$. There are
no $\bF$ and no $\bD$ field strength superfields in type-C$_\pp$ gauging.

Based on the last remark of the previous subsection, the canonical
dimension of $a,b,C_\mm$ and $C_\pp$ from~~\eCpA{}--\eCpCp{} is the same as
that of the $\G$'s in~\eGCp{}; these are the gauge bosons. The
$\a_-,\ra_+,\b_\pm,\g^\pp_-$ and $\g^\mm_+$ from~\eCpA{}--\eCpCp{} have the
canonical dimensions of the $\g$'s in~\eGCp{}; these are the gauginos.
Notice that the type-C spin-($\pm\frc32$) gauginos, $\g^\pp_-,\g^\mm_+$,
accompany the spin-($\pm1$) gauge bosons $C_\mm,C_\pp$, and should not be
confused with gravitini which are superpartners of the spin-($\pm2$)
gravitons.

The definition of $\g^\pp_-$ in~\eCpCm{b} and $\g^\mm_+$ in~\eCpCp{b}
differs from that in (the conjugates of)~\eGCp{} by a numerical
multiplicative factor, and the addition of two terms depending on
the lowest components of $\bGB_\pm$. The latter discrepancy vanishes in the
appropriate Wess-Zumino-like gauge, and the numerical multipliers are
inessential. The precise relation of the component fields~\eCpA{}, \eCpB{},
\eCpCm{} and~\eCpCp{} to those in~\eGCp{} is left to the diligent Reader;
we proceed using the gauge covariant component fields~\eCpA{}, \eCpB{},
\eCpCm{} and~\eCpCp{}.

\newsec{Lagrangian Densities}\noindent
\seclab\sLagD
All the numerous Lagrangian densities listed in Ref.~\refs{\rHSS} admit a
straightforward extension to include minimal coupling type gauge
interactions: the fermionic integration must be redefined so as to use the
covariant derivatives~\eCDs{} instead the ordinary ones.
Section~\sIntro\ and Appendix~A contain all the necessary conventions and
notation.

\subsec{Kinetic terms for gauge fields}\noindent
\subseclab\ssKinG
It is straightforward to write down (flat) kinetic terms for the pure type-A
and type-B gauge fields~\refs{\rTwJim}:
\eqn\eToyAB{ -\inv8\int\rd^4\vs~\|\bA\|^2~,\qquad\hbox{and}\qquad
           +\inv8\int\rd^4\vs~\|\bB\|^2~, }
where the relative sign ensures the correct sign for the bosonic terms.

It is immediately obvious that no such (flat) kinetic terms exist for the
type-C gauge superfields! This may sound alarming, as it is sharply
counterintuitive when compared with the familiar situation in
3+1-dimensional spacetime. However, as we have just seen in Eqs.~\eCpCm{}
and~\eCpCp{}, the type-C gauge multiplets contain spin-$(\pm1,\pm\frc32)$
fields, neither of which is {\it supposed\/} to be a propagating degree of
freedom in 1+1-dimensional spacetime. A quantum treatment would have to
introduce appropriate (and propagating) spin-$(0,\pm\inv2)$ ghost degrees
of freedom, which then would carry the dynamics of this type of symmetry
gauging, and the study of which we defer to a later effort.

For illustrative purposes, consider however the pure type-A gauging of an
abelian symmetry, say $U(1)$, with respect to which the $\bA,\bAb$
themselves are chargeless, and the integration measure in the first of the
two integrals in~\eToyAB\ reduces to that one in Eq.~\eDI.

After straightforward $D$-algebra, we obtain for a single type-A $U(1)$
gauge multiplet:
\eqn\eAbA{{\eqalign{
 \[3-\inv8\int\rd^4\vs~\|\bA\|^2
 &=\inv4\big[(\vd_\mm\B{a})(\vd_\pp a)+(\vd_\pp\B{a})(\vd_\mm a)\big]
  +\inv2\rD_\bA^2+\inv2\rF_\bA^2\cr
 &\]3+\frc{i}4\Big[\big(\ab_-\dvd_\pp\a_-\big)
  +\big({\B\ra}_+\dvd_\mm\ra_+\big)\Big]~, }}}
 Identifying $\rF_\bA=\inv2\e^{ab}{\cal F}_{ab}^{(A)}$ and using that
$\e^{ab}\e_{cd}=(\d^{ab}_{dc}-\d^{ab}_{cd})$ since $\e^{01}=1=\e_{10}$, the
last bosonic term becomes $-\inv4{\cal F}_{ab}^{(A)}\,{\cal F}^{ab}_{(A)}$,
which is the standard Lagrangian density for a Yang-Mills gauge boson,
given in terms of its field strength. Furthermore, one can write
\eqn\eXXX{ {\cal F}^{(A)}_{ab}~=~(\vd_a{\cal A}_b-\vd_b{\cal A}_a)~, }
but the Reader should be cautioned that the gauge vector potential
${\cal A}_a$ carries no physical degree of freedom in 1+1 dimensions. Its
only use is for identification with the components of the 3+1-dimensional
gauge vector potential `along' the 1+1-dimensional sub-spacetime; the
`transversal' components form the lowest components of the complex
superfield, $\bA$, a scalar from the 1+1-dimensional point of view.
 In fact, it is not an accident that
$-\inv4{\cal F}_{ab}^{(A)}\,{\cal F}^{ab}_{(A)}$ emerges, in~\eToyAB, as a
half of the norm-squared of the `auxiliary' component field of $\bA$, with
$\inv2\rD^2_\bA$ being the other half.

 Up to a few numerical differences in convention, this is in complete
agreement with Ref.~\rTwJim, and the dimensional reduction results obtained
from Refs.~\refs{\rGGRS,\rWB,\rPW,\rBK}.

The second of the two Lagrangians in~\eToyAB\ is equally easily evaluated
for a single type-B $U(1)$ gauge multiplet:
\eqn\eAbB{{\eqalign{
 \[3\inv8\int\rd^4\vs~\|\bB\|^2
 &=\inv4\big[(\vd_\mm\B{b})(\vd_\pp b)+(\vd_\pp\B{b})(\vd_\mm b)\big]
  +\inv2\rD_\bB^2+\inv2\rF_\bB^2\cr
 &\]3+\frc{i}4\Big[\big(\bb_-\dvd_\pp\b_-\big)
  +\big(\bb_+\dvd_\mm\b_+\big)\Big]~, }}}
Up to some numerical differences of convention, this is again in agreement
with the corresponding result of Ref.~\rTwJim.

 Again, writing somewhat formally $\rF_\bB=\inv2\e^{ab}{\cal F}_{ab}^{(B)}$,
the last bosonic term becomes the standard Yang-Mills Lagrangian density,
$-\inv4{\cal F}_{ab}^{(B)}\,{\cal F}^{ab}_{(B)}$. This, however, is not the
result of dimensional reduction of any 3+1-dimensional Yang-mills gauge
multiplet and the introduction of a gauge vector potential, ${\cal B}_a$
through writing ${\cal F}^{(B)}_{ab}=(\vd_a{\cal B}_b-\vd_b{\cal B}_a)$ is
merely for analogy. As mentioned in \SS\,\ssB, the definition~\eTRs{b}
permits us to identify, in a 3+1-dimensional framework prior to dimensional
reduction to 1+1-dimensions, the lowest components of the superfield
$\bB,\bBb$ as gauge fields covariantizing central charges, were such to
have been introduced.

For nonabelian gauge symmetry, the component field expansions of the
simple Lagrangian density terms~\eToyAB\ become rather more involved that
their abelian versions~\eAbA\ and~\eAbB. In addition, even for pure type-A
and type-B gauge multiplets, one can also write down a `superpotential'
term:
\eqn\eSPG{ \int\rd^2\sT~\Tr\S(\bA)+\hc,~\quad\hbox{and}\quad
           \int\rd^2\vs~\Tr W(\bB)+\hc, }
owing to the fact that in pure type-A gauging $\bA$ is gauge-covariantly
twisted-chiral, and in pure type-B gauging $\bB$ is gauge-covariantly
chiral. The superpotentials, $\S$ and $W$, are arbitrary analytic functions
of their respective arguments, and `$\Tr$' makes the Lagrangian density
terms gauge-invariant. Note that this annihilates all linear terms for all
nonabelian factors in $\cG_A\8\cG_B$. For the abelian, $U(1)$, factors a
linear (Fayet-Illiopoulos) term is permitted by gauge invariance, and has
been given a pivotal r\^ole in the gauged linear
$\s$-model of Refs.~\refs{\rPhases} and subsequent work.

Also, the `kinetic' terms~\eToyAB\ may well be generalized into
\eqn\eGenKinAB{ \int\rd^4\vs~\Tr K(\bA,\bAb;\bB,\bBb)~, }
where $K$ is a general, real, function of its arguments, just as first
given in Ref.~\rGHR; gauge-invariance is ensured by taking its trace. This
of course does not in general produce the `standard' Yang-Mills terms of the
type $-\inv4{\cal F}_{ab}\,{\cal F}^{ab}$, \ie,
$\inv2[\Imm(A)^2+\Imm(B)^2]$, which may be regarded as `flat'. Instead,
Eq.~\eGenKinAB\ yields an $\bA,\bB$-dependent set of terms the nonlinearity
of which is governed by the choice of the function $K$~\eGenKinAB. Of
course, the {\it choice\/} of the Lagrangian, and the adherence to the
`standard' one, is governed only by the intended application. From the
intrinsically 1+1-dimensional point of view, there is no compelling reason
in general not to consider the non-linear and rather general Lagrangian
density term~\eGenKinAB. Moreover, the D-term~\eGenKinAB\ may well be made
dependent on all available fields:
\eqn\eGenKin{ \int\rd^4\vs~\Tr K(\bA,\bAb;\bB,\bBb;\bC;\bCb;\3)~, }
where the ellipses stand for any other available superfield, including
those used to represent `matter'. This of course makes the kinetic terms
for all involved component fields depend, in general, on all of the fields
involved. Most applications will only need a special case of this general
form, but we leave that to the decision of the interested Reader, depending
on the intended application.

\subsec{General terms for gauge fields}\noindent
\subseclab\ssGenG
In `unpure' gauging types, such as the type-AC$_\mm$, discussed briefly in
\SS\,\ssMixG, the gauge superfield $\bA$ is no longer gauge-covariantly
twisted-chiral, and neither is then the integrand of the first Berezin
integral in~\eSPG. Instead, the superfield and so also
$\Tr\big(\S(\bA)\big)$ are annihilated only by $\N_+$. The {\it triple\/}
Berezin integral
\eqn\eLmbd{ \inv2\int\rd\vsb^-\rd^2\sT~\Tr\big(\S(\bA)\big) }
would be supersymmetric and gauge-invariant, but its result has
spin\ft{Recall that $\int\rd\vsb^-\simeq\bN_-$, so it has spin
$+\inv2$, not $-\inv2$ as the na\"{\ii}ve reading of the integration
measure in Eq.~\eLmbd\ would suggest.}
$+\inv2$, and so is not suitable for adding to the Lagrange density, which
has spin 0.
 Of course, {\it if\/} the considered model also includes a
spin-$(-\inv2)$ superfield which is also annihilated by $\N_+$|say, an
antichiral $\FB_+$, then
\eqn\eXXX{ \inv2\int\rd\vsb^-\rd^2\sT~\Tr\big(\S(\bA)\FB_+\big)
 = \inv4\Tr\big(\bN_-[\N_-,\bN_+]\S(\bA)\FB_+\big)\Zp }
is a candidate term for the Lagrangian density. Since $\bC_\pp{\id}0$ in
type-AC$_\mm$ gauging, $\N_+^{~2}=0$, and with a suitable twisted-antichiral
superfield, $\XB'$, we may replace $\FB_+$ in the above expressions freely
with a linear combination of $\N_+\XB'$ and $\FB_+$. Since also
$[\N_+,\bA]{=}0$, the $\N_+$ may be passed to the left and this addition,
upon including also the Hermitian conjugate, becomes proportional to the
D-term
\eqn\eAdd{ \inv4\Tr\big(\N^4\,\S(\bA)\XB'\big)\Zp~. }
{\it Typically\/}, models are {\it chosen\/} to include the most general
D-term (on the premise that quantum corrections will typically generate such
terms anyway), so that this addition is already accounted for as a special
case of a (D-)term in the total Lagrangian. However, under special
circumstances (when quantum corrections will either not be considered or
are sufficiently restricted by (additional) symmetry), it is perfectly
possible to include {\it no\/} general D-term in the Lagrangian, whence the
term~\eAdd\ is indeed a new addition.

\subsec{Interaction with matter}\noindent
\subseclab\ssIntM
Ref.~\rHSS\ has uncovered a great many candidate terms|each
(2,2)-supersymmetric all by itself|for the Lagrangian density involving
`matter' represented by the constrained superfields~\eHSF{}, \eQSF{}
and~\eNMF{}.

It is straightforward to turn all these Lagrangian density terms into their
gauge-invariant counterparts, provided a trace over the gauge group action
is taken in addition to fermionic integration, as illustrated in
Eq.~\eGenKin. This trace automatically projects out gauge-noninvariant
choices of the Lagrangian densities. Consider, for example, a collection of
$n$ chiral superfields, $\F^i$, coupled to a type-A gauged $SU(n)$ symmetry
with respect to which the $\F^i$ transform as the fundamental
representation. Then, the analytic function $W(\F)$ appearing in the
gauge-invariant $F$-term
\eqn\eGauW{ \inv2\int\rd^2\vs~\Tr\,W(\F) =
             \inv4\Tr\big([\N_-,\N_+]W(\F)\big)\Zp }
is restricted by the trace operation in front to
\eqn\eXXX{ \Tr\,W(\F) = \sum_{k=0}^\infty w_k (\F^{[n]})^k~,
 \qquad
 \F^{[n]}\define\Tr\big(\e_{i_1\cdots i_n}\F^{i_1}{\cdots}\F^{i_n}\big)~. }
This follows on realizing that powers of $\F^{[n]}$ are the only holomorphic
(chiral) $SU(n)$-invariants one can make from $\F^i$. Were the gauge group
chosen to be the $SO(n)\subset SU(n)$ subgroup, $W(\F)$ would become
\eqn\eXXX{ \Tr\,W(\F)=\sum_{k,l=0}^\infty w_{k,l}(\F^{[n]})^k(\F^{(2)})^l~,
 \qquad
  \F^{(2)}\define\Tr\big(\F^i\d_{ij}\F^j\big)~, }
where the Kronecker $\d_{ij}$ is the $SO(n)$-invariant metric. Note that in
both cases the initial (constant) term in the series is irrelevant, as it is
annihilated in the Berezin integration~\eGauW.

Keeping such gauge group dependent restrictions in mind, we conclude that
all the candidate Lagrangian density terms listed in~\rHSS\ can be used,
and that all couplings of the matter (super)fields to the gauge
(super)fields stem from the covariantization of the Berezin integration
process, \ie, from replacing the superderivatives in the
superdifferential equivalents of the various Berezin integrals with the
gauge-covariant superderivatives. Of course, one should also add the
gauge-kinetic terms~\eGenKin. Moreover, the gauge superfields may also
be included in the construction of {\it all\/} Lagrangian density terms,
\eg, the superpotential function in~\eGauW, may also be allowed to depend
on $\bB$, these being chiral. Since the $\F$ are $\bB$-chargeless, 
$[\bB,\F]{=}0$, and the superpotential factorizes into a product of
the (type-A gauge invariant) $W(\F)$ and a (type-B gauge invariant)
function of $\bB$. 

With that in mind, we now turn to a sample calculation.

\topic{The general gauge-invariant D-term}
Consider a model built from the following collection of gauge-covariantly
constrained superfields:
\eqna\eXXX
 $$\twoeqsalignno{
 \F^\m,&~\m=1,\3,N_c~,\quad&&\hbox{satisfying Eqs.}~\eHSF{a}~; &\eXXX{a}\cr
 \X^\a,&~\a=1,\3,N_t~,\quad&&\hbox{satisfying Eqs.}~\eHSF{c}~; &\eXXX{b}\cr
 \L^a ,&~ a=1,\3,N_L~,\quad&&\hbox{satisfying Eqs.}~\eHSF{e}~; &\eXXX{c}\cr
 \Y^i ,&~ i=1,\3,N_R~,\quad&&\hbox{satisfying Eqs.}~\eHSF{f}~. &\eXXX{d}\cr
}$$
This information suffices to work out the projections by all of the
superderivative operators given in Fig.~1 in
\SS\,\sCmps, which in turn suffices to determine the component field
Lagrangian densities in all cases. The Reader may wish to use the
intermediate results collected in appendix~C.

As an example, we present here the general D-term, for which we use the
gauge-covariant form of~\eDI:
\eqn\eCDI{ \inv4\int\rd^4\vs~K =
 \inv{32}\Tr\big[\{[\N_-,\bN_-],[\N_+,\bN_+]\}K\big]\Zp~.  }
The so defined Berezin integration is Hermitian; if $K{=}K^\dagger$
is a Hermitian function of its arguments, so is the resulting term in the
Lagrangian density. Furthermore, this Berezin $\rd^4\vs$-integral
is {\it even\/} under parity, and {\it odd\/} under the `mirror map',
$\bC_+$ of Ref.~\rHSS---as it should~\rGHR. This latter property we find the
crucial reason for adopting~\eCDI\ over the choice of Ref.~\rTwJim, which
transcribed into our notation becomes
\eqn\eJim{ \int\rd^4\vs~K ~\tooo{~{\rm Ref.}~\rTwJim~}~
           \inv{32}\Tr\big[\{[\N_-,\N_+],[\bN_+,\bN_-]\}K\big]\Zp~. }
Under $\bC_+$, this D-term transforms into
$\inv{32}\Tr\big[\{[\N_-,\bN_+],[\N_+,\bN_-]\}K\big]\Zp$, which would have
had to have been {\it subtracted\/} for the required {\it anti\/}symmetry
with respect to mirror symmetry.

Terms for the Lagrangian density such as~\eCDI\ (and all the many other
found in Ref.~\rHSS) may be expanded `for the general case' as follows. Let
$\W$ denote a string of all the superfields on which $K$ depends; we assume
nothing about (anti)commutivity of $\W$. For example, we may set
$\W=(\F^\m,\FB^\mb,\X^\a,\XB^\ab,\L^a,\LB^{\B{a}},\Y^i,\YB^\bi,\3)$,
including all the `species' of superfields on which $K$ may depend in
addition to what was specified above. Then, expressions like
$(\cO_1\W^1)(\cO_2\W^2)K_{12}$ simply abbreviate the sum over all
`species' 
\eqn\eXXX{{\eqalign{&(\cO_1\W^1)(\cO_2\W^2)K_{12}\define\cr
 &\]3\matrix{
 \llap{=~}(\cO_1\F^\m)(\cO_2\F^\n)K_{\1\m\n} &+& \cdots
 &+& (\cO_1\F^\m)(\cO_2\YB^\bi)K_{\1\m\bi} &+& \cdots
  \cr \noalign{\vglue1mm}
 \vdots & & \ddots & & \vdots & & \ddots
  \cr \noalign{\vglue1mm}
 (\cO_1\YB^\bi)(\cO_2\F^\n)K_{\1\bi\n} &+& \cdots
 &+& (\cO_1\YB^\bi)(\cO_2\YB^\bj)K_{\1\bi\bj} &+& \cdots
  \cr \noalign{\vglue1mm}
 \vdots & & \ddots & & \vdots & & \ddots \cr }\cr }}}
The numerical index on $\W$ is a {\it multi-index\/}, taking values in the
{\it array if indices\/}, $\m_1,\nb_1,\a_1,\bb_1,\ldots$, each of which in
turn takes on its appropriate range of values, counting the corresponding
superfields. The indices on the (multiple) superderivative operators,
$\cO_1,\cO_2$, simply distinguish one from another.
 The extreme compactness of this notation ought to be obvious.

Without any assumption regarding (anti)commutivity of $\W$, we expand
the integrand of~\eCDI\ as follows:
\eqna\eLng{\eightpoint
 $$\eqalignno{
 &\[4\Big(\big\{[\N_-,\bN_-],[\N_+,\bN_+]\big\}K\Big)\cr
 &\[4=\Big(\big\{[\N_-,\bN_-],[\N_+,\bN_+]\big\}\W^1\Big)K_1  &\eLng{a}\cr
 &\[3+\big[([\N_+,\bN_+]\W^1)([\N_-,\bN_-]\W^2)
       +([\N_-,\bN_-]\W^1)([\N_+,\bN_+]\W^2)\big]K_{12}         &\eLng{b}\cr
 &\[3-2\big[(\bN_+\bN_-\W^1)(\N_+\N_-\W^2)+(\bN_-\bN_+\W^1)(\N_-\N_+\W^2)\cr
 &+(\N_+\N_-\W^1)(\bN_+\bN_-\W^2)+(\N_-\N_+\W^1)(\bN_-\bN_+\W^2)
                                                    \big]K_{12} &\eLng{c}\cr
 &\[3+2\big[(\N_+\bN_-\W^1)(\bN_+\N_-\W^2)+(\bN_-\N_+\W^1)(\N_-\bN_+\W^2)\cr
 &+(\bN_+\N_-\W^1)(\N_+\bN_-\W^2)+(\N_-\bN_+\W^1)(\bN_-\N_+\W^2)
                                                    \big]K_{12} &\eLng{d}\cr
 &\[3+(-)^{\p_1}\big[2(\N_+[\N_-,\bN_-]\W^1)(\bN_+\W^2)
       +2(\N_-[\N_+,\bN_+]\W^1)(\bN_-\W^2)\cr
 &\]3+([\N_-,\bN_-]\N_+\W^1)(\bN_+\W^2)+([\N_+,\bN_+]\N_-\W^1)(\bN_-\W^2)\cr
 &\]3-(\bN_+\W^1)([\N_-,\bN_-]\N_+\W^2)-(\bN_-\W^1)([\N_+,\bN_+]\N_-\W^2)
                                                    \big]K_{12} &\eLng{e}\cr
 &\[3-(-)^{\p_1}\big[2(\bN_+[\N_-,\bN_-]\W^1)(\N_+\W^2)
       +2(\bN_-[\N_+,\bN_+]\W^1)(\N_-\W^2)\cr
 &\]3+([\N_-,\bN_-]\bN_+\W^1)(\N_+\W^2)+([\N_+,\bN_+]\bN_-\W^1)(\N_-\W^2)\cr
 &\]3-(\N_+\W^1)([\N_-,\bN_-]\bN_+\W^2)-(\N_-\W^1)([\N_+,\bN_+]\bN_-\W^2)
                                                    \big]K_{12} &\eLng{f}\cr
 &\[3+(-)^{\p_2}\big[([\N_+,\bN_+]\W^1)(\N_-\W^2)(\bN_-\W^3)
       -([\N_+,\bN_+]\W^1)(\bN_-\W^2)(\N_-\W^3)\cr
 &\]3+(\N_-\W^1)(\bN_-\W^2)([\N_+,\bN_+]\W^3)
       -(\bN_-\W^1)(\N_-\W^2)([\N_+,\bN_+]\W^3)\cr
 &\]3+([\N_-,\bN_-]\W^1)(\N_+\W^2)(\bN_+\W^3)
       -([\N_-,\bN_-]\W^1)(\bN_+\W^2)(\N_+\W^3)\cr
 &\]3+(\N_+\W^1)(\bN_+\W^2)([\N_-,\bN_-]\W^3)
       -(\bN_+\W^1)(\N_+\W^2)([\N_-,\bN_-]\W^3)\big]K_{123}     &\eLng{g}\cr
 &\[3+(-)^{\p_2}2\big[[(\N_-\N_+\W^1)(\bN_+\W^2)
       +(-)^{\p_1}(\bN_+\W^1)(\N_-\N_+\W^2)](\bN_-\W^3)\cr
 &\]3+[(\N_+\N_-\W^1)(\bN_-\W^2)
       +(-)^{\p_1}(\bN_-\W^1)(\N_+\N_-\W^2)](\bN_+\W^3)\cr
 &\]3+[(\bN_+\bN_-\W^1)(\N_-\W^2)
       +(-)^{\p_1}(\N_-\W^1)(\bN_+\bN_-\W^2)](\N_+\W^3)\cr
 &\]3+[(\bN_-\bN_+\W^1)(\N_+\W^2)
       +(-)^{\p_1}(\N_+\W^1)(\bN_-\bN_+\W^2)](\N_-\W^3)
                                                   \big]K_{123} &\eLng{h}\cr
 &\[3-(-)^{\p_2}2\big[[(\N_-\bN_+\W^1)(\N_+\W^2)
       +(-)^{\p_1}(\N_+\W^1)(\N_-\bN_+\W^2)](\bN_-\W^3)\cr
 &\]3+[(\bN_-\N_+\W^1)(\bN_+\W^2)
       +(-)^{\p_1}(\bN_+\W^1)(\bN_-\N_+\W^2)](\N_-\W^3)\cr
 &\]3+[(\N_+\bN_-\W^1)(\N_-\W^2)
       +(-)^{\p_1}(\N_-\W^1)(\N_+\bN_-\W^2)](\bN_+\W^3)\cr
 &\]3+[(\bN_+\N_-\W^1)(\bN_-\W^2)
       +(-)^{\p_1}(\bN_-\W^1)(\bN_+\N_-\W^2)](\N_+\W^3)
                                                   \big]K_{123} &\eLng{i}\cr
 &\[3+(-)^{\p_2{\cdot}\p_3}
      \big[[(\N_-\W^1)(\bN_-\W^2)-(\bN_-\W^1)(\N_-\W^2)]
        [(\N_+\W^3)(\bN_+\W^4)-(\bN_+\W^3)(\N_+\W^4)]\cr
 &\]5+[(\N_+\W^1)(\bN_+\W^2)-(\bN_+\W^1)(\N_+\W^2)]
        [(\N_-\W^3)(\bN_-\W^4)-(\bN_-\W^3)(\N_-\W^4)]\big]K_{1234}~,~~~~~
                                                                &\eLng{j}\cr
}$$}%
where $(-)^{\p_1}\define(-)^{\p(\W^1)}=\pm1$ when $\W_1$ is
(anti)commuting; see appendix~A.

Most frequently, the $\W$'s simply stand for a string of bosonic
(commuting) superfields, $(-)^{\p_1}={\cdots}=(-1)^{\p_4}=+1$, and the
multiple derivatives of $K(\W)$ satisfy the usual {\it symmetrization\/}:
$K_{12}=K_{21}$, $K_{123}=K_{(123)}$ and $K_{1234}=K_{(1234)}$, and the
(anti)commutivity of (multiple) derivatives $(\cO\W)$ stems solely from the
(anti)commutivity of the (multiple) derivative  $\cO$. This permits the
combination of many of the terms in~\eLng{}, and results in:
\eqna\eQck{
 $$\eqalignno{
 &\[4\Big(\big\{[\N_-,\bN_-],[\N_+,\bN_+]\big\}K\Big)\cr
 &\[4=\Big(\big\{[\N_-,\bN_-],[\N_+,\bN_+]\big\}\W^1\Big)K_1    &\eQck{a}\cr
 &\[3+2\big[([\N_+,\bN_+]\W^1)([\N_-,\bN_-]\W^2)\big]K_{12}     &\eQck{b}\cr
 &\[3-4\big[(\bN_+\bN_-\W^1)(\N_+\N_-\W^2)+(\bN_-\bN_+\W^1)(\N_-\N_+\W^2)
                                                    \big]K_{12} &\eQck{c}\cr
 &\[3+4\big[(\N_+\bN_-\W^1)(\bN_+\N_-\W^2)+(\bN_-\N_+\W^1)(\N_-\bN_+\W^2)
                                                    \big]K_{12} &\eQck{d}\cr
 &\[3+2\big[(\{\N_+,[\N_-,\bN_-]\}\W^1)(\bN_+\W^2)
           +(\{\N_-,[\N_+,\bN_+]\}\W^1)(\bN_-\W^2)\big]K_{12}   &\eQck{e}\cr
 &\[3-2\big[(\{\bN_+,[\N_-,\bN_-]\}\W^1)(\N_+\W^2)
           +(\{\bN_-,[\N_+,\bN_+]\}\W^1)(\N_-\W^2)\big]K_{12}   &\eQck{f}\cr
 &\[3+4\big[(\N_-\W^1)(\bN_-\W^2)([\N_+,\bN_+]\W^3)
           +([\N_-,\bN_-]\W^1)(\N_+\W^2)(\bN_+\W^3)\big]K_{123} &\eQck{g}\cr
 &\[3+4\big[([\N_-,\N_+]\W^1)(\bN_+\W^2)(\bN_-\W^3)
           +([\bN_+,\bN_-]\W^1)(\N_-\W^2)(\N_+\W^3)\big]K_{123} &\eQck{h}\cr
 &\[3-4\big[([\N_-,\bN_+]\W^1)(\N_+\W^2)(\bN_-\W^3)
           +([\N_+,\bN_-]\W^1)(\N_-\W^2)(\bN_+\W^3)\big]K_{123} &\eQck{i}\cr
 &\[3+8(\N_-\W^1)(\bN_-\W^2)(\N_+\W^3)(\bN_+\W^4)K_{1234}~.     &\eQck{j}\cr
}$$}

All Lagrangian densities obtained in this way involve a final $\vs,\vsb=0$
projection, which can be performed on the expansions~\eLng{} and~\eQck{}
term by term and factor by factor, using the projections collected in
appendix~C. The final insertion of these projections and the collection of
(very many) terms for the general D-term~\eCDI, and any other of the many
possible terms~\rHSS, we leave to the diligent Reader.

\newsec{Summary, Outlook and Conclusions}\noindent
\seclab\sSOC
The above analysis shows how to couple matter fields, represented by
constrained superfields~\ft{Recall that the quartoid superfields~\eQSF{}
cannot couple to any of the gauge fields by means of  `minimal
coupling'.}~\eHSF{} and~\eNMF{}, to `minimally coupled' gauge (super)fields.

Considering the most general gauge-covariant extension of the supersymmetry
algebra~\eTRs{} and the consistency requirements (appendix~B), we find that
1+1-dimensional theories admit {\it four\/} distinct types of symmetry
gauging: \SS~\ssA--\ssCp. Furthermore, allowing for a duplication among
gauge superfield components, there also exist additional, `mixed' types of
symmetry gauging; see \SS~\ssMixG.

Given the wealth of candidate Lagrangian density terms listed in Ref.~\rHSS,
this uncovers a vast arena in which to generalize the results of
Refs.~\refs{\rPhases,\rHPS,\rTwJim,\rGGW}. The methods of
Refs.~\refs{\rWAB,\rBeast} then are well suited to explore the geometry of
the target spaces as related to the quantum dynamics in these models.

The `projection method' described above makes it possible to
straightforwardly gauge-covariantize all of the candidate Lagrangian
density terms listed in Ref.~\rHSS. The process reduces to the following
five steps:
\item{1.} Select a candidate Lagrangian density term from Ref.~\rHSS.
Above, we chose $\int\rd^4\vs~K$.
\item{2.} Gauge-covariantize the chosen term by substituting
gauge-covariant superderivatives and applying the overall trace operation.
Above, this produces the D-term~\eCDI
\item{3.} Expand the (multiple) superderivatives; to this end, the
identities in appendix~A are useful. Above, this resulted in the
expansions~\eLng{} and~\eQck{}.
\item{4.} Perform the $\vs,\vsb=0$ projections; to this end, the
intermediate results in appendix~C are useful. This final (and voluminous)
collection of the results for the general D-term~\eCDI, and any other of the
many possible terms~\rHSS, we leave to the interested Reader.
\item{5.} The previous two steps in the calculation are more easily
performed using the gauge group index notation~\eCovInd--\eTTrInd. The
results can then be recast into the perhaps neater `matrix' notation,
tidying the results further by using the cyclicity of the overall trace
operation.

For each of these candidate Lagrangian density terms, supersymmetry is most
easily proven upon the non-unitary decovariantization described in
\SS~\ssDeCov. On the other hand, the gauge-covariant formalism used
throughout makes gauge invariance manifest.

The combination of the above results and techniques, and the large number of
candidate Lagrangian density terms~\rHSS\ guarantees an unsuspected wealth
of various (2,2)-supersymmetric gauged $\s$-models in 1+1-dimensional
spacetime. This provides a vast number of generalizations to the models of
Refs.~\refs{\rPhases,\rHPS,\rTwJim,\rGGW}, and a stable framework for
exploring these generalizations.

 %
\vfill\eject
\appendix{A}{Absolutely Basic Conventions (ABC)}\noindent
We follow Ref.~\refs{\rHSS}'s adaptation of Wess and Bagger's
definitions~\rWB. The $-,+$ (spinorial) indices actually indicate spin:
$\j^-{=}\j_+$ has spin $-\inv2$, whereas $\j^+{=}-\j_-$ has spin $+\inv2$.
Since $\j{\cdot}\c\define\j^\a\c_\a$, but
$\jb{\cdot}\cb\define\jb_\ad\cb^\ad$,
it follows that
\eqna\eSqr
 $$ \twoeqsalignno{
 \j^2  &=~[\j^+,\j^-]~=~2\j^+\j^-~,  \quad&\quad
 \c^2  &=~[\c_+,\c_-]~=~2\c_+\c_-~,  &\eSqr{a}\cr
 \noalign{\vskip-3mm\noindent but\vskip-2mm}
 \cb^2 &=~[\cb^-,\cb^+]~=~2\cb^-\cb^+~, \quad&\quad
 \jb^2 &=~[\jb_-,\jb_+]~=~2\jb_-\jb_+~,  &\eSqr{b}\cr
 }$$
and where we used $\e^{-+}=1=\e_{+-}$; we also use $\e^{01}=1=\e_{10}$.

The formulae~\eSqr{} explain the order of the superderivatives in~\eHFI{};
for example,
\eqn\eXXX{ \int\rd^2\vs~(\vs^2) = \inv2[D_-,D_+](2\vs^+\vs^-)\Zp=2~. }
Owing to the functional equivalence of the Berezin integration to
derivative, spinorial Dirac delta-functions are particularly simple,
\eg, $\d(\vs^+)\id\vs^+$. Multiple delta-functions are of course products
of simple ones, and we merely have to choose the order and the sign. Fixing
$\d^4(\vs)\define\vsb^-\vsb^+\vs^+\vs^-$ so that
$\int\rd^4\vs\>\d^4(\vs)=1$, we set
\eqn\eXXX{ \d^2(\vs)=\vs^+\vs^-~,\quad
            \d^2(\sT)=\vsb^+\vs^-~,\quad
             \d^2(\sR)=\vsb^-\vs^-~, }
\eqn\eXXX{ \d^2(\vsb)=\vsb^-\vsb^+~,\quad
            \d^2(\sW)=\vsb^-\vs^+,\quad
             \d^2(\sL)=\vsb^+\vs^+~, }
so that Eqs.~\eHFI{} follow from~\eDI.\ping

Switching to gauge-covariant (super)derivatives, the following {\it
operatorial\/} identity may be of help. Let $\N_i$ range
over the gauge-covariant superderivatives~\eCDs{}. Then:
\eqn\eHfHf{ \N_1\N_2=\inv2[\N_1,\N_2]+\inv2\{\N_1,\N_2\}~, }
so that
\eqn\eXXX{ [\N_1,\N_2]=2\N_1\N_2-\{\N_1,\N_2\}=\{\N_1,\N_2\}-2\N_2\N_1~. }
Then, for example
\eqna\eIds
 $$\eqalignnotwo{
 [\N_-,\bN_-]&\id2(\N_-\bN_- -i\N_\mm)&\id-2(\bN_-\N_- -i\N_\mm)~,
&\eIds{a}\cr
 [\N_+,\bN_+]&\id2(\N_+\bN_+ -i\N_\pp) &\id-2(\bN_+\N_+ -i\N_\pp)~,
&\eIds{b}\cr
 }$$
and so on. Similarly,
\eqna\eXXX
 $$\eqalignno{
 [\N_-,\bN_-]\N_-&\id 2i\N_\mm \N_--2\bN_-\bCb_\mm~, &\eXXX{a}\cr
 [\N_-,\bN_-]\bN_-&\id 2\N_-\bC_\mm-2i\N_\mm\bN_-~, &\eXXX{b}\cr
 [\N_+,\bN_+]\N_+&\id 2i\N_\pp \N_+-2\bN_+\bCb_\pp~, &\eXXX{c}\cr
 [\N_+,\bN_+]\bN_+&\id 2\N_+\bC_\pp-2i\N_\pp\bN_+~. &\eXXX{d}\cr
 }$$
\ping

When using gauge-covariant derivatives to calculate gauge-covariant Berezin
integrals|or, indeed, simply apply to superfields, the (anti)commutator
notation is more precise, albeit also more cumbersome. Let
\eqn\eXXX{\p(\BM{X})\define\cases{0\cr1\cr}\quad\hbox{if \BM{X} is }
                       \cases{\hbox{commutative,}\cr\noalign{\vglue1mm}
                              \hbox{anticommutative,}\cr}}
so that
\eqn\eACm{ \big[\,\BM{X}\,,\,\BM{Y}\,\big\}~\define~
 \BM{XY}-(-)^{\p(\BM{X})\cdot\p(\BM{Y})}\BM{YX} }
is the graded commutator: a commutator unless both arguments are
anticommutative in which case it is an anticommutator.
 Then,
\eqna\eBrz
 $$\eqalignno{
 \int\rd\vs^1\,\BM{X}\define D_1\BM{X}\Zp
 &\to~[\N_1,\BM{X}\}\Zp~, &\eBrz{a}\cr
 \int\rd\vs^1\rd\vs^2\,\BM{X}\define \inv2[D_1,D_2]\BM{X}\Zp
 &\to~\inv2\big[\N_{[1},[\N_{2]},\BM{X}\}\big\}\Zp~, &\eBrz{b}\cr
}$$
and so on, iterating these two expressions. Note that the index on the
$\N$'s is meant to encode also the conjugation bar. Before switching to the
nested (anti)commutator notation, the expression~\eDI{} merely needed
covariantization:
 $$ \int\rd^4\vs~(\3) ~\define~
 \inv8\big\{[\N_-,\bN_-],[\N_+,\bN_+]\big\}(\3)\Zp ~\define~(\N^4\3)\Zp~,
 \eqno\eCDI $$
Finally, to obtain the gauge-covariant D-term projector corresponding to
the Berezin $\rd^4\vs$-integration, the substitution~\eBrz{b} is iterated
twice.

 The full $\rd^4\vs$ integration measure actually corresponds to the full,
totally antisymmetrized product of the four superderivatives. However, it
is not necessary to calculate with all the $4!$ permutations of
$\N_-\bN_-\N_+\bN_+$, as many are equivalent up to total derivatives. This
leads to various {\it choices\/} of calculationally convenient subsets of
the $4!$ quartic superderivatives for the assignment
$\int\rd^4\vs[\3]\to(\N^4\3)|$. Several such choices are readily found in
the literature, typically obtained through dimensional reduction from
3+1-dimensional $N{=}1$ supersymmetry. However, our discussion of
1+1-dimensional $(2,2)$-supersymmetric gauge theories follows the intrinsic
approach of Ref.~\rHSS, disregarding as much as possible dimensional
reduction. We then require the assignment~\eCDI\ to satisfy the usual
requirement of Hermiticity, and evenness under parity. In addition, we
require that $\N^4$ should be odd with respect to the `mirror map',
$\bC_+$~\rHSS. The latter requirement would necessitate a doubling of terms
for the Ref.~\rTwJim's choice, but leaves~\eCDI\ as it is. This choice
provides for greater symmetry between type-A and type-B gauged models with
chiral and twisted-chiral superfields, although it also causes a few minor
differences when comparing with the framework set-up (but not the results
presented) in Ref.~\rTwJim.\ping

The work with nested (anti)commutators is simplified a good deal by using
the `derivative' (operatorial) identities:
\eqna\eXXX
 $$\eqalignno{
 \big[\BM{XY},\BM{Z}\big\}
 &\id\BM{X}\big[\BM{Y},\BM{Z}\big\} + (-)^{\p(\BM{Y})\cdot\p(\BM{Z})}
            \big[\BM{X},\BM{Z}\big\}\BM{Y}~, &\eXXX{a}\cr
 \noalign{\vglue-2mm\noindent and\vglue-2mm}
 \big[\BM{X},\BM{YZ}\big\}
 &\id\big[\BM{X},\BM{Y}\big\}\BM{Z} + (-)^{\p(\BM{X})\cdot\p(\BM{Y})}
            \BM{X}\big[\BM{Y},\BM{Z}\big\}~. &\eXXX{b}\cr
 }$$
In particular, then
\eqn\eXXX{ \big[\N,\BM{XY}\big\}
 \id\big[\N,\BM{X}\big\}\BM{Y} + (-)^{\p(\BM{X})}
            \BM{X}\big[\N,\BM{Y}\big\}~. }
Iterating this for superderivatives $\N_1,\N_2$, we obtain
\eqn\eXXX{{\eqalign{
 &\big[\N_1,[\N_2,\BM{XY}\}\big\}\cr
 &\id\big[\N_1,[\N_2,\BM{X}\}\big\}\BM{Y}
 + (-)^{\p(\BM{X})}[\N_{[1},\BM{X}\}[\N_{2]},\BM{Y}\}
 + \BM{X}\big[\N_1,[\N_2,\BM{Y}\}\big\}~. }}}
where bracketing indices indicates their antisymmetrization, as
in~\eFinW{d}; it stems from the anticommutivity of the $\N$'s. 
When \BM{X} and \BM{Y} are simple (multiplicative) {\it commuting\/}
superfields and $\N_1,\N_2$ superderivatives, this reduces to
\eqn\eXXX{ (\N_1\N_2\BM{XY})
 =(\N_1\N_2\BM{X})\BM{Y} + (-)^{\p(\BM{X})} 2(\N_1\BM{X})(\N_2\BM{Y})
  + \BM{X}(\N_1\N_2\BM{Y})~, }
in agreement with Eq.~(A.11) of Ref.~\rHSS.

The (anti)commutator analogue of the `chain rule' is
\eqn\eXXX{ [\N,f(\BM{X})\}~\id~[\N,\BM{X}\}\,f'(\BM{X})~. }
Iterating this, we obtain
\eqn\eXXX{ \big[\N_1,[\N_2,f(\BM{X})\}\big\}
 \id~\big[\N_1,[\N_2,\BM{X}\}\big\}\,f'(\BM{X})
  -[\N_2,\BM{X}\}[\N_1,\BM{X}\}f''(\BM{X})~, }
which, after antisymmetrizing of the superderivatives, shows how to apply
the expressions~\eBrz{b} for a two-fold gauge-covariant Berezin integration.
When $\BM{X}^i$ are simple (multiplicative) commuting superfields and
$\N_1,\N_2$ superderivatives, this reduces to
\eqn\eXXX{
 (\N_1\N_2\>f(\BM{X}))
 \id(\N_1\N_2\BM{X}^i)\,f_{\1i}(\BM{X})
  -(\N_2\BM{X}^i)(\N_1\BM{X}^j)f_{\1ij}(\BM{X})~. }
Note that the order of superderivatives becomes reversed in the second
term:$(\N_1\BM{X}^j)$ is not trivial to `pass back to the left', as
the (superderivatives of) $\BM{X}^i$ may not (anti)commute.

Finally, the Jacobi identities (see appendix~B) imply the following two
identities:
\eqna\eXXX
 $$\eqalignno{ \{\N_1,[\N_2,\BM{X}]\}
  &= -\{\N_2,[\N_1,\BM{X}]\} + [\{\N_1,\N_2\},\BM{X}]~,  &\eXXX{a}\cr
  &= \inv2\{\N_{[1},[\N_{2]},\BM{X}]\}+\inv2[\{\N_1,\N_2\},\BM{X}]~, 
                                                         &\eXXX{b}\cr
}$$
the latter of which is also seen as an application of the operatorial
identity~\eHfHf.

A combined iteration of the above identities typically simplifies most of
the calculations significantly. Even so, however, the considerable
technical tedium makes one wish for a mechanization of these calculations.
Short of this, we present the details of the calculations involving
the (graded) Jacobi identities and component field projections in the next
two appendices, and several sample results are presented within the body of
the article; we hope this will suffice for the interested Reader to master
the technique.

A final remark is in order, relating to Eq.~\eTTr. The original definitions
in \SS~\ssDefs\ imply a `matrix' representation of Lie algebra-valued
objects, which is especially important in case of nonabelian symmetries.
That is, the gauge-covariant derivatives $\N$, and all gauge (super)fields
should be regarded as square matrices. On `column-vectors' \BM{X}, these
act from the left, but on `row-vectors' $\bXB$, they should act from the
right. To avoid a meticulous indication of this leftward action, we adopt a
`mixed' notation in which various derivatives act always from the left,
whereas Lie algebra valued objects act as appropriate. Thus (suppressing
spacetime, \ie, spin indices):
\eqn\eXXX{ \N\BM{X}=D\BM{X}-i\GB\BM{X}~, \quad\hbox{but}\quad
            \N\bXB=D\bXB-i\bXB\GB~, }
which also follows from Eq.~\eTTr. For the record, in the explicit
gauge group index notation, this is
\eqn\eXXX{ \N_\b^\a\BM{X}^\b=D\BM{X}^\a-i\GB^\a_\b\BM{X}^\b~,
            \quad\hbox{but}\quad
             \N^\b_\a\bXB_\b=D\bXB_\b-i\bXB_\b\GB^\b_\a~, }
where now the ordering of the superfields, both being bosonic, no longer
matters.

As Lagrangian density terms must be gauge-invariant, the Fermionic
integration is also meant to implicitly include a trace over the gauge group
action.
 But then, owing to the cyclicity of the trace operator and using Eq.~\eTTr,
the product rule yields, e.g.,
\eqn\eChn{{\eqalign{\Tr\big[\N\BM{X}\bXB\big]
 &= \Tr\big[\bXB\7{\to}{\N}\BM{X}\big]
   +\Tr\big[\bXB\7{\from}{\N}\BM{X}\big]~,\cr
 &= \Tr\big[\bXB(\N\BM{X})\big]
   +\Tr\big[(\N\T\bXB\T)\T\,\BM{X}\big]~,\cr
   }}}
where both terms are separately gauge-invariant, as they transform into
\eqn\eXXX{ \Tr\big[\bXB\CG^{-1}\CG\7{\to}{\N}\CG^{-1}\CG\BM{X}\big]
   +\Tr\big[\bXB\CG^{-1}\CG\7{\from}{\N}\CG^{-1}\CG\BM{X}\big]~.  }
Higher order superderivative analogues of this are not difficult to obtain
iterating the chain rule~\eChn. This proves their gauge-invariance and also
allows explicit calculations in terms of the projections listed in
appendix~C.

The only alternative to this procedure is the use of the (de)covariantizing
non-unitary transformation~\eDeCov. This lets the Berezin integration to be
calculated as a projection using gauge-noncovariant superderivatives
$D_\mp,\Db_\mp$ which act on all superfields standardly, from the left. The
Berezin integrand, in turn, becomes covariantized through the explicit
insertion of factors such as $\bCH\CH=\ex{-2\CV}$ in~\eStdNrm, in case of
the `flat' kinetic term for a chiral superfield. Such covariantizing
factors would have to be determined separately for every other candidate
term for the Lagrangian density. To the best of our knowledge, the use of
this gauge non-covariant notation and framework is prevalent in the
existing literature. In spite of the slight notational awkwardness
described above, we choose instead the gauge covariant notation and
framework for its universality.

\appendix{B}{Graded Jacobi Identities}\noindent 
The general, graded Jacobi (cyclic) {\it identity\/}
$\big[\BM{X}_{<i},[\BM{X}_j,\BM{X}_{k>}\}\big\}\id0$, is a property of the
binary operation~\eACm, and does not depend on the (derivative,
multiplicative or otherwise) nature of $\BM{X}_i$; $<\3>$ here indicates a
summation over cyclic permutations of the enclosed indices. It is therefore
used to verify the consistency of the algebra~\eTRs{}.

The Jacobi identities involving three spinorial $\N$'s are:
\eqna\eJfff
 $$\eqalignno{
 0&\: 3\big[\N_-,\{\N_-,\N_-\}\big] = 6[\N_-,\bCb_\mm]~;       &\eJfff{a}\cr
 0&\: \big[\bN_-,\{\N_-,\N_-\}\big]
      +\big[\N_-,\{\N_-,\bN_-\}\big]
      +\big[\N_-,\{\bN_-,\N_-\}\big]~,\cr
  &=2[\bN_-,\bCb_\mm]-4i\bWb_\Mm~;                           &\eJfff{b}\cr
 0&\: \big[\N_+,\{\N_-,\N_-\}\big]
      +\big[\N_-,\{\N_-,\N_+\}\big]
      +\big[\N_-,\{\N_+,\N_-\}\big]~,\cr
  &=2[\N_+,\bCb_\mm]+2[\N_-,\bBb]~;                            &\eJfff{c}\cr
 0&\: \big[\bN_+,\{\N_-,\N_-\}\big]
      +\big[\N_-,\{\N_-,\bN_+\}\big]
      +\big[\N_-,\{\bN_+,\N_-\}\big]~,\cr
  &=2[\bN_+,\bCb_\mm]+2[\N_-,\bAb]~;                           &\eJfff{d}\cr
 0&\: \big[\bN_-,\{\N_-,\bN_-\}\big]
      +\big[\N_-,\{\bN_-,\bN_-\}\big]
      +\big[\bN_-,\{\bN_-,\N_-\}\big]~,\cr
  &=-4i\bW_\Mm+2[\N_-,\bC_\mm]~;                             &\eJfff{e}\cr
 0&\: \big[\N_+,\{\N_-,\bN_-\}\big]
      +\big[\N_-,\{\bN_-,\N_+\}\big]
      +\big[\bN_-,\{\N_+,\N_-\}\big]~,\cr
  &=-2i\bWb_-+[\N_-,\bA]+[\bN_-,\bBb]~;                        &\eJfff{f}\cr
 0&\: \big[\bN_+,\{\N_-,\bN_-\}\big]
      +\big[\N_-,\{\bN_-,\bN_+\}\big]
      +\big[\bN_-,\{\bN_+,\N_-\}\big]~,\cr
  &=-2i\bW_-+[\N_-,\bB]+[\bN_-,\bAb]~;                         &\eJfff{g}\cr
 0&\: \big[\bN_+,\{\N_-,\bN_+\}\big]
      +\big[\N_-,\{\bN_+,\bN_+\}\big]
      +\big[\bN_+,\{\bN_+,\N_-\}\big]~,\cr
  &=2[\bN_+,\bAb]+2[\N_-,\bC_\pp]~;                            &\eJfff{h}\cr
 0&\: 3\big[\bN_-,\{\bN_-,\bN_-\}\big] = 6[\bN_-,\bC_\mm]~;    &\eJfff{i}\cr
 0&\: \big[\bN_+,\{\bN_-,\bN_-\}\big]
      +\big[\bN_-,\{\bN_-,\bN_+\}\big]
      +\big[\bN_-,\{\bN_+,\bN_-\}\big]~,\cr
  &=2[\bN_+,\bC_\mm]+2[\bN_-,\bB]~;                            &\eJfff{j}\cr
 }$$
and the ten identities~\eJfff{k\?t}, omitted here, but obtained
from~\eJfff{a\?j} upon a `$-{\iff}+$' swap in the subscripts (note that
this changes the signs of the $\bW$'s, swaps $\bA{\iff}\bAb$ but leaves
$\bB,\bBb$ intact).

The Jacobi identities involving one vectorial and two spinorial $\N$'s are:
\eqna\eJffb
 $$\eqalignno{
 0&\:[\N_\mm,\{\N_-,\N_-\}]+\{\N_-,[\N_-,\N_\mm]\}+\{\N_-,[\N_-,\N_\mm]\}\cr
  &=2[\N_\mm,\bCb_\mm]-2\{\N_-,\bWb_\Mm\}~,                  &\eJffb{a}\cr
 0&\:[\N_\mm,\{\N_-,\bN_-\}]+\{\N_-,[\bN_-,\N_\mm]\}
                           +\{\bN_-,[\N_-,\N_\mm]\}\cr
  &=-\{\N_-,\bW_\Mm\}-\{\bN_-,\bWb_\Mm\}~,                 &\eJffb{b}\cr
 0&\:[\N_\mm,\{\N_-,\N_+\}]+\{\N_-,[\N_+,\N_\mm]\}
                           +\{\N_+,[\N_-,\N_\mm]\}\cr
  &=[\N_\mm,\bBb]-\{\N_-,\bWb_-\}-\{\N_+,\bWb_\Mm\}~,        &\eJffb{c}\cr
 0&\:[\N_\mm,\{\N_-,\bN_+\}]+\{\N_-,[\bN_+,\N_\mm]\}
                           +\{\bN_+,[\N_-,\N_\mm]\}\cr
  &=[\N_\mm,\bAb]-\{\N_-,\bW_-\}-\{\bN_+,\bWb_\Mm\}~,        &\eJffb{d}\cr
 0&\:[\N_\mm,\{\bN_-,\bN_-\}]+\{\bN_-,[\bN_-,\N_\mm]\}
                           +\{\bN_-,[\bN_-,\N_\mm]\}\cr
  &=2[\N_\mm,\bC_\mm]-2\{\bN_-,\bW_\Mm\}~,                   &\eJffb{e}\cr
 0&\:[\N_\mm,\{\bN_-,\N_+\}]+\{\bN_-,[\N_+,\N_\mm]\}
                           +\{\N_+,[\bN_-,\N_\mm]\}\cr
  &=[\N_\mm,\bA]-\{\bN_-,\bWb_-\}-\{\N_+,\bW_\Mm\}~,         &\eJffb{f}\cr
 0&\:[\N_\mm,\{\bN_-,\bN_+\}]+\{\bN_-,[\bN_+,\N_\mm]\}
                           +\{\bN_+,[\bN_-,\N_\mm]\}\cr
  &=[\N_\mm,\bB]-\{\bN_-,\bW_-\}-\{\bN_+,\bW_\Mm\}~,         &\eJffb{g}\cr
 0&\:[\N_\mm,\{\N_+,\N_+\}]+\{\N_+,[\N_+,\N_\mm]\}
                           +\{\N_+,[\N_+,\N_\mm]\}\cr
  &=2[\N_\mm,\bCb_\pp]-2\{\N_+,\bWb_-\}~,                      &\eJffb{h}\cr
 0&\:[\N_\mm,\{\N_+,\bN_+\}]+\{\N_+,[\bN_+,\N_\mm]\}
                           +\{\bN_+,[\N_+,\N_\mm]\}\cr
  &=2\bF-\{\N_+,\bW_-\}-\{\bN_+,\bWb_-\}~,                     &\eJffb{i}\cr
 0&\:[\N_\mm,\{\bN_+,\bN_+\}]+\{\bN_+,[\bN_+,\N_\mm]\}
                           +\{\bN_+,[\bN_+,\N_\mm]\}\cr
  &=2[\N_\mm,\bC_\pp]-2\{\bN_+,\bW_-\}~,                       &\eJffb{j}\cr
}$$
and the ten identities~\eJffb{k\?t}, omitted here, but easily obtained
from~\eJffb{a\?j} upon a `$-{\iff}+$' swap in the subscripts (recall that
this changes the signs of the $\bW$'s and of $\bF$, swaps $\bA{\iff}\bAb$
but leaves $\bB,\bBb$ intact). Notice that in Eq.~\eJffb{b},
\eqn\eXXX{ [\N_\mm,\{\N_-,\bN_-\}] =2i[\N_\mm,\N_\mm]\id0~, }
since the $\GB^j_\mm$ in $\N_\mm=\vd_\mm-i\GB^j_\mm T_j$ are commutative.
This sets the `usual' bosonic $\cG_\mm$-field strength,
$\bF_{\mm\mm}$, to zero. Similarly, $\bF_{\pp\pp}\id0$ for $\cG_\pp$.

There are four Jacobi identities involving two bosonic and one fermionic
$\N$:
\eqna\eJffb
 $$\eqalignno{
 0&\:\big[\N_\mm,[\N_\pp,\N_-]\big]+\big[\N_\pp,[\N_-,\N_\mm]\big]
        +\big[\N_-,[\N_\mm,\N_\pp]\big]\cr
  &=-i[\N_\mm,\bWb_+]-i[\N_\pp,\bWb_\Mm]+i[\N_-,\bF]~,      &\eJffb{a}\cr
 0&\:\big[\N_\mm,[\N_\pp,\bN_-]\big]+\big[\N_\pp,[\bN_-,\N_\mm]\big]
        +\big[\bN_-,[\N_\mm,\N_\pp]\big]\cr
  &=-i[\N_\mm,\bW_+]-i[\N_\pp,\bW_\Mm]+i[\bN_-,\bF]~,       &\eJffb{b}\cr
 0&\:\big[\N_\mm,[\N_\pp,\N_+]\big]+\big[\N_\pp,[\N_+,\N_\mm]\big]
        +\big[\N_+,[\N_\mm,\N_\pp]\big]\cr
  &=-i[\N_\mm,\bWb_+^\mm]-i[\N_\pp,\bWb_-]+i[\N_+,\bF]~,      &\eJffb{c}\cr
 0&\:\big[\N_\mm,[\N_\pp,\bN_+]\big]+\big[\N_\pp,[\bN_+,\N_\mm]\big]
        +\big[\bN_+,[\N_\mm,\N_\pp]\big]\cr
  &=-i[\N_\mm,\bW_+^\mm]-i[\N_\pp,\bW_-]+i[\bN_+,\bF]~.       &\eJffb{d}\cr
}$$
All these are identically satisfied upon substituting one of Eqs.~\eFinW{}
and some of the Eqs.~\eWs{}. Finally the Jacobi identities involving only
bosonic $\N$'s are trivially satisfied as there is only two bosonic $\N$'s.

\appendix{C}{Component Projections}\noindent
The following is a collection of identities and component field projections
which the diligent Reader may find useful.

\topic{Superderivatives of the fermionic field strength superfields}
Below are listed a half of the non-zero superderivatives of the spinorial
field strength superfields; the other half is implied by Hermitian
conjugation.

\ytem Type-A:
$\{\N_-,\bW_-\}=+\big[\N_\mm,\bAb\big]$,
$\{\N_+,\bW_-\}=\Bfb+\frc i4\big[\bAb,\bA\big]$,\\
$\{\N_+,\bW_+\}=-\big[\N_\pp,\bA\big]$,
$\{\N_-,\bW_+\}=\Bf+\frc i4\big[\bAb,\bA\big]$.
\ytem Type-B:
$\{\bN_-,\bW_-\}=-\big[\N_\mm,\bB\big]$,
$\{\N_+,\bW_-\}=\Bf+\frc i4\big[\bBb,\bB\big]$,\\
$\{\bN_+,\bW_+\}=+\big[\N_\pp,\bB\big]$,
$\{\N_-,\bW_+\}=\Bf-\frc i4\big[\bBb,\bB\big]$.
\ytem Type-C$_\mm$:
$\{\N_-,\bW_\Mm\}=+\frc i2\big[\bCb_\mm,\bC_\mm\big]$,
$\{\bN_+,\bW_\Mm\}=-\big[\N_\mm,\bC_\mm\big]$.
\ytem Type-C$_\mm$:
$\{\N_+,\bW^\mm_+\}=-\frc i2\big[\bCb_\pp,\bC_\pp\big]$,
$\{\bN_+,\bW^\mm_+\}=-\big[\N_\pp,\bC_\pp\big]$.

Upon the $\vs,\vsb=0$ projection, this provides (differential) relations
among the component fields of the gauge superfields. However, the
above (super)differential relations among the gauge superfields also hold
without any projection.

\topic{Component projections of matter superfields}
In sections~\sCmps\ and~\sLagD, we need the non-zero gauge-covariant
projections of a gauge-covariantly chiral superfield~\eHSF{a}, including for
completeness those listed in~\eCps{a}:\\
$\F|=\f$; ~$\N_\mp\F|=\sqrt2\j_\mp$;
~$[\N_-,\N_+]\F|=4F$,
~$\bN_-\N_+\F|=a\f$,
~$\bN_+\N_-\F|=\B{a}\f$,\\
~$[\N_-,\bN_-]\F|=-2i(\N_\mm\f)$,
~$[\N_+,\bN_+]\F|=-2i(\N_\pp\f)$;\\
$[\N_-,\bN_-]\N_+\F|=2\sqrt2(a\j_-+\a_-\f)-2\sqrt2i(\N_\mm\j_+)$,
~$\bN_+[\N_-,\bN_-]\F|=-\sqrt2\ab_-\f$,\\
$[\N_+,\bN_+]\N_-\F|=2\sqrt2(\B{a}\j_+-\B\ra_+\f)-2\sqrt2i(\N_\pp\j_-)$,
~$\bN_-[\N_+,\bN_+]\F|=+\sqrt2\ra_+\f$,\\
~$\N_+[\N_-,\bN_-]\F|=\sqrt2[\a_-\f-2i(\N_\mm\j_+)]$,
~$\N_-[\N_+,\bN_+]\F|=-\sqrt2[\B\ra_+\f+2i(\N_\pp\j_-)]$;\\
and finally, $\{[\N_-,\bN_-],[\N_+,\bN_+]\}\F|
=-8(\Box\f)+8\rD_\bA\f+4(\ab_-\j_+-\ra_+\j_-)$.\\
Note: since $\{\N_-,\N_+\}\F=\bBb\F=0$, then also $\N_-\N_+\F|=2F$ and
$\N_+\N_-\F|=-2F$.

The corresponding projections of a gauge-covariantly antichiral superfield
are obtained by Hermitian conjugation, using Eq.~\eTTr:\\
$\FB|=\fb$; ~$\bN_\mp\FB|=\sqrt2\jb_\mp$;
~$([\bN_-,\bN_+]\FB\T)\T|=4\Fb$,
~$(\N_-\bN_+\FB\T)\T|=\fb\B{a}$,
~$(\N_+\bN_-\FB\T)\T|=\fb a$,\\
~$([\bN_-,\N_-]\FB\T)\T|=-2i(\N_\mm\fb)$,
~$([\bN_+,\N_+]\FB\T)\T|=-2i(\N_\pp\fb)$;\\
$([\bN_-,\N_-]\bN_+\FB\T)\T|=2\sqrt2(\jb_-\B{a}+\fb\B\a_-)
 -2\sqrt2i(\N_\mm\jb_+)$,
~$(\N_+[\bN_-,\N_-]\FB\T)\T|=-\sqrt2\fb\a_-$,\\
$([\bN_+,\N_+]\bN_-\FB\T)\T|=2\sqrt2(\jb_+a-\fb\ra_+)
 -2\sqrt2i(\N_\pp\jb_-)$,
~$(\N_-[\bN_+,\N_+]\FB\T)\T|=+\sqrt2\fb\B\ra_+$,\\
~$(\bN_+[\bN_-,\N_-]\FB\T)\T|=\sqrt2[\fb\B\a_-{-}2i(\N_\mm\jb_+)]$,
~$(\bN_-[\bN_+,\N_+]\FB\T)\T|=-\sqrt2[\fb\ra_+{+}2i(\N_\pp\jb_-)]$;\\
and $\big(\{[\bN_-,\N_-],[\bN_+,\N_+]\}\FB\T\big)\T|
=-8(\Box\fb)+8\fb\rD_\bA+4(\jb_+\a_--\jb_-\B\ra_+)$.\\
As with $\F$, we also have $\bN_+\bN_-\FB|=2\Fb$ and
$\bN_-\bN_+\FB|=-2\Fb$.

The needed projections of a gauge-covariantly twisted-chiral superfield
are:\\
$\X|=x$; ~$\N_-\X|=\sqrt2\x_-$, ~$\bN_+\X|=\sqrt2\c_+$;
~$\N_+\N_-\X|=\B{b}x$,
~$\bN_-\bN_+\X|=bx$,\\
~$[\N_-,\bN_+]\X|=4X$,
~$[\N_-,\bN_-]\X|=-2i(\N_\mm x)$,
~$[\N_+,\bN_+]\X|=2i(\N_\pp x)$;\\
$[\N_-,\bN_-]\bN_+\X|=-2\sqrt2i(\N_\mm\c_+)+2\sqrt2(b\x_-+\b_-x)$,
~$\N_+[\N_-,\bN_-]\X|=-\sqrt2\bb_-x$,\\
$[\N_+,\bN_+]\N_-\X|=2\sqrt2i(\N_\pp\x_-)-2\sqrt2(\B{b}\c_+-\bb_+x)$,
~$\bN_-[\N_+,\bN_+]\X|=-\sqrt2\b_+x$,\\
~$\bN_+[\N_-,\bN_-]\X|=\sqrt2[\b_-x{-}2i(\N_\mm\c_+)]$,
~$\N_-[\N_+,\bN_+]\X|=\sqrt2[\bb_+x{+}2i(\N_\pp\x_-)]$;\\
and finally, $\{[\N_-,\bN_-],[\N_+,\bN_+]\}\X|
=+8(\Box x)-8\rD_\bB x+4(\b_+\x_--\b_-\c_+)$.\\
In analogy to $\F,\FB$, we also have $\N_-\bN_+\X|=2X$ and
$\bN_+\N_-\X|=-2X$.

The needed projections of a gauge-covariantly twisted-antichiral
superfield are:\\
$\XB|=\bx$; ~$\bN_-\XB|=\sqrt2\xb_-$, ~$\N_+\XB|=\sqrt2\cb_+$;
~$(\bN_+\bN_-\XB\T)\T|=\bx b$,
~$(\N_-\N_+\XB\T)\T|=\bx\B{b}$,\\
~$([\bN_-,\N_+]\XB\T)\T|=4\Xb$,
~$([\bN_-,\N_-]\XB\T)\T|=-2i(\N_\mm\bx)$,
~$([\bN_+,\N_+]\XB\T)\T|=2i(\N_\pp\bx)$;\\
$([\bN_-,\N_-]\N_+\XB\T)\T|=
  -2\sqrt2i(\N_\mm\cb_+)+2\sqrt2(\xb_-\B{b}+\bx\bb_-)$,
~$(\bN_+[\bN_-,\N_-]\XB\T)\T|=-\sqrt2\bx\b_-$,\\
$([\bN_+,\N_+]\bN_-\XB\T)\T|=
  2\sqrt2i(\N_\pp\xb_-)-2\sqrt2(\cb_+b-\bx\b_+)$,
~$(\N_-[\bN_+,\N_+]\XB\T)\T|=-\sqrt2\bx\bb_+$,\\
~$(\N_+[\bN_-,\N_-]\XB\T)\T|=\sqrt2[\bx\bb_-{-}2i(\N_\mm\cb_+)]$,
~$(\bN_-[\bN_+,\N_+]\XB\T)\T|=\sqrt2[\bx\b_+{+}2i(\N_\pp\xb_-)]$;\\
and $(\{[\bN_-,\N_-],[\bN_+,\N_+]\}\XB\T)\T|
=+8(\Box\bx)-8\bx\rD_\bB+4(\xb_-\bb_+-\cb_+\bb_-)$.\\
Again, we also have $\N_+\bN_-\XB|=2\Xb$ and $\bN_-\N_+\XB|=-2\Xb$.

The needed projections of a gauge-covariant lefton are:\\
$\L|=\ell$; ~$\N_+\L|=\sqrt2\l_+$, ~$\bN_+\L|=\sqrt2\bl_+$;
~$[\N_+,\bN_+]\L|=4L_\pp$,\\
~$\N_-\N_+\L|=\B{b}\ell$, ~$\bN_-\N_+\L|=a\ell$,
~$\N_-\bN_+\L|=\B{a}\ell$, ~$\bN_-\bN_+\L|=b\ell$;\\
$[\N_-,\bN_-]\N_+\L|=\sqrt2(\a_-+\bb_-)\ell$,
~$\N_-[\N_+,\bN_+]\L|=2\sqrt2(\B{b}\bl_+-\B{a}\l_+)
                                       +\sqrt2(\B{\ra}_+-\bb_+)\ell$,\\
$[\N_-,\bN_-]\bN_+\L|=\sqrt2(\ab_-+\b_-)\ell$,
~$\bN_-[\N_+,\bN_+]\L|=2\sqrt2(a\bl_+-b\l_+)+\sqrt2(\ra_+-\b_+)\ell$,\\
$\{[\N_-,\bN_-],[\N_+,\bN_+]\}\L|
=(\{a,\B{a}\}-\{b,\B{b}\})\ell+2[(\a_-+\bb_-)\bl_+-(\ab_-+\b_-)\l_+]$.

The needed projections of a gauge-covariant righton are:\\
$\Y|=r$; ~$\N_-\Y|=\sqrt2\r_-$, ~$\bN_-\Y|=\sqrt2\vr_-$;
~$[\N_-,\bN_-]\Y|=4R_\mm$,\\
~$\N_+\N_-\Y|=\B{b}r$, ~$\bN_+\N_-\Y|=\B{a}r$,
~$\N_+\bN_-\Y|=ar$, ~$\bN_+\bN_-\Y|=br$;\\
$[\N_+,\bN_+]\N_-\Y|=\sqrt2(\bb_+-\B{\ra}_+)r$,
~$\N_+[\N_-,\bN_-]\Y|=2\sqrt2(\B{b}\vr_--a\r_-)-\sqrt2(\a_-+\bb_-)r$,\\
$[\N_+,\bN_+]\bN_-\Y|=\sqrt2(\b_+-\ra_+)r$,
~$\bN_+[\N_-,\bN_-]\Y|=2\sqrt2(\B{a}\vr_--b\r_-)-\sqrt2(\B\a_-+\b_-)r$,\\
$\{[\N_-,\bN_-],[\N_+,\bN_+]\}\Y|
=(\{a,\B{a}\}-\{b,\B{b}\})r-2[(\B\ra_+-\bb_+)\vr_--(\ra_+-\b_+)\r_-]$.

When evaluating candidate Lagrangian density terms other then~\eCDI,
additional projections may be necessary. Given the above list, the
derivation of such additional projections should pose no significant problem
for the interested Reader.

 %
\vfill
\bigskip\noindent{\it Acknowledgments\/}:
 We are indebted to S.J.~Gates, Jr.,
 for helpful and encouraging discussions, and to the generous support
 of the Department of Energy through the grant DE-FG02-94ER-40854.

\vfill

\bigskip\listrefs

 %
\bye